\pdfoutput=1

\documentclass[12pt,a4paper]{article}

\usepackage{ifthen} 
\newboolean{pdflatex}
\setboolean{pdflatex}{true} 

\newboolean{articletitles}
\setboolean{articletitles}{true} 

\newboolean{uprightparticles}
\setboolean{uprightparticles}{false} 

\newboolean{inbibliography}
\setboolean{inbibliography}{false} 

\usepackage{bold-extra} 

\usepackage{microtype}
\usepackage{lineno}  
\usepackage{xspace} 
\usepackage{caption} 

\usepackage{graphicx}  
\usepackage{color}
\usepackage{colortbl}
\graphicspath{{./figs/}} 

\usepackage{amsmath} 
\usepackage{amssymb}
\usepackage{amsfonts}
\usepackage{upgreek} 

\newcommand*\patchAmsMathEnvironmentForLineno[1]{%
\expandafter\let\csname old#1\expandafter\endcsname\csname #1\endcsname
\expandafter\let\csname oldend#1\expandafter\endcsname\csname
end#1\endcsname
 \renewenvironment{#1}%
   {\linenomath\csname old#1\endcsname}%
   {\csname oldend#1\endcsname\endlinenomath}%
}
\newcommand*\patchBothAmsMathEnvironmentsForLineno[1]{%
  \patchAmsMathEnvironmentForLineno{#1}%
  \patchAmsMathEnvironmentForLineno{#1*}%
}
\AtBeginDocument{%
\patchBothAmsMathEnvironmentsForLineno{equation}%
\patchBothAmsMathEnvironmentsForLineno{align}%
\patchBothAmsMathEnvironmentsForLineno{flalign}%
\patchBothAmsMathEnvironmentsForLineno{alignat}%
\patchBothAmsMathEnvironmentsForLineno{gather}%
\patchBothAmsMathEnvironmentsForLineno{multline}%
\patchBothAmsMathEnvironmentsForLineno{eqnarray}%
}

\usepackage{hyperref}    
\usepackage[all]{hypcap} 

\def\babar  {\mbox{BaBar}\xspace}

\def\MagUp {\mbox{\em Mag\kern -0.05em Up}\xspace}

\ifthenelse{\boolean{uprightparticles}}%
{

 \def\Peta        {\ensuremath{\upeta}\xspace}

 \def\Ppi         {\ensuremath{\uppi}\xspace}

 \def\Pchi        {\ensuremath{\upchi}\xspace}                 
 \def\Ppsi        {\ensuremath{\uppsi}\xspace}

 \def\PDelta      {\ensuremath{\Delta}\xspace}                 
 \def\PXi      {\ensuremath{\Xi}\xspace}                 
 \def\PLambda      {\ensuremath{\Lambda}\xspace}                 
 \def\PSigma      {\ensuremath{\Sigma}\xspace}                 
 \def\POmega      {\ensuremath{\Omega}\xspace}                 
 \def\PUpsilon      {\ensuremath{\Upsilon}\xspace}                 
 
 \def\PB      {\ensuremath{\mathrm{B}}\xspace}                 
                  
 \def\PD      {\ensuremath{\mathrm{D}}\xspace}                 
                  
 \def\PF      {\ensuremath{\mathrm{F}}\xspace}

 \def\PJ      {\ensuremath{\mathrm{J}}\xspace}                 
 \def\PK      {\ensuremath{\mathrm{K}}\xspace}

 \def\PP      {\ensuremath{\mathrm{P}}\xspace}

 \def\Pb      {\ensuremath{\mathrm{b}}\xspace}                 
 \def\Pc      {\ensuremath{\mathrm{c}}\xspace}

 \def\Pi      {\ensuremath{\mathrm{i}}\xspace}

 \def\Pp      {\ensuremath{\mathrm{p}}\xspace}

 \def\Ps      {\ensuremath{\mathrm{s}}\xspace}

}
{

 \def\Peta        {\ensuremath{\eta}\xspace}

 \def\Ppi         {\ensuremath{\pi}\xspace}

 \def\Pchi        {\ensuremath{\chi}\xspace}                 
 \def\Ppsi        {\ensuremath{\psi}\xspace}                 
                  
 \mathchardef\PDelta="7101
 \mathchardef\PXi="7104
 \mathchardef\PLambda="7103
 \mathchardef\PSigma="7106
 \mathchardef\POmega="710A
 \mathchardef\PUpsilon="7107
                  
 \def\PB      {\ensuremath{B}\xspace}                 
                  
 \def\PD      {\ensuremath{D}\xspace}                 
                  
 \def\PF      {\ensuremath{F}\xspace}

 \def\PJ      {\ensuremath{J}\xspace}                 
 \def\PK      {\ensuremath{K}\xspace}

 \def\PP      {\ensuremath{P}\xspace}

 \def\Pb      {\ensuremath{b}\xspace}                 
 \def\Pc      {\ensuremath{c}\xspace}

 \def\Pi      {\ensuremath{i}\xspace}

 \def\Pp      {\ensuremath{p}\xspace}

 \def\Ps      {\ensuremath{s}\xspace}

}

\makeatletter
\ifcase \@ptsize \relax
  \newcommand{\miniscule}{\@setfontsize\miniscule{4}{5}}
\or
  \newcommand{\miniscule}{\@setfontsize\miniscule{5}{6}}
\or
  \newcommand{\miniscule}{\@setfontsize\miniscule{5}{6}}
\fi
\makeatother

\DeclareRobustCommand{\optbar}[1]{\shortstack{{\miniscule (\rule[.5ex]{1.25em}{.18mm})}
  \\ [-.7ex] $#1$}}

\def\squark    {{\ensuremath{\Ps}}\xspace}

\def\cquark    {{\ensuremath{\Pc}}\xspace}

\def\bquark    {{\ensuremath{\Pb}}\xspace}

\def\pion   {{\ensuremath{\Ppi}}\xspace}
\def\piz    {{\ensuremath{\pion^0}}\xspace}

\def\pip    {{\ensuremath{\pion^+}}\xspace}
\def\pim    {{\ensuremath{\pion^-}}\xspace}
\def\pipm   {{\ensuremath{\pion^\pm}}\xspace}
\def\pimp   {{\ensuremath{\pion^\mp}}\xspace}

\def\kaon    {{\ensuremath{\PK}}\xspace}
  \def\Kbar    {{\kern 0.2em\overline{\kern -0.2em \PK}{}}\xspace}

\def\KorKbar    {\kern 0.18em\optbar{\kern -0.18em K}{}\xspace}
\def\Kz      {{\ensuremath{\kaon^0}}\xspace}

\def\Kp      {{\ensuremath{\kaon^+}}\xspace}
\def\Km      {{\ensuremath{\kaon^-}}\xspace}
\def\Kpm     {{\ensuremath{\kaon^\pm}}\xspace}
\def\Kmp     {{\ensuremath{\kaon^\mp}}\xspace}
\def\KS      {{\ensuremath{\kaon^0_{\rm\scriptscriptstyle S}}}\xspace}
\def\KL      {{\ensuremath{\kaon^0_{\rm\scriptscriptstyle L}}}\xspace}

\def\Kstar   {{\ensuremath{\kaon^*}}\xspace}
\def\Kstarb  {{\ensuremath{\Kbar{}^*}}\xspace}

\newcommand{\etaz}{\ensuremath{\Peta}\xspace}
\newcommand{\etapr}{\ensuremath{\Peta^{\prime}}\xspace}

  \def\Dbar    {{\kern 0.2em\overline{\kern -0.2em \PD}{}}\xspace}
\def\D       {{\ensuremath{\PD}}\xspace}

\def\DorDbar    {\kern 0.18em\optbar{\kern -0.18em D}{}\xspace}
\def\Dz      {{\ensuremath{\D^0}}\xspace}
\def\Dzb     {{\ensuremath{\Dbar{}^0}}\xspace}
\def\Dp      {{\ensuremath{\D^+}}\xspace}
\def\Dm      {{\ensuremath{\D^-}}\xspace}

\def\Dstar   {{\ensuremath{\D^*}}\xspace}

\def\Ds      {{\ensuremath{\D^+_\squark}}\xspace}
\def\Dsp     {{\ensuremath{\D^+_\squark}}\xspace}
\def\Dsm     {{\ensuremath{\D^-_\squark}}\xspace}

\def\B       {{\ensuremath{\PB}}\xspace}
\def\Bbar    {{\ensuremath{\kern 0.18em\overline{\kern -0.18em \PB}{}}}\xspace}

\def\BorBbar    {\kern 0.18em\optbar{\kern -0.18em B}{}\xspace}
\def\Bz      {{\ensuremath{\B^0}}\xspace}
\def\Bzb     {{\ensuremath{\Bbar{}^0}}\xspace}
\def\Bu      {{\ensuremath{\B^+}}\xspace}
\def\Bub     {{\ensuremath{\B^-}}\xspace}
\def\Bp      {{\ensuremath{\Bu}}\xspace}
\def\Bm      {{\ensuremath{\Bub}}\xspace}
\def\Bpm     {{\ensuremath{\B^\pm}}\xspace}

\def\Bd      {{\ensuremath{\B^0}}\xspace}
\def\Bs      {{\ensuremath{\B^0_\squark}}\xspace}
\def\Bsb     {{\ensuremath{\Bbar{}^0_\squark}}\xspace}
\def\Bdb     {{\ensuremath{\Bbar{}^0}}\xspace}

\def\jpsi     {{\ensuremath{{\PJ\mskip -3mu/\mskip -2mu\Ppsi\mskip 2mu}}}\xspace}

\def\chiczero {{\ensuremath{\Pchi_{\cquark 0}}}\xspace}

  \def\Y#1S{\ensuremath{\PUpsilon{(#1S)}}\xspace}

\def\proton      {{\ensuremath{\Pp}}\xspace}

\def\Xires       {{\ensuremath{\PXi}}\xspace}

\def\Lz          {{\ensuremath{\PLambda}}\xspace}
\def\Lbar        {{\ensuremath{\kern 0.1em\overline{\kern -0.1em\PLambda}}}\xspace}
\def\LorLbar    {\kern 0.18em\optbar{\kern -0.18em \PLambda}{}\xspace}

\def\Lb      {{\ensuremath{\Lz^0_\bquark}}\xspace}

\def\Xibm    {{\ensuremath{\Xires^-_\bquark}}\xspace}

\def\ra                 {\ensuremath{\rightarrow}\xspace}
\def\to                 {\ensuremath{\rightarrow}\xspace}

\def\CP                {{\ensuremath{C\!P}}\xspace}

\def\AT#1     {\ensuremath{A_{\mathrm{T}}^{#1}}\xspace}           

\def\C#1      {\ensuremath{\mathcal{C}_{#1}}\xspace}                       
\def\Cp#1     {\ensuremath{\mathcal{C}_{#1}^{'}}\xspace}                    
\def\Ceff#1   {\ensuremath{\mathcal{C}_{#1}^{\mathrm{(eff)}}}\xspace}        
\def\Cpeff#1  {\ensuremath{\mathcal{C}_{#1}^{'\mathrm{(eff)}}}\xspace}       
\def\Ope#1    {\ensuremath{\mathcal{O}_{#1}}\xspace}                       
\def\Opep#1   {\ensuremath{\mathcal{O}_{#1}^{'}}\xspace}                    

\def\Fbar    {{\kern 0.2em\overline{\kern -0.2em \PF}{}}\xspace}
\def\Pbar    {{\kern 0.18em\overline{\kern -0.18em \PP}{}}\xspace}

\newcommand{\tev}{\ifthenelse{\boolean{inbibliography}}{\ensuremath{~T\kern -0.05em eV}\xspace}{\ensuremath{\mathrm{\,Te\kern -0.1em V}}}\xspace}
\newcommand{\gev}{\ensuremath{\mathrm{\,Ge\kern -0.1em V}}\xspace}
\newcommand{\nbspgev}{\ensuremath{\mathrm{Ge\kern -0.1em V}}\xspace}
\newcommand{\mev}{\ensuremath{\mathrm{\,Me\kern -0.1em V}}\xspace}
\newcommand{\kev}{\ensuremath{\mathrm{\,ke\kern -0.1em V}}\xspace}
\newcommand{\ev}{\ensuremath{\mathrm{\,e\kern -0.1em V}}\xspace}
\newcommand{\gevc}{\ensuremath{{\mathrm{\,Ge\kern -0.1em V\!/}c}}\xspace}
\newcommand{\nbspgevc}{\ensuremath{{\mathrm{Ge\kern -0.1em V\!/}c}}\xspace}
\newcommand{\mevc}{\ensuremath{{\mathrm{\,Me\kern -0.1em V\!/}c}}\xspace}
\newcommand{\gevcc}{\ensuremath{{\mathrm{\,Ge\kern -0.1em V\!/}c^2}}\xspace}
\newcommand{\nbspgevcc}{\ensuremath{{\mathrm{Ge\kern -0.1em V\!/}c^2}}\xspace}
\newcommand{\gevgevcccc}{\ensuremath{{\mathrm{\,Ge\kern -0.1em V^2\!/}c^4}}\xspace}
\newcommand{\nbspgevgevcccc}{\ensuremath{{\mathrm{Ge\kern -0.1em V^2\!/}c^4}}\xspace}
\newcommand{\mevcc}{\ensuremath{{\mathrm{\,Me\kern -0.1em V\!/}c^2}}\xspace}

\def\fm   {\ensuremath{\rm \,fm}\xspace}

\def\sec  {\ensuremath{\rm {\,s}}\xspace}

\def\ghz  {\ensuremath{{\rm \,GHz}}\xspace}

\def\gsim{{~\raise.15em\hbox{$>$}\kern-.85em
          \lower.35em\hbox{$\sim$}~}\xspace}
\def\lsim{{~\raise.15em\hbox{$<$}\kern-.85em
          \lower.35em\hbox{$\sim$}~}\xspace}

\def\sPlot{\mbox{\em sPlot}\xspace}

\def\evtgen     {\mbox{\textsc{EvtGen}}\xspace}

\def\minuit     {\mbox{\textsc{Minuit}}\xspace}

\def\root       {\mbox{\textsc{Root}}\xspace}

\def\laura       {\mbox{\textsc{Laura$^{++}$}}\xspace}

\def\cpp        {\mbox{\textsc{C\raisebox{0.1em}{{\footnotesize{++}}}}}\xspace}

\def\gbsps      {\ensuremath{{\rm \,Gbits/s}}\xspace}
\def\gbytes     {\ensuremath{{\rm \,Gbytes}}\xspace}

\def\tell1  {TELL1\xspace}
\def\ukl1   {UKL1\xspace}

\newcommand{\eg}{\mbox{\itshape e.g.}\xspace}
\newcommand{\ie}{\mbox{\itshape i.e.}\xspace}

\def\mpr        {\ensuremath{m^\prime}\xspace}
\def\thpr       {\ensuremath{\theta^\prime}\xspace}
\def\Kstarbsubz  {{\ensuremath{\Kbar{}^*_0}}\xspace}
\def\jfit        {\mbox{\textbf{\textit{J}}\hspace{-0.15em}\textsc{fit}}\xspace}

\usepackage{tikz}
\usetikzlibrary{shapes,arrows}

\usepackage{listings}
\usepackage{array}
\newcolumntype{L}[1]{>{\raggedright\let\newline\\\arraybackslash\hspace{0pt}}p{#1}}
\newcolumntype{C}[1]{>{\centering\let\newline\\\arraybackslash\hspace{0pt}}p{#1}}
\newcolumntype{R}[1]{>{\raggedleft\let\newline\\\arraybackslash\hspace{0pt}}p{#1}}

\usepackage{cite} 
\usepackage{mciteplus}

\begin{document}

\renewcommand{\thefootnote}{\fnsymbol{footnote}}
\setcounter{footnote}{1}

\begin{titlepage}
\pagenumbering{roman}

\vspace*{0.5cm}

\begin{center}
  \textbf{\huge 
    \laura\ : a Dalitz plot fitter
  }
\end{center}

\vspace*{\fill}

\begin{center}
John Back,
Tim Gershon,
Paul Harrison,
Thomas Latham,\footnote{Corresponding author, email \href{mailto:T.Latham@warwick.ac.uk}{T.Latham@warwick.ac.uk}}\\
Daniel O'Hanlon,\footnote{Now at Sezione INFN di Bologna, Bologna, Italy}
Wenbin Qian\\
Department of Physics, University of Warwick, Coventry, United Kingdom \\

\vspace*{0.3cm}
Pablo del~Amo~Sanchez\\
LAPP, Universit\'{e} Savoie Mont-Blanc, CNRS/IN2P3, Annecy-Le-Vieux, France\\

\vspace*{0.3cm}
Daniel Craik\\
Massachusetts Institute of Technology, Cambridge, United States of America

\vspace*{0.3cm}
Jelena Ilic\\
STFC Rutherford Appleton Laboratory, Didcot, United Kingdom\\

\vspace*{0.3cm}
Juan~Martin Otalora~Goicochea\\
Universidade Federal do Rio de Janeiro (UFRJ), Rio de Janeiro, Brazil\\

\vspace*{0.3cm}
Eugenia Puccio\\
Stanford University, Stanford, United States of America \\

\vspace*{0.3cm}
Rafael Silva~Coutinho\\
Physik-Institut, Universit{\"a}t Z{\"u}rich, Z{\"u}rich, Switzerland

\vspace*{0.3cm}
Mark Whitehead\footnote{Now at I. Physikalisches Institut, RWTH Aachen University, Aachen, Germany}\\
European Organization for Nuclear Research (CERN), Geneva, Switzerland

\end{center}

\begin{center}
\today
\end{center}

\vspace*{\fill}

\begin{abstract}
  \noindent
  The Dalitz plot analysis technique has become an increasingly important method in heavy flavour physics. 
  The \laura\ fitter has been developed as a flexible tool that can be used for Dalitz plot analyses in different experimental environments.
  Explicitly designed for three-body decays of heavy-flavoured mesons to spinless final state particles, it is optimised in order to describe all possible resonant or nonresonant contributions, and to accommodate possible \CP violation effects.
\end{abstract}

\vspace*{\fill}

\end{titlepage}


\renewcommand{\thefootnote}{\arabic{footnote}}
\setcounter{footnote}{0}

\tableofcontents

\clearpage


\pagestyle{plain} 
\setcounter{page}{1}
\pagenumbering{arabic}

\section{Introduction}
\label{sec:introduction}

Decays of unstable heavy particles to multibody final states can in general occur through several different intermediate resonances.  
Each decay channel can be represented quantum-mechanically by an amplitude, and the total density of decays across the phase space is represented by the square of the coherent sum of all contributing amplitudes.
Interference effects can lead to excesses or deficits of decays in regions of phase space where different resonances overlap.
Investigations of such dynamical effects in multibody decays are of great interest to test the Standard Model of particle physics and to investigate resonant structures. 

The Dalitz plot (DP)~\cite{Dalitz:1953cp,Fabri:1954zz} was introduced originally to describe the phase space of $\KL \to \pi\pi\pi$ decays, but is relevant for the decay of any spin-zero particle to three spin-zero particles, $P \to d_{1} d_{2} d_{3}$.
In such a case, energy and momentum conservation give
\begin{equation}
  m_P^2 + m_{d_{1}}^2 + m_{d_{2}}^2 + m_{d_{3}}^2 = 
  m^2(d_{1}d_{2}) + m^2(d_{2}d_{3}) + m^2(d_{3}d_{1}) \, ,
\end{equation}
where $m(d_id_j)$ is the invariant mass obtained from the two-body combination of the $d_i$ and $d_j$ four momenta.
Consequently, assuming that the masses of $P$, $d_1$, $d_2$ and $d_3$ are all known, any two of the $m^2(d_id_j)$ values --- subsequently referred to as Dalitz-plot variables --- are sufficient to describe fully the kinematics of the decay in the $P$ rest frame.
This can also be shown by considering that the 12 degrees of freedom corresponding to the four-momenta of the three final-state particles are accounted for by two DP variables, the three $d_i$ masses, four constraints due to energy--momentum conservation in the $P \to d_{1} d_{2} d_{3}$ decay, and three co-ordinates describing a direction in space which carries no physical information about the decay since all particles involved have zero spin.

\begin{figure}[!htb]
  \centering
  \includegraphics[width=0.49\textwidth]{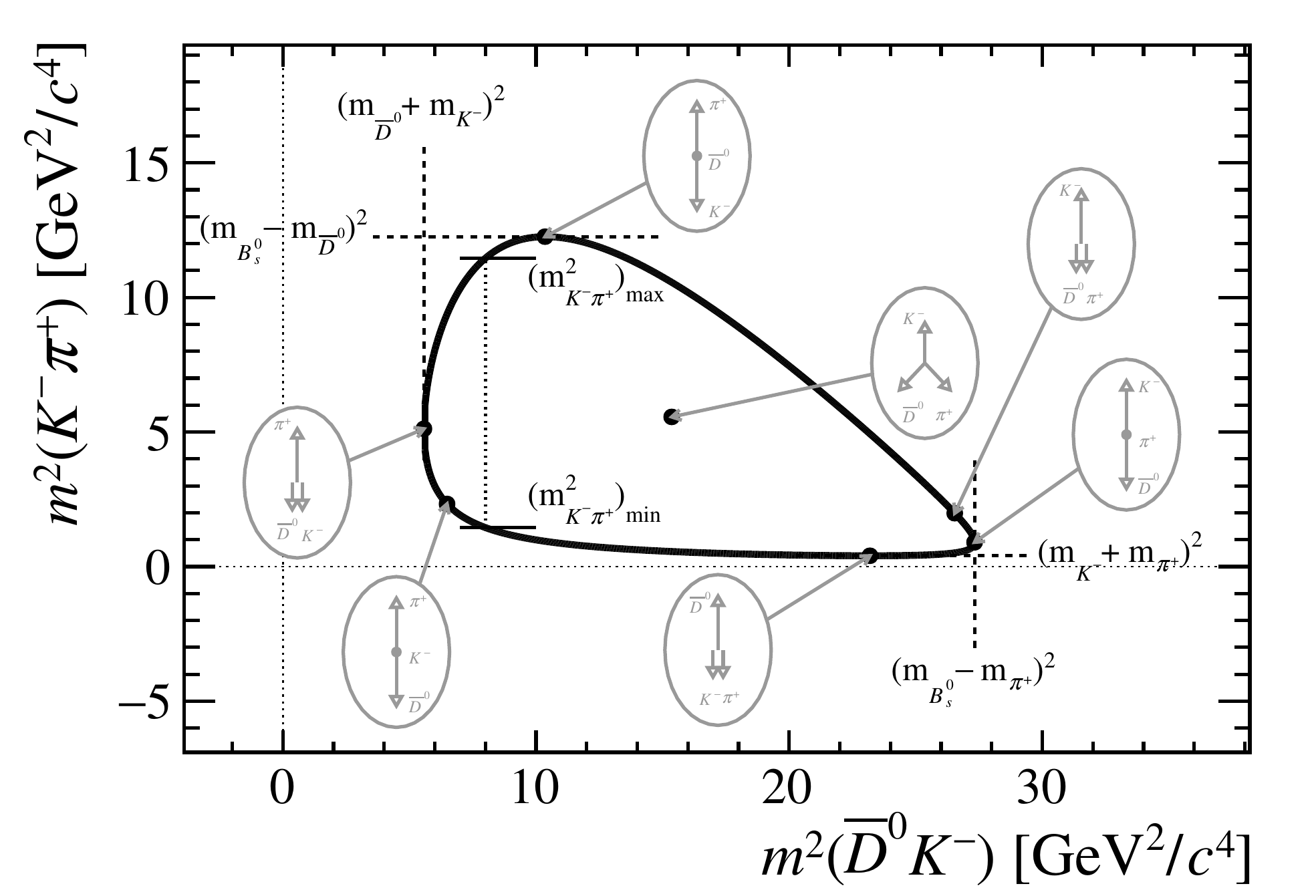}
  \includegraphics[width=0.49\textwidth]{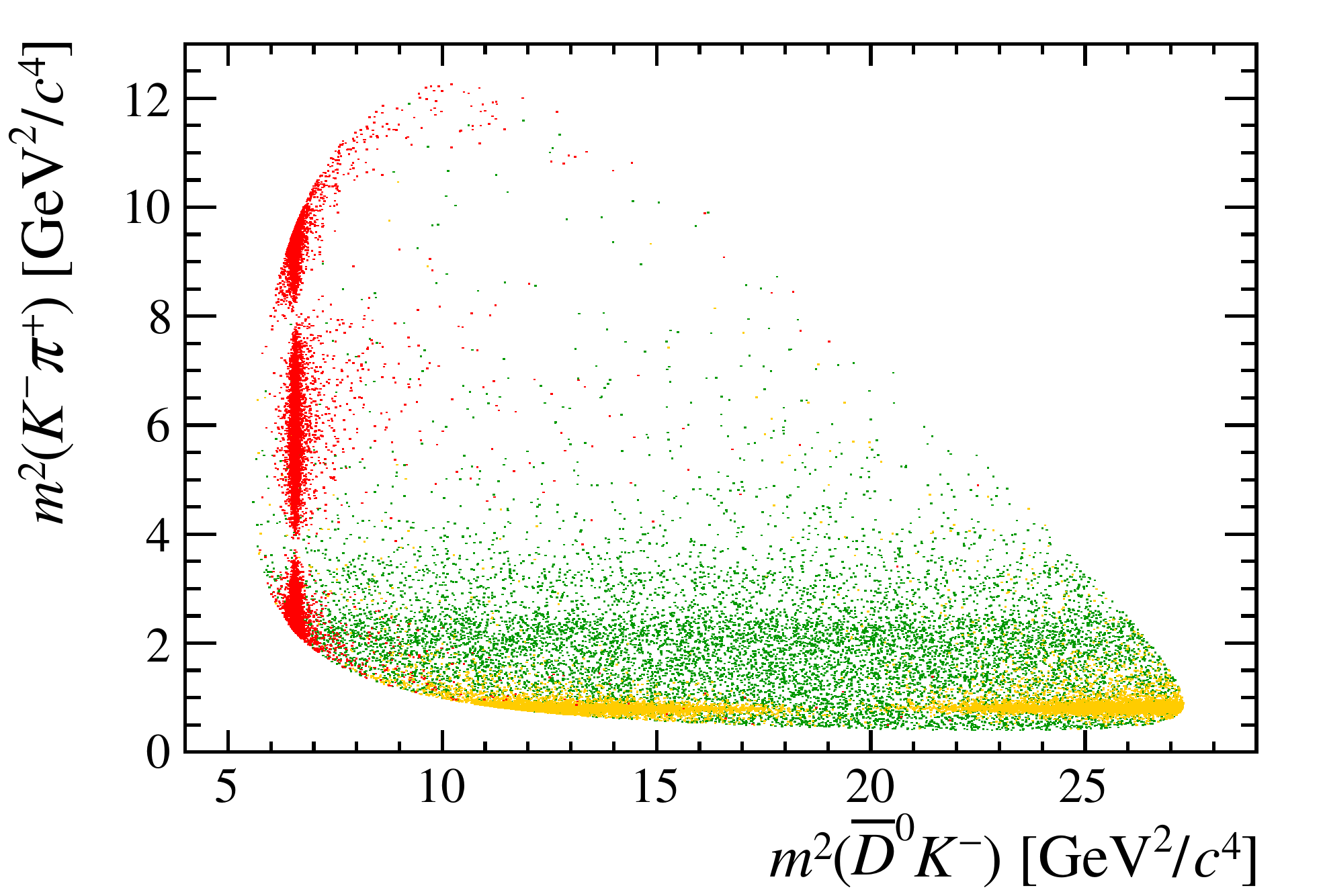}\\
  \includegraphics[width=0.32\textwidth]{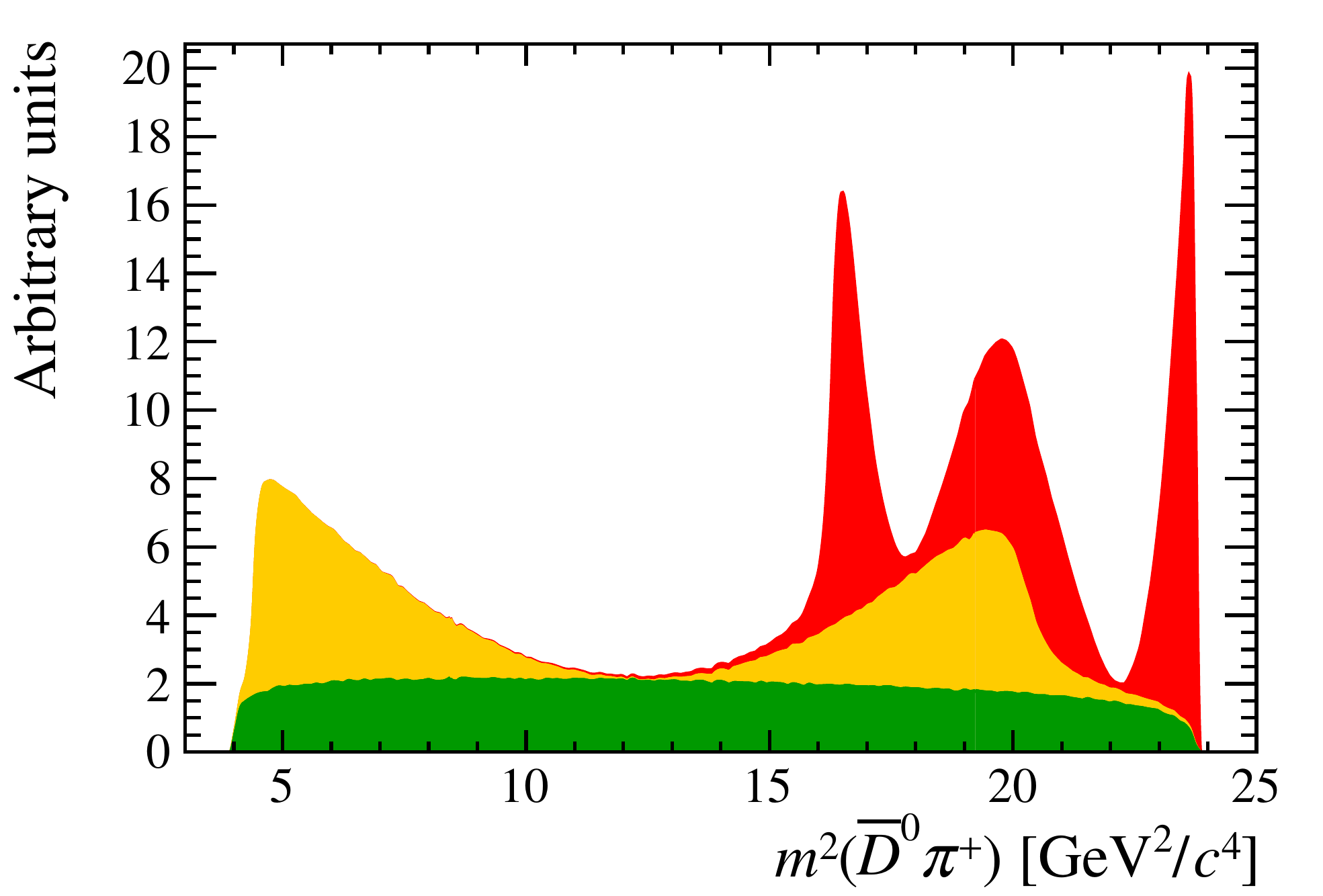}
  \includegraphics[width=0.32\textwidth]{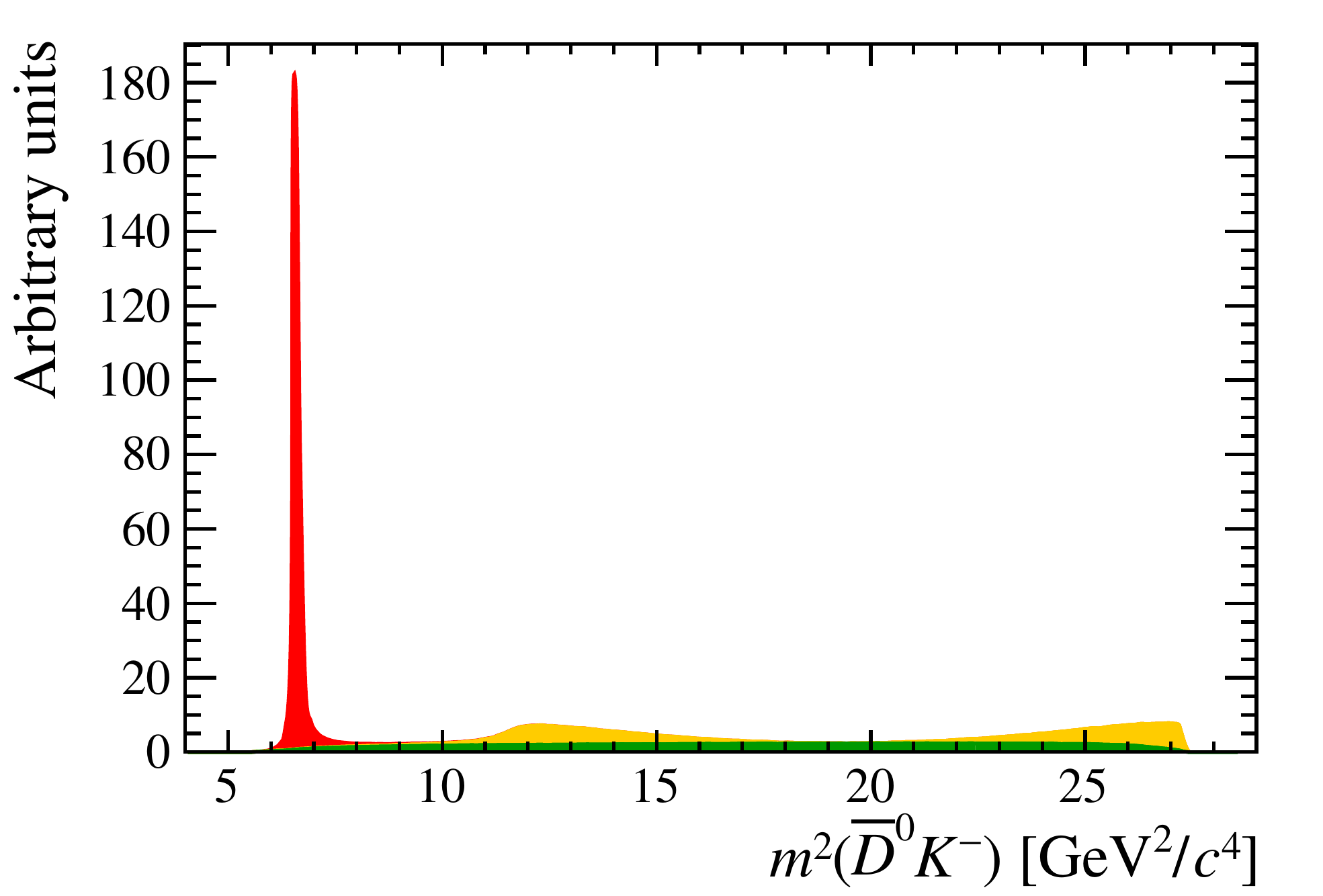}
  \includegraphics[width=0.32\textwidth]{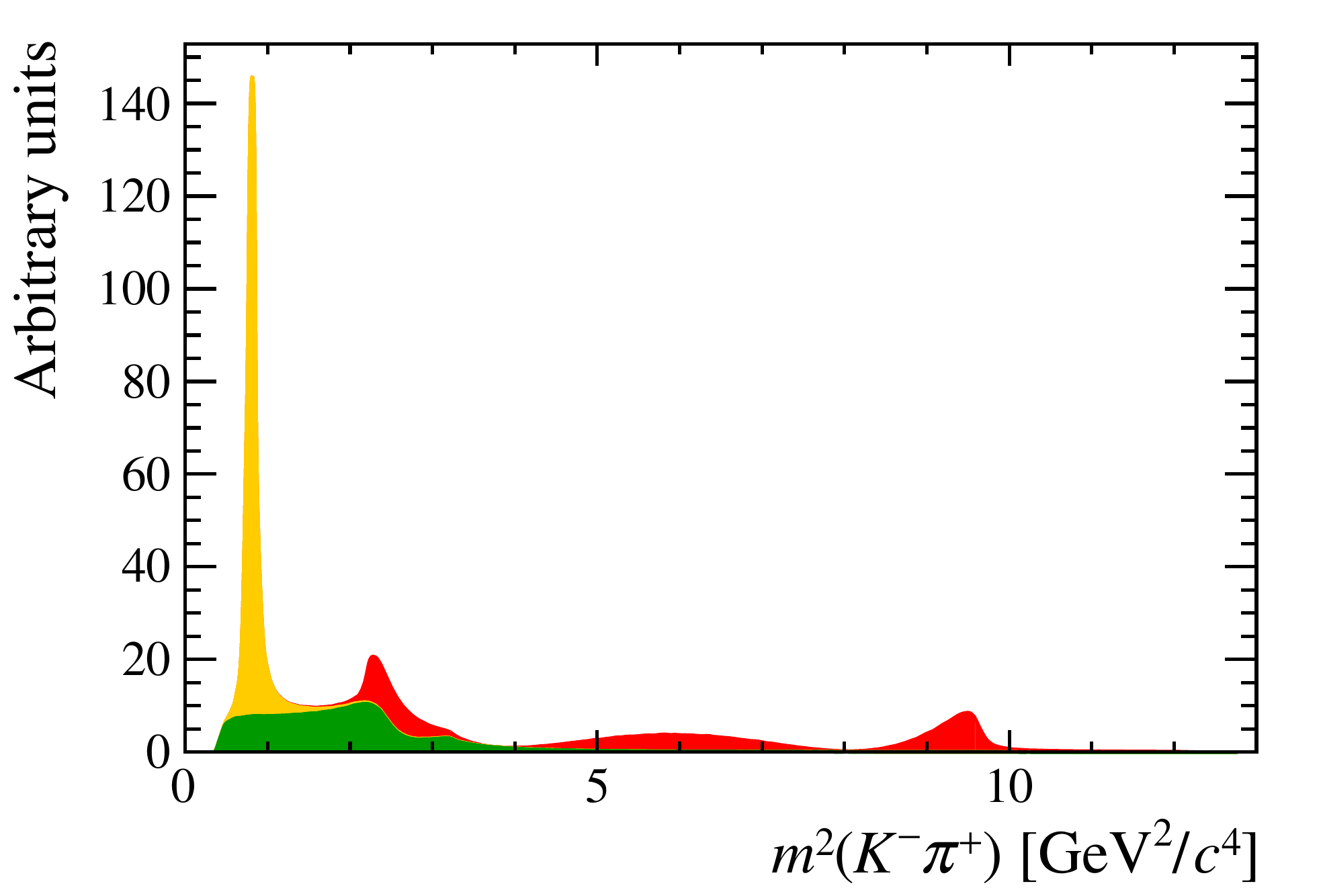}
  \caption{
    (Top left) kinematic boundaries of the three-body phase space for the decay $\Bs\to\Dzb\Km\pip$.
    The insets indicate the configuration of the final-state particle momenta in the parent rest frame at various different DP positions.
    (Top right) examples of the resonances which may appear in the Dalitz
    plot for this decay: (red) $\D^*_{s2}(2573)^-$, (orange) $\kaon^*(892)^0$,
    (green) $\kaon\pion$ S-wave.
    (Bottom) projections of this DP onto the squares of the invariant
    masses (from left to right): $m^2_{\Dzb\pip}$, $m^2_{\Dzb\Km}$, $m^2_{\Km\pip}$. 
  }
  \label{fig:dpexample}
\end{figure}

A Dalitz plot is then the visualisation of the phase space of a particular three-body decay in terms of the two DP variables.\footnote{
  The phrase ``Dalitz plot'' is often used more broadly in the literature.
  In particular, it can be used to describe the projection onto two of the two-body invariant mass combinations of a three-body decay even when one or more of the particles involved has non-zero spin.
}
Analysis of the distribution of decays across a DP can reveal information about the underlying dynamics of the particular three-body decay, since the differential rate is
\begin{equation}
  d\Gamma = \frac{1}{(2\pi)^3}\frac{1}{32\,m_P^3}\left| {\cal A} \right|^2 dm^2(d_{1}d_{3}) \, dm^2(d_{2}d_{3}) \, ,
\end{equation}
where ${\cal A}$ is the amplitude for the three-body decay.
Thus, any deviation from a uniform distribution is due to the dynamical structure of the amplitude.
Examples of the kinematic boundaries of a DP, and of resonant structures that may appear in this kind of decay, are shown in Fig.~\ref{fig:dpexample}.

The Dalitz-plot analysis technique, usually implemented with model-dependent descriptions of the amplitudes involved, has been used to understand hadronic effects in, for example, the $\pi^0\pi^0\pi^0$ system produced in $p\bar{p}$ annihilation~\cite{Amsler:1995bf}.
Recently, it has also been used to study three-body $\eta_c$ decays~\cite{Lees:2014iua,Lees:2015zzr}.
However, DP analyses have become particularly popular to study multibody decays of the heavy-flavoured $D$ and $B$ mesons.
Not only do the relatively large masses of these particles provide a broad kinematic range in which resonant structures can be studied but, since the decays are mediated by the weak interaction, there may be \CP-violating differences between the DP distributions for particle and antiparticle. 
Studying these differences can test the Standard Model mechanism for \CP violation: if the asymmetries are not consistent with originating from the single complex phase in the Cabibbo-Kobayashi-Maskawa (CKM) quark mixing matrix~\cite{Cabibbo:1963yz,Kobayashi:1973fv} then contributions beyond the Standard Model must be present.

Until around the year 2000, most DP analyses of charm decays were focussed on understanding hadronic structures at low $\pi\pi$ or $K\pi$ mass.
In particular, pioneering analyses of $D \to K\pi\pi$ decays were carried out by experiments such as MARK-II, MARK-III, E687, ARGUS, E691 and CLEO~\cite{Schindler:1980ws,Adler:1987sd,Anjos:1992kb,Albrecht:1993jn,Frabetti:1994di,Kopp:2000gv}.
These analyses revealed the existence of a broad structure in the $K\pi$ S-wave that could not be well described with a Breit--Wigner lineshape.  
In later analyses, it was shown that this contribution could be modelled in a quasi-model-independent way, in which the partial wave is fitted using splines to describe the magnitude and phase as a function of $m(K\pi)$~\cite{Aitala:2005yh}. 
Subsequent uses of this approach include further studies of the $K\pi$ S-wave~\cite{Bonvicini:2008jw,Link:2009ng,delAmoSanchez:2010fd,Lees:2015zzr} as well as the $\Kp\Km$~\cite{Aubert:2007dc} and $\pip\pim$~\cite{Aubert:2008ao} S-wave{s}, in various processes.
Similarly, DP analyses of decays such as $\Dp \to \pip\pip\pim$~\cite{Aitala:2000xu,Muramatsu:2002jp,Link:2003gb,Bonvicini:2007tc} indicated the existence of a broad low-mass $\pi\pi$ S-wave known as the $\sigma$ pole~\cite{Pelaez:2015qba}.

With the advent of the $e^+e^-$ \B-factory experiments, \babar~\cite{Aubert:2001tu,TheBABAR:2013jta} and Belle~\cite{Abashian:2000cg}, DP analyses of \B meson decays became feasible.  
The method was used to obtain insights into charm resonances through analyses of $\Bp\to \Dm\pip\pip$~\cite{Abe:2003zm,Aubert:2009wg} and $\Bz \to \Dzb\pip\pim$~\cite{Kuzmin:2006mw,delAmoSanchez:2010ad} decays.
Studies of \B meson decays to final states without any charm or charmonium particles also became possible~\cite{Garmash:2004wa,Aubert:2005sk,Aubert:2005ce}.
Once baseline DP models were established, it was then possible to search for \CP violation effects, with results including the first evidence for \CP violation in the $\Bp \to \rho(770)^0 \Kp$ decay~\cite{Garmash:2005rv,Aubert:2008bj}.
Moreover, analyses that accounted for possible dependence of the \CP violation effect with decay time as well as with DP position were carried out for both $D$~\cite{Asner:2005sz,Abe:2007rd} and $B$ decays~\cite{Aubert:2007jn,Kusaka:2007dv,Kusaka:2007mj,Aubert:2007sd,:2008wwa,Aubert:2009me,Nakahama:2010nj,Lees:2013nwa}.

With the availability of increasingly large data samples at these experiments and, more recently, at the Large Hadron Collider experiments (in particular, LHCb~\cite{Alves:2008zz}), more detailed studies of these and similar decays become possible.  
In addition, many ideas for DP analyses have been proposed, since they provide interesting possibilities to provide insight into hadronic structures, to measure \CP violation effects and to test the Standard Model.
These include methods to determine the angles $\alpha$, $\beta$ and $\gamma$ of the CKM Unitarity Triangle with low theoretical uncertainty from, respectively $\Bz \to \pip\pim\piz$~\cite{Snyder:1993mx}, $\Bz \to D\pip\pim$~\cite{Charles:1998vf,Latham:2008zs} and $\Bz \to D\Kp\pim$ decays~\cite{Gershon:2008pe,Gershon:2009qc}, among many other potential analyses.

Thus, it has become increasingly important to have a publicly available Dalitz-plot analysis package that is flexible enough both to be used in a range of experimental environments and to describe many possible different decays and types of analyses.
Such a package should be well validated and have excellent performance characteristics, in particular in terms of speed since complicated amplitude fits can otherwise have unacceptable CPU requirements.
This motivated the creation, and ongoing development, of the \laura\ package, which is described in the remainder of the paper.
\laura\ is written in the \cpp\ programming language and is intended to be as close as possible to being a standalone package, with a sole external dependency on the \root\ package~\cite{Brun:1997pa}.
In particular, \root\ is used to handle data file input/output, histogrammed quantities, and the minimisation of negative log-likelihood functions with \minuit~\cite{James:1975dr}.
Further documentation and code releases (distributed under the Apache
License, Version 2.0~\cite{apache-license}) are available from
\begin{center}
  \href{http://laura.hepforge.org/}{\tt http://laura.hepforge.org/} \, .
\end{center}
The description of the software given in this paper corresponds to that
released in \laura\ version {\tt v3r4}.

In Sec.~\ref{sec:DalitzGeneralities}, a brief summary of the Dalitz-plot analysis formalism is given, and the conventions used in \laura\ are set out.
Section~\ref{sec:expt-effects} describes effects that must also be taken into account when performing an experimental analysis.
Sections~\ref{sec:signal},~\ref{sec:eff-resol} and~\ref{sec:backgrounds} then contain discussions of, respectively, the implementation of the signal model, efficiency and resolution effects, and the background components in \laura, including explicit classes and methods with high-level details given in Appendices.
These elements are then put together in Sec.~\ref{sec:workflow}, where the overall work flow in \laura\ is described.  
The performance of the software is discussed in Sec.~\ref{sec:performance}, ongoing and planned future developments are briefly mentioned in Sec.~\ref{sec:future-devel}, and a summary is given in Sec.~\ref{sec:summary}.

\section{Dalitz-plot analysis formalism}
\label{sec:DalitzGeneralities}

Given two variables that describe the Dalitz plot of the $P \to d_{1} d_{2} d_{3}$ decay, all other kinematic quantities can be uniquely determined for fixed initial- and final-state (subsequently referred to as parent and daughter) particle masses.
The convention adopted in \laura is that the DP is described in terms of $m^2_{13} \equiv m^2(d_{1}d_{3})$ and $m^2_{23} \equiv m^2(d_{2}d_{3})$.
Hence, these two variables are required to be present in any input data provided to \laura.

The description of the complex amplitude is based on the isobar model~\cite{Fleming:1964zz,Morgan:1968zza,Herndon:1973yn}, which describes the total amplitude as a coherent sum of $N$ amplitudes from resonant or nonresonant intermediate processes.\footnote{
  Alternative descriptions of parts of the amplitude model are also available in \laura, and are discussed in later sections.}
This means that the total amplitude is given by
\begin{equation}
\label{eq:amp}
{\cal A}\left(m^2_{13}, m^2_{23}\right) = \sum_{j=1}^{N} c_j F_j\left(m^2_{13}, m^2_{23}\right) \,,
\end{equation}
where $c_j$ are complex coefficients, discussed further in Sec.~\ref{sec:sigmodel}, giving the relative contribution of decay channel $j$.
There are several different choices used in the literature to express the resonance dynamics contained within the $F_j\left(m^2_{13},m^2_{23}\right)$ terms.
Here, one common approach, which is the default in \laura, is outlined; other possibilities are discussed in Appendices~\ref{sec:res-formulae} and~\ref{sec:angular-formulae}.
For a resonance in $m_{13}$, the dynamics can be written as 
\begin{equation}
  \label{eq:ResDynEqn}
  F\left(m^2_{13}, m^2_{23}\right) = {\cal N} \times
  R\left(m_{13}\right) \times T(\vec{p},\vec{q}) \times X(|\vec{p}|\,r^P_{\rm BW}) \times X(|\vec{q}|\,r^R_{\rm BW}) \, ,
\end{equation}
where the functions $R$ and $T$ describe the invariant mass and angular dependence of the amplitude, the $X$ functions are form factors, and ${\cal N}$ is a normalisation constant.
In Eq.~(\ref{eq:ResDynEqn}) only the kinematic --- \ie\ DP --- dependence has been specified; the functions may also depend on properties of the resonance such as mass, width and spin.
The arguments $\vec{q}$ and $\vec{p}$ are the momentum of one of the resonance decay products ($d_3$ in this case, see Sec.~\ref{sec:HelicityConvention} for further information) and that of the so-called ``bachelor'' particle (\ie\ the particle not associated with the decay of the resonance; $d_2$ in this case), both evaluated in the rest frame of the resonance.
The parameters $r^P_{\rm BW}$ and $r^R_{\rm BW}$ are characteristic meson radii described below.
The resonance dynamics are normalised in \laura\ such that the integral over the DP of the squared magnitude of each term is unity
\begin{equation}
\int\!\!\int_{\rm DP} \, \left|F_j\left(m^2_{13},m^2_{23}\right)\right|^2 \, dm^2_{13} \, dm^2_{23} = 1 \,.
\end{equation}
Although not strictly necessary, as only the total probability density function (PDF) needs to be normalised, this allows a meaningful comparison of the values of the $c_j$ coefficients.

\subsection{Resonance lineshapes}
\label{sec:lineshapes}

In Eq.~(\ref{eq:ResDynEqn}), the function $R\left(m_{13}\right)$ is the resonance mass term.
The detailed forms for all available shapes in \laura\ are given in App.~\ref{sec:res-formulae}.
Here the most commonly used relativistic Breit--Wigner (RBW) lineshape is given as an example
\begin{equation}
\label{eq:RelBWEqn}
R(m) = \frac{1}{(m_0^2 - m^2) - i\, m_0 \Gamma(m)} \,,
\end{equation}
where $m_0$ is the nominal mass of the resonance and the dependence of the decay width of the resonance on $m$ is given by
\begin{equation}
\label{eq:GammaEqn}
\Gamma(m) = \Gamma_0 \left(\frac{q}{q_0}\right)^{2L+1}
\left(\frac{m_0}{m}\right) X^2(q\,r^R_{\rm BW}) \,,
\end{equation}
where $\Gamma_0$ is the nominal width of the resonance and $q_0$ denotes the value of $q$ when $m = m_0$.
In Eq.~(\ref{eq:GammaEqn}), $L$ is the orbital angular momentum between the resonance daughters.
Note that since all the initial- and final-state particles have zero spin, this quantity is the same as the spin of the resonance and is also the same as the orbital angular momentum between the resonance and the bachelor.

It is relevant to note that Eq.~(\ref{eq:RelBWEqn}) can be written
\begin{equation}
  \label{eq:RelBWEqn2}
  m_0 \Gamma(m) R(m) = \frac{m_0 \Gamma(m)}{(m_0^2 - m^2) - i\, m_0 \Gamma(m)}
  \equiv \sin \phi \exp{ i \phi} \,,
\end{equation}
where $\tan \phi = \frac{m_0 \Gamma(m)}{m_0^2 - m^2}$.
This shows the characteristic phase rotation of a resonance as $m^2$ increases from far below to far above $m_0^2$.

\subsection{Angular distributions and Blatt--Weisskopf form factors}
\label{sec:angular}

Using the Zemach tensor formalism~\cite{Zemach:1963bc,Zemach:1968zz}, the angular probability distribution terms $T(\vec{p},\vec{q})$ are given by
\begin{eqnarray}
\label{eq:ZTFactors}
L = 0 \ : \ T(\vec{p},\vec{q}) & = & \phantom{-}\,1\,,\\
L = 1 \ : \ T(\vec{p},\vec{q}) & = & -\,2\,\vec{p}\cdot\vec{q}\,,\\
L = 2 \ : \ T(\vec{p},\vec{q}) & = & \phantom{-}\,\frac{4}{3} \left[3(\vec{p}\cdot\vec{q}\,)^2 - (p\,q)^2\right]\,,\\
L = 3 \ : \ T(\vec{p},\vec{q}) & = & -\,\frac{24}{15} \left[5(\vec{p}\cdot\vec{q}\,)^3 - 3(\vec{p}\cdot\vec{q}\,)(p\,q)^2\right]\,,\\
L = 4 \ : \ T(\vec{p},\vec{q}) & = & \phantom{-}\,\frac{16}{35} \left[35(\vec{p}\cdot\vec{q}\,)^4 - 30(\vec{p}\cdot\vec{q}\,)^2(p\,q)^2 + 3(p\,q)^4\right]\,,\\
L = 5 \ : \ T(\vec{p},\vec{q}) & = & -\,\frac{32}{63} \left[63(\vec{p}\cdot\vec{q}\,)^5 - 70(\vec{p}\cdot\vec{q}\,)^3(p\,q)^2 + 15(\vec{p}\cdot\vec{q}\,)(p\,q)^4\right]~,
\label{eq:ZTFactors-end}
\end{eqnarray}
where $q \equiv |\vec{q}\,|$ and $p \equiv |\vec{p}\,|$.
These have the form of the Legendre polynomials $P_L(\cos\theta)$,
where $\theta$ is the ``helicity'' angle between $\vec{p}$ and $\vec{q}$, multiplied by the appropriate power of $-2\,p\,q$.
These factors act to suppress the amplitude at low values of the break-up momentum in either the decay of the parent or the resonance --- the so-called ``angular momentum barrier''.
However, these factors on their own would cause the amplitude to continue to grow with increasing break-up momentum even once the barrier was exceeded.
The terms $X(z)$, where $z=p\,r^P_{\rm BW}$ or $q\,r^R_{\rm BW}$, are Blatt--Weisskopf form factors~\cite{blatt-weisskopf}, which act to cancel this behaviour once above the barrier.
They are given by 
\begin{eqnarray}
\label{eq:BWFormFactors}
L = 0 \ : \ X(z) & = & 1\,, \\
L = 1 \ : \ X(z) & = & \sqrt{\frac{1 + z_0^2}{1 + z^2}}\,, \\
L = 2 \ : \ X(z) & = & \sqrt{\frac{z_0^4 + 3z_0^2 + 9}{z^4 + 3z^2 + 9}}\,,\\
L = 3 \ : \ X(z) & = & \sqrt{\frac{z_0^6 + 6z_0^4 + 45z_0^2 + 225}{z^6 + 6z^4 + 45z^2 + 225}}\,,\\
L = 4 \ : \ X(z) & = & \sqrt{\frac{z_0^8 + 10z_0^6 + 135z_0^4 + 1575z_0^2 + 11025}{z^8 + 10z^6 + 135z^4 + 1575z^2 + 11025}}\,,\\
L = 5 \ : \ X(z) & = & \sqrt{\frac{z_0^{10} + 15z_0^8 + 315z_0^6 + 6300z_0^4 + 99225z_0^2 + 893025}{z^{10} + 15z^8 + 315z^6 + 6300z^4 + 99225z^2 + 893025}}\,,
\label{eq:BWFormFactors-end}
\end{eqnarray}
where $z_0$ represents the value of $z$ when $m = m_0$.
The radius of the barrier, denoted $r^P_{\rm BW}$ or $r^R_{\rm BW}$ where the superscript indicates that the parameter is associated with either the parent or resonance in the decay chain, is usually taken to be $4.0\gev^{-1} \approx 0.8\fm$~\cite{Aubert:2005ce}.
Alternative descriptions of the angular distributions and Blatt--Weisskopf form factors are given in App.~\ref{sec:angular-formulae}.

\subsection{Fit fractions}
\label{sec:FFs}

In the absence of any reconstruction effects, the DP PDF would be
\begin{equation}
\label{eq:SigDPLike}
{\cal{P}}_{\rm phys}\left(m^2_{13}, m^2_{23}\right) =
\frac
{\left|{\cal A}\left(m^2_{13}, m^2_{23}\right)\right|^2}
{\int\!\!\int_{\rm DP}~{\left|{\cal A}\left(m^2_{13}, m^2_{23}\right)\right|^2}~dm^2_{13}\,dm^2_{23}} \, .
\end{equation}
In a real experiment, the variation of the efficiency across the DP
and the contamination from background processes must be taken into account; these details are discussed
in Sections~\ref{sec:expt-effects},~\ref{sec:eff-resol} and~\ref{sec:backgrounds}.

Typically, the primary results --- \ie\ the values obtained directly in the fit to data --- of a DP analysis include the complex amplitude coefficients, given by $c_j$ in Eq.~(\ref{eq:amp}), that describe the relative contributions of each intermediate process.
These results are dependent on a number of factors, including the amplitude formalism, choice of normalisation and phase convention used in each DP analysis. 
This makes it difficult to make useful comparisons between complex coefficients obtained from different analyses using different software.
Fit fractions provide a convention-independent method to make meaningful comparisons of results from different fits. 
The fit fraction is defined as the integral of a single decay amplitude squared divided by that of the coherent matrix element squared for the complete DP, 
\begin{equation}
{\it FF}_j =
\frac
{\int\!\!\int_{\rm DP}\left|c_j F_j\left(m^2_{13}, m^2_{23}\right)\right|^2~dm^2_{13}\,dm^2_{23}}
{\int\!\!\int_{\rm DP}\left|{\cal A}\left(m^2_{13}, m^2_{23}\right)\right|^2~dm^2_{13}\,dm^2_{23}} \, .
\label{eq:fitfraction}
\end{equation}
The sum of these fit fractions is not necessarily unity due to the potential presence of net constructive or destructive interference.
Such effects can be quantified by defining interference fit fractions (for $i<j$ only) as
\begin{equation}
  {\it FF}_{ij} =
  \frac
  {\int\!\!\int_{\rm DP} 2 \, {\rm Re}\left[c_ic_j^* F_i\left(m^2_{13}, m^2_{23}\right)F_j^*\left(m^2_{13}, m^2_{23}\right)\right]~dm^2_{13}\,dm^2_{23}}
  {\int\!\!\int_{\rm DP}\left|{\cal A}\left(m^2_{13}, m^2_{23}\right)\right|^2~dm^2_{13}\,dm^2_{23}} \, .
  \label{eq:intfitfraction}
\end{equation}
The interference fit fractions describe the net interference between the amplitudes of two intermediate processes.
Interference effects between different partial waves in a given two-body combination cancel when integrated over the helicity angle.
Therefore, non-zero interference fit fractions should arise only between contributions in the same partial wave of one two-body combination, or between contributions in different two-body combinations.
Large interference fit fractions, or equivalently a sum of fit fractions very different from unity, can often be an indication of inadequate modelling of the Dalitz plot.

\subsection{Helicity angle convention}
\label{sec:HelicityConvention}

\begin{figure}[!htb]
\centering
\includegraphics[width=0.32\textwidth]{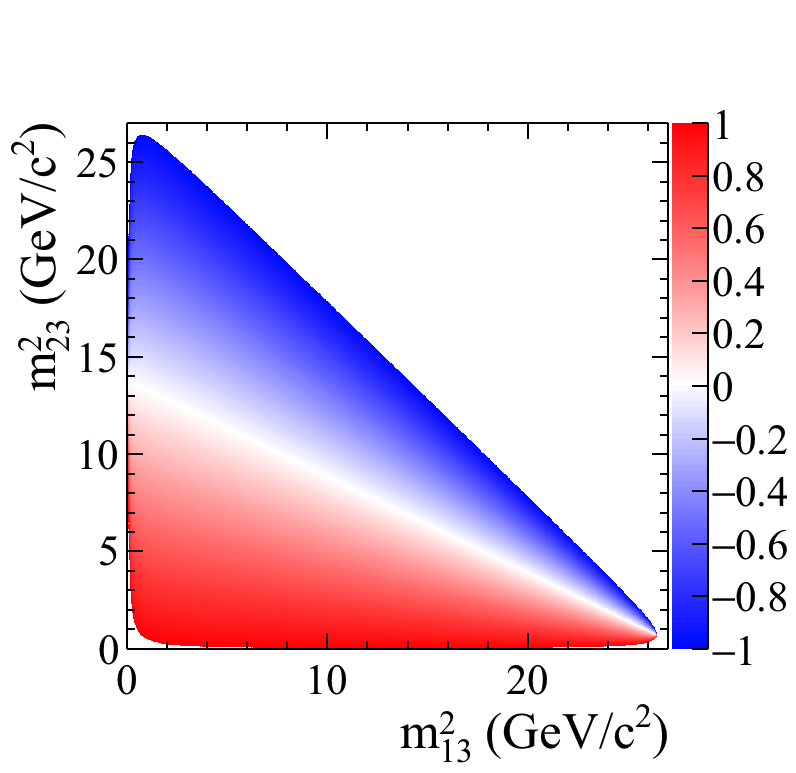}
\includegraphics[width=0.32\textwidth]{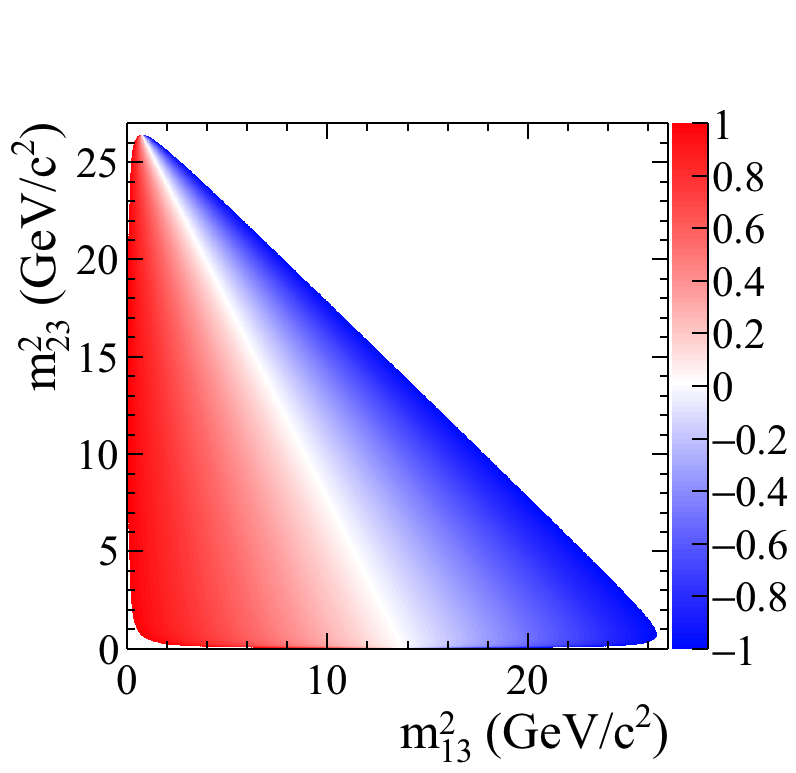}
\includegraphics[width=0.32\textwidth]{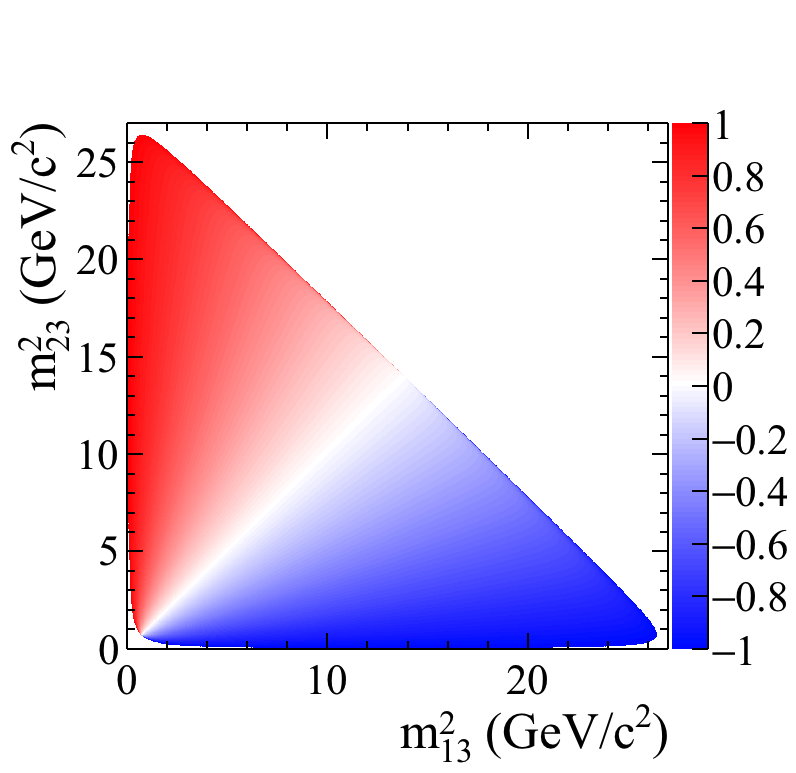}
\caption{
  Values of the cosine of the helicity angles
  (left) $\theta_{13}$,
  (middle) $\theta_{23}$, and
  (right) $\theta_{12}$
  as functions of DP position.
  The kinematic boundary of this DP corresponds to that for the $\Bzb \to \pip\pim\piz$ decay.
}
\label{fig:helicity-convention}
\end{figure}

In the formalism just described, there is a choice as to which
of the two resonance daughters the momentum $\vec{q}$ should be attributed
(and hence to attribute the momentum $-\vec{q}$ to the other).
This choice, although arbitrary, will affect the values of the measured
phases and hence it is important that it is documented to allow comparisons
between results.
The convention used in \laura is as follows:
\begin{itemize}
\setlength\itemsep{0mm}
\item $\theta_{12}$ is defined as the angle between $d_1$ and $d_3$ in the
rest frame of the $d_1$ and $d_2$ system, \ie $\vec{q}$ is the momentum of
$d_1$;
\item $\theta_{23}$ is defined as the angle between $d_3$ and $d_1$ in the
rest frame of the $d_2$ and $d_3$ system, \ie $\vec{q}$ is the momentum of
$d_3$;
\item $\theta_{13}$ is defined as the angle between $d_3$ and $d_2$ in the
rest frame of the $d_1$ and $d_3$ system, \ie $\vec{q}$ is the momentum of
$d_3$.
\end{itemize}
This convention is illustrated in Fig.~\ref{fig:helicity-convention}.
One important point to note is that it is not a cyclic permutation.
Rather it is designed such that for decays where $d_1$ and $d_2$ are
identical particles the formalism is already symmetric under their
exchange, as required.

For decays of neutral particles to flavour-conjugate final states
containing two charged daughters, \eg $\BorBbar\to\pip\pim\KS$, there is a
further complication that must be considered.
In the example given, if one chooses for the \B decay that $d_1$ would be
\pip, $d_2$ would be \pim and $d_3$ would be \KS, one should then define
the \Bbar decay using the conjugate particles, \ie $d_1$ would be \pim,
$d_2$ would be \pip and $d_3$ would be \KS.
In practice, however, one often has an untagged data sample that contains both
\B and \Bbar decays that are not distinguished and so a single definition of
the DP must be used.
(Flavour-tagged analyses are discussed in Sec.~\ref{sec:future-devel}.)
Consequently, the amplitude model must account for the incorrect particle
assignments for one of the flavours.
Due to the choice of convention in \laura this can be handled in a
straightforward manner as long as the self-conjugate particle (the \KS in
the example given) is assigned to be $d_3$.
Under these circumstances, the relation
$F\left(m^2_{13}, m^2_{23}\right) = \Fbar\left(m^2_{23}, m^2_{13}\right)$
can be restored simply by multiplying by $-1$ the cosine of the helicity
angle $\theta_{12}$ in the amplitude calculations for either the particle
or antiparticle decay --- we choose to do this for the particle decay (the
\B decay in the example given).
So, in the considered example, a contribution from $\Bd\to\rho(770)^0\KS$
would have its helicity angle definition inverted with respect to that of
$\Bdb\to\rho(770)^0\KS$.

\section{Experimental effects}
\label{sec:expt-effects}

In order to extract physics results from reconstructed $P \rightarrow d_1 d_2 d_3$ candidates 
using real experimental data, several effects need to be taken into account.
One major concern will be that any backgrounds that fake the signature of signal decays need to be removed, which is usually done by imposing various selection criteria that exploit differences in the kinematics and topologies between signal and background events.

The effect of applying selection criteria invariably means that the probability of reconstructing signal decays will not be 100\% and furthermore could vary
as a function of DP position.
Along the boundaries, at least one of the daughter particles has low momentum,
which typically reduces the reconstruction efficiency compared to decays at or
near the DP centre.
To account for this effect properly, the signal PDF, originally defined in
Eq.~(\ref{eq:SigDPLike}), needs to be modified to
\begin{equation}
\label{eq:SigDPLikeEff}
{\cal{P}}_{\rm sig}\left(m^2_{13}, m^2_{23}\right) =
\frac
{\left|{\cal A}\left(m^2_{13}, m^2_{23}\right)\right|^2 \, \epsilon\left(m^2_{13}, m^2_{23}\right)}
{\int\!\!\int_{\rm DP}~{\left|{\cal A}\left(m^2_{13}, m^2_{23}\right)\right|^2 \, \epsilon\left(m^2_{13}, m^2_{23}\right)}~dm^2_{13}\,dm^2_{23}} \, ,
\end{equation}
where the signal efficiency $\epsilon\left(m^2_{13}, m^2_{23}\right)$ is defined as the fraction of signal decays at the given DP position that are retained after all selection criteria have been applied.

For certain modes, there can be decay channels that can mimic the properties of the signal mode under study. 
For example, there may be significant backgrounds to the charmless decay $\Bm \rightarrow \Km \pip \pim$ from the decay modes $\Bm \rightarrow \Dz \pim, \Dz \rightarrow \Km \pip$ or $\Bm \to \chi_{c0} \Km, \chi_{c0} \to \pip \pim$. 
Such backgrounds can be removed, or at least suppressed, by applying a ``veto'', which means excluding candidates that lie within, typically, three to five widths on either side of the mass peak.
Alternatively, they can be accounted for within the signal model.

Another issue that needs to be considered is the effect of finite experimental resolution in the determination of the momentum of the parent $P$ and its daughter particles. 
This leads to imperfect measurements of the invariant mass-squared combinations of the daughters, and also causes uncertainty on the invariant mass of the reconstructed $P$ candidate. 
To avoid creating a DP with a fuzzy boundary, the mass of $P$ can be fixed to its expected value, with adjustments made to the four-momenta of its daughter particles to ensure momentum--energy conservation.
This will in general improve the resolution of the measurement of the DP co-ordinates, however there may still be significant effects related to events migrating from one region of the DP to another, especially near the corners of the kinematic boundary.
This effect is usually ignored if the size of the migration/resolution is smaller than the width of the narrowest resonance under consideration, or if the largest migration probability is below 10\% or so (although in the latter case, it is likely that systematic uncertainties on the physics results would need to be evaluated).
If particularly narrow resonances contribute to the decay or if the final state particles under study suffer from significant misreconstruction effects (as is often the case for decays involving neutral pions), these effects can be taken into account rather generically by adding a ``self cross-feed'' component to the signal PDF.
In this component, the true PDF is smeared by the resolution function $w_{\rm scf}(s_{\rm reco},s_{\rm true})$, given in terms of the reconstructed and true DP positions $s_{\rm reco} \equiv (m^2_{13}, m^2_{23})$ and $s_{\rm true}$.
The total signal PDF is then 
\begin{equation}
\label{eq:scflike}
{\cal P}_{\rm sig-scf}(s_{\rm reco}) = [1 - f_{\rm scf}(s_{\rm reco})] \, {\cal P}_{\rm sig}(s_{\rm reco}) + 
\int f_{\rm scf}(s_{\rm true}) \, w_{\rm scf}(s_{\rm reco},s_{\rm true}) \, {\cal P}_{\rm sig}(s_{\rm true})\,ds_{\rm true}\,,
\end{equation} 
where for the first component the resolution is negligible (equivalent to $w_{\rm scf}$ being a delta function), and the level of the second is determined by the self cross-feed fraction $f_{\rm scf}(s_{\rm true})$, \ie\ the fraction of reconstructed events with true DP position $s_{\rm true}$ that are misreconstructed.
The integral is over all true DP positions, although in practice only those with non-zero values of $w_{\rm scf}$ for the given $s_{\rm reco}$ need to be included.
The fraction $f_{\rm scf}$ can vary between 0 and 1, which correspond to the cases that resolution is negligible or that it must be considered for all signal events.
The map $w_{\rm scf}$ is sufficiently flexible to account for the fact that resolution may be more important to consider in some regions of the DP than others.

Despite imposing selection requirements in order to select signal candidates, there 
can remain significant fractions of various backgrounds in the DP analysis sample.
This means that an extended likelihood function $\cal{L}$ needs to be
employed in order to include these additional contributions:
\begin{equation}
\label{eq:likelihood}
{\cal L} = e^{-N} \prod_j^{N_c}\left[ \sum_k{N_k {\cal P}_k^j} \right] \,,
\end{equation}
where $N$ is equal to $\sum_k{N_k}$, $N_k$ is the yield for the event category $k$
(signal or background), $N_c$ is the total number of candidates in the data sample,
and ${\cal P}^j_k$ is the PDF for the category $k$ for event $j$, which consists of the product of the DP PDF and any other (uncorrelated) PDFs that are used to discriminate between signal and background.
The function $-2\,{\rm ln}\,{\cal L}$ is minimised in an unbinned fit to the data in order to extract all of the parameters.

Amplitude analyses often feature multiple solutions, which are local minima of the negative log-likelihood function.
For example, in the case that two broad overlapping resonances appear in the same partial wave, it may be possible to have solutions with either constructive or destructive interference with similar log-likelihood values.  
In order to find the true global minimum, the fit should be repeated many times with randomised initial values of the free parameters. 
The best solution can then be found by taking that with the minimum negative log-likelihood.
In addition to providing confidence that the result obtained corresponds to the global minimum, this procedure is helpful to understand the sensitivity of the data to rejecting the secondary solutions.

\section{Implementation of the signal component}
\label{sec:signal}

In this section we begin to describe the code structure of the \laura
package by first outlining the classes and methods used to build up the
total DP amplitude of the signal, given in Eq.~(\ref{eq:amp}).
Furthermore, we describe how this is normalised in order to form the signal
PDF defined in Eq.~(\ref{eq:SigDPLike}).

\subsection{Particle definitions and kinematics}
\label{sec:daughters}

The most fundamental parts of the code define the properties of the parent
particle $P$ and its three daughters $d_1$, $d_2$ and $d_3$ and their
associated kinematics.
Allowed types for $P$ are \Bp, \Bm, \Bz, \Bzb, \Bs, \Bsb, \Dp, \Dm, \Dz,
\Dzb, \Dsp and \Dsm, while the possible daughters types are \pip, \pim,
\piz, \Kp, \Km, \KS, \etaz, \etapr, \Dp, \Dm, \Dz, \Dzb, \Dsp and \Dsm.\footnote{
  Other scalar or pseudoscalar possibilities for parent or daughter particles can be trivially added.
} 
The information on the decay that is to be modelled is encapsulated within
the \texttt{LauDaughters} class, which is constructed by providing the
names or PDG codes~\cite{PDG2016} of the parent and daughters.
The particle properties are retrieved using the \texttt{LauDatabasePDG}
singleton class, which extracts and supplements information from the \root
\texttt{TDatabasePDG} particle property class.

The \texttt{LauDaughters} object then instantiates a \texttt{LauKinematics}
instance, supplying to it the masses of the parent and its daughters.
Instances of \texttt{LauKinematics} are used throughout the \laura code to
calculate and store all of the required kinematic variables for a given
position in the DP (usually supplied as $m^2_{13}$ and $m^2_{23}$).
These kinematic variables include the two-body invariant masses and
helicity angles, the momenta of the daughters in the parent rest frame and
in each of the two-body rest frames.
In addition, there is the option to calculate the co-ordinates of the
so-called ``square Dalitz plot''.

\subsubsection{Square Dalitz plot}
\label{sec:sqdp}

Since, particularly in \B decays, signal events tend to populate regions close to the kinematic boundaries of the DP, it can be convenient to use a co-ordinate transformation into the so-called square Dalitz plot (SDP)~\cite{Aubert:2005sk}.
The SDP is defined by variables \mpr and \thpr that have validity ranges between 0 and 1 and are given by
\begin{equation}
\label{eq:sqdp-vars}
\mpr \equiv \frac{1}{\pi} 
\cos^{-1}\left(2\frac{m_{12} - m^{\rm min}_{12}}{m^{\rm max}_{12} - m^{\rm min}_{12}} - 1 \right)
\hspace{10mm}{\rm and}\hspace{10mm}
\thpr \equiv \frac{1}{\pi}\theta_{12}\,,
\end{equation}
where $m^{\rm max}_{12} = m_{P} - m_{d_3}$ and $m^{\rm min}_{12} = m_{d_1}
+ m_{d_2}$ are the kinematic limits of $m_{12}$ allowed in the $P \to d_1
d_2 d_3$ decay, while $\theta_{12}$ is the helicity angle between $d_1$ and
$d_3$ in the $d_1d_2$ rest frame, as explained in Sec.~\ref{sec:HelicityConvention}.
Similar to how a choice of DP variables must be made, the SDP can be defined in several different ways.
The expressions of Eq.~(\ref{eq:sqdp-vars}) correspond to the choice used in
\laura, which must be employed consistently whenever a SDP is used.

To transform between DP and SDP representations, it is necessary to ensure
correct normalisation.
This is achieved by including the determinant of the Jacobian of the
transformation, which is given by
\begin{equation}
\label{eq:sqdp-jacobian}
\left| J \right| = 4 \, p \, q \, m_{12} \frac{\partial m_{12}}{\partial \mpr} \frac{\partial \cos\theta_{12}}{\partial \thpr} \,,
\end{equation}
where $p$ and $q$ are evaluated in the $d_1d_2$ rest frame and the partial
derivatives evaluate to
\begin{eqnarray}
\nonumber
\frac{\partial m_{12}}{\partial \mpr} &=& -\frac{\pi}{2} \sin(\pi\mpr) \left( m^{\rm max}_{12} - m^{\rm min}_{12} \right) \,, \\
\frac{\partial \cos\theta_{12}}{\partial \thpr} &=& -\pi\sin(\pi\thpr) \,.
\label{eq:sqdp-derivatives}
\end{eqnarray}

\begin{figure}[!htb]
  \centering
  \includegraphics[width=0.49\textwidth]{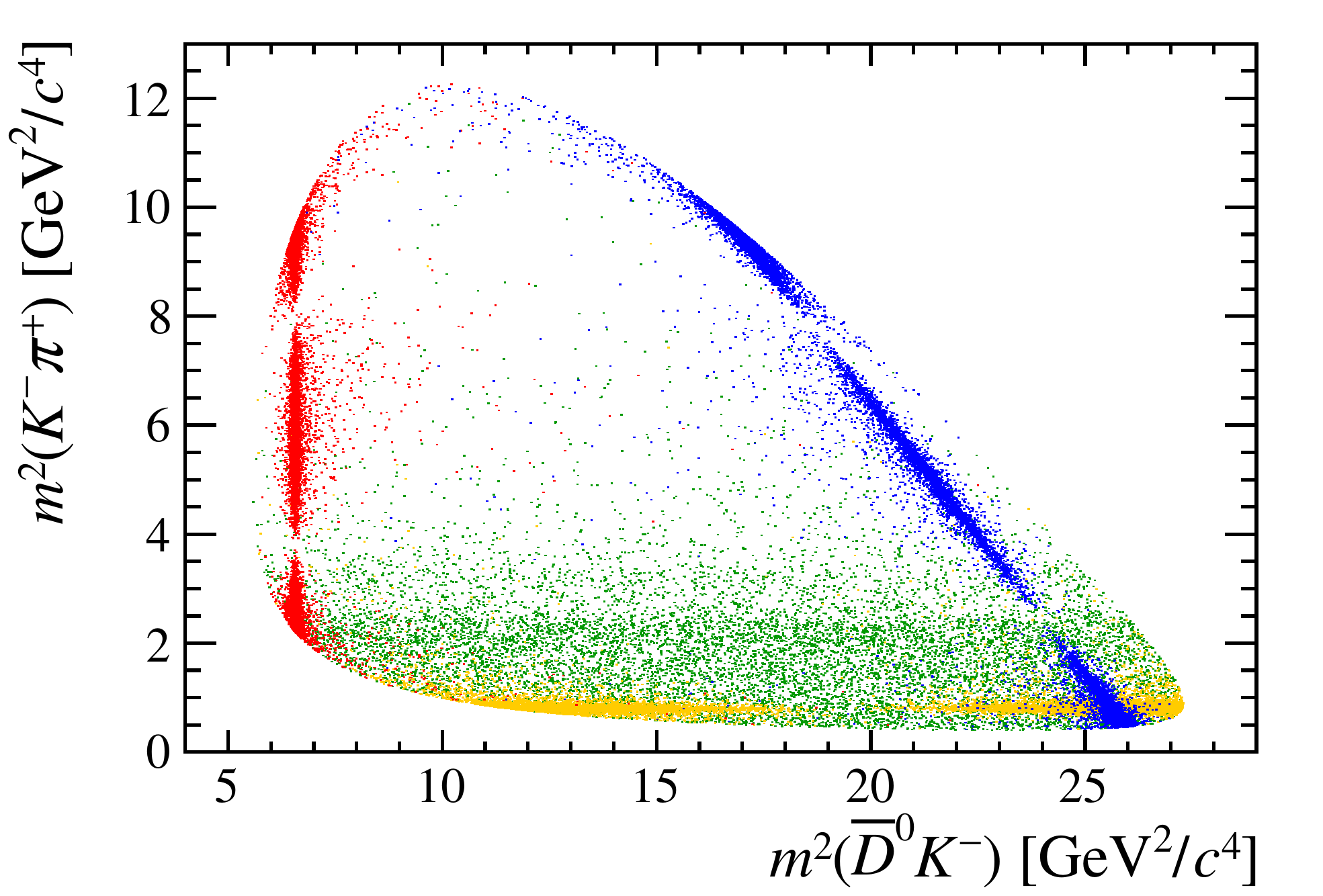}
  \includegraphics[width=0.49\textwidth]{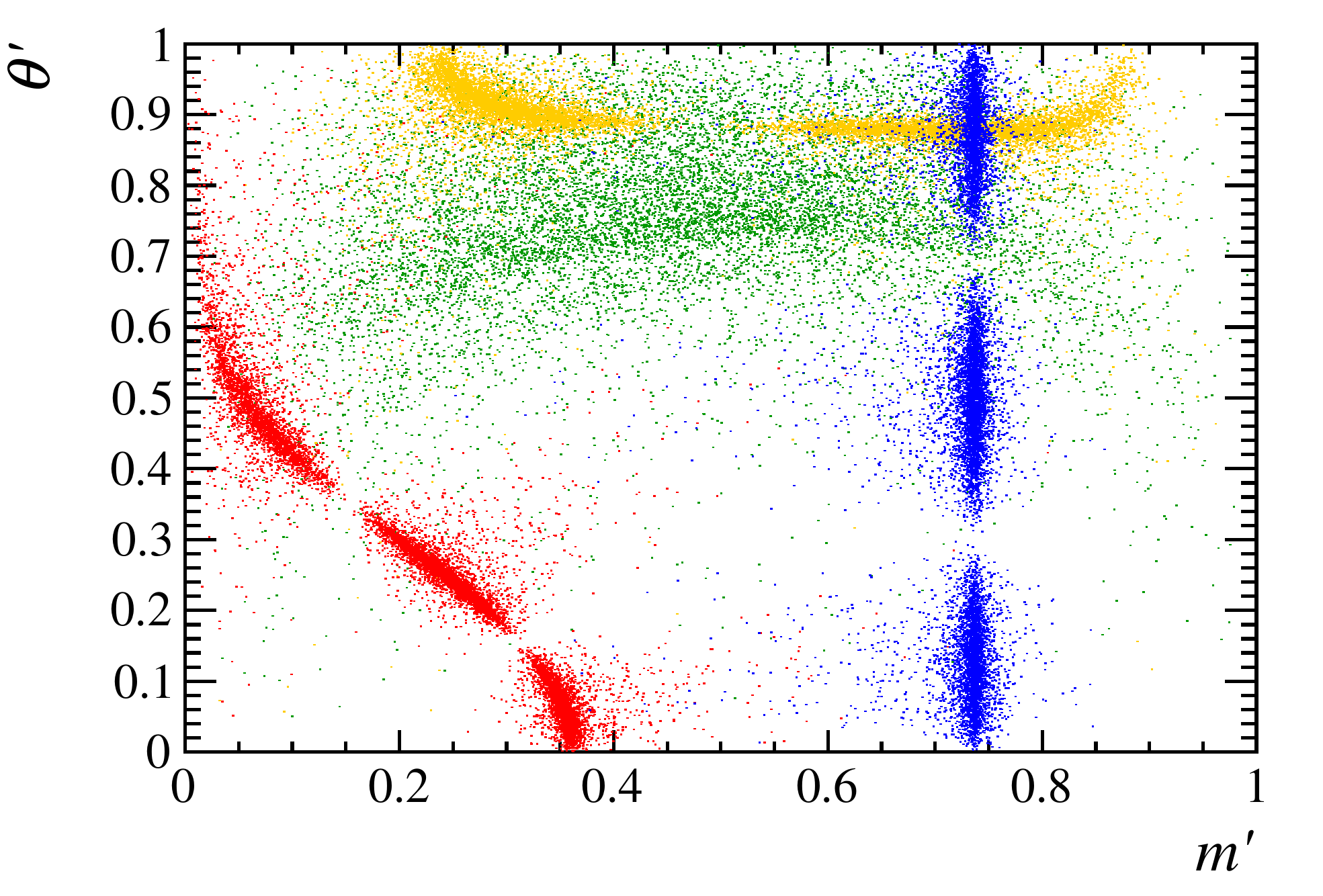}
  \caption{
    Illustration of the transformation between conventional and square Dalitz
    plot representations, for resonances in the $\Bs \to \Dzb\pip\Km$ decay
    (here the final-state particles are ordered following the $d_1d_2d_3$
    convention of \laura).
    Compared to the same DP shown in Fig.~\ref{fig:dpexample}, a fake
    $\Dzb\pip$ resonance, with parameters corresponding to those of the
    $\D^*_2(2460)^+$ state (blue points) has been added in order to better
    visualise the transformation in relevant DP regions. 
  }
  \label{fig:sdpexample}
\end{figure}

The SDP coordinate system has several advantages that apply whenever there
is a need to bin the phase space, which are illustrated in Fig.~\ref{fig:sdpexample}.
Firstly, the regions near the kinematic boundary are spread out, which
means that these regions where the signal is often concentrated and where
also there can be rapid variation in efficiency and background
distributions can be treated with a much finer resolution, even while
maintaining a uniform binning.
Secondly, the kinematic boundary is perfectly aligned with the bin edges.

\subsection{Isobar dynamics and resonances}
\label{sec:isobar-dynamics}

Once the kinematics of a particular decay mode have been established, the structure of the signal DP model can be by defined by creating a \texttt{LauIsobarDynamics} object.
Components of the model are specified using the \texttt{addResonance}
member function, which requires:
\begin{itemize}
\setlength\itemsep{0mm}
\item the name of the resonance,
\item an integer that specifies which of the daughters is the bachelor particle, and hence in which invariant mass spectrum this resonance will appear (1 for $m_{23}$, 2 for $m_{13}$, 3 for $m_{12}$ or 0 for some nonresonant models),
\item an enumeration to select the form of the dynamical amplitude.
\end{itemize}
Appendix~\ref{sec:resNames} contains lists of the names of the allowed resonances along with their nominal mass, width, spin, charge and Blatt--Weisskopf barrier radius.
This information is all automatically retrieved from \texttt{LauResonanceInfo} records that are stored in the \texttt{LauResonanceMaker} class. 
Appendix~\ref{sec:resNames} also provides information on how to account for a state that is not already included in \laura, and how to change the nominal values of the properties of any resonance.
Appendix~\ref{sec:res-formulae} gives details of all the dynamical amplitude forms that are currently implemented in the package and in Table~\ref{tab:resForms} supplies the corresponding enumeration types.
Examples of usage are given in Sec.~\ref{sec:setup}.

The signal model may also include contributions that do not interfere with the other resonances in the DP.
These may arise due to decays that proceed via intermediate long-lived, \ie\ negligible natural width, states; an example is the contribution from $\Bm \to \Dz \pim$, $\Dz \to \Km\pip$ in $\Bm \to \Km\pip\pim$ decays.
Experimentally, such components can be considered either as signal or background, and selection requirements (\eg, based on the consistency of the three daughter tracks originating from the same vertex position) may be used to suppress them, but in certain cases some contribution will remain.
Within \laura, the user can choose how to treat such contributions.
When considered as part of the signal model, non-interfering components can be added with the \texttt{addIncoherentResonance} member function, which has the same number and type of arguments as the \texttt{addResonance} function.
In this case the form of the dynamical amplitude should be specified as \texttt{GaussIncoh}, and the width should be changed to correspond to the experimental resolution.
Note that the resolution for incoherent contributions is handled in a different way to the approach described in Sec.~\ref{sec:expt-effects} and Sec.~\ref{sec:resol}.
As part of the signal model, a non-interfering component will contribute to
the denominator of the fit fractions, but it is simple for the user to subtract it from the results since there is no interference with other components. 

When building the model, \laura\ performs a simple check of charge conservation, while angular momentum is conserved by construction.  
However, the onus is on the user to make sure that the strong decays of resonances included in the model respect conservation of parity and flavour quantum numbers, since these are not checked by the code.
In order to help with this, a summary of all resonances used in the model
is printed out during the initialisation.

The complex dynamical amplitudes $R(m)$ of the various resonance forms are
defined using classes that inherit from the abstract base class
\texttt{LauAbsResonance}.
For example, the relativistic Breit--Wigner lineshape is defined within the
\texttt{LauRelBreitWignerRes} class.
All such classes implement the \texttt{resAmp} member function that
returns a \texttt{LauComplex} class that represents the amplitude at the given
value of the relevant two-body invariant mass.
The \texttt{LauAbsResonance} base class implements the calculation of the
angular distribution factor.
Those amplitude forms that require the calculation of the Blatt--Weisskopf
factors make use of the \texttt{LauBlattWeisskopfFactor} helper class.

\subsection{Symmetry}
\label{sec:symmetry}

If the decay of $P$ contains two identical daughters, such as in $\Bp
\rightarrow \pip \pip \pim$, then the DP will be symmetric.
As mentioned in Sec.~\ref{sec:HelicityConvention}, the identical particles should be positioned as $d_1$ and $d_2$.
This situation is automatically detected by \texttt{LauDaughters} and the
information propagated to the amplitude model.
In this case, it is required only to define the resonances for the pair
$d_{1}d_{3}$; the amplitude is automatically symmetrised by
\texttt{LauIsobarDynamics} by flipping the invariant-mass squared variables
$m^2_{13} \leftrightarrow m^2_{23}$, recalculating the amplitude and
summing.

When all of the daughters are identical, for example in $\Bz \rightarrow \KS
\KS \KS$, it is again only needed to define the resonances for one of the
pairs (usually $d_{1}d_{3}$).
For this fully symmetric case, \texttt{LauIsobarDynamics} automatically
performs the necessary symmetrisation of the amplitude by cyclically
rotating the invariant-mass squared variables ($m^2_{12} \rightarrow
m^2_{23}$, $m^2_{23} \rightarrow m^2_{13}$, $m^2_{13} \rightarrow
m^2_{12}$) and flipping them ($m^2_{13} \leftrightarrow m^2_{23}$).

\subsection{Normalisation of signal model}

Various integrals of the dynamical amplitude across the DP need to
be calculated in order to normalise the signal PDF given by
Eq.~(\ref{eq:SigDPLike}), as well as for calculating the fit fractions for
individual resonances defined in Eqs.~(\ref{eq:fitfraction})
and~(\ref{eq:intfitfraction}). 
Since the $c_j$ coefficients are constant across the DP, only the
amplitude terms $F_j$ need to be integrated.
In general, these integrals cannot be found analytically, so
Gauss--Legendre quadrature is used to evaluate them numerically.
This is achieved by dividing the DP into an unequally spaced
grid whose points correspond to the abscissa co-ordinates from the
quadrature procedure.
The granularity of the grid is chosen to ensure sufficiently precise
integration, as discussed in more detail below.
The $F_j$ terms are then multiplied by the quadrature weights and summed
over all grid points that lie within the DP boundary.
To remove the quadratic variable dependence, and to improve
numerical precision, the DP area element $dm^2_{13}\,dm^2_{23}$ is
replaced by $dm_{13}\,dm_{23}$ multiplied with the Jacobian factor
$4\,m_{13}\,m_{23}$.
This means that the normalisation of the total amplitude $\cal{A}$
is given by
\begin{equation}
\label{eq:AmpNorm}
{\int\!\!\int_{\rm DP}~{|{\cal A}|^2}~dm^2_{13}\,dm^2_{23}}
\approx \sum_{a=1}^{N_a}\sum_{b=1}^{N_b} {4h_a h_b w_a w_b |{\cal{A}}|^2} \,,
\end{equation}
where $w_a$ ($w_b$) are the weights for the Gauss--Legendre quadrature abscissa 
values $h_a$ ($h_b$), which correspond to the grid points along the 
$m_{13}$ ($m_{23}$) axis, and the amplitude $\cal{A}$ is evaluated for all abscissas inside the DP kinematic boundary.
An equivalent expression is used for normalisation of the experimental signal PDF of Eq.~(\ref{eq:SigDPLikeEff}).
The number of points $N_a$ ($N_b$) is set by dividing the $m_{13}$
($m_{23}$) mass range by a default ``bin width'' $\delta m$ of 5\mevcc,
which can be changed using the \texttt{setIntegralBinWidths} function in
\texttt{LauIsobarDynamics}, giving $\sim 1000$ integration bins along each
mass axis for $B$ decays.
It is important to realise that $\delta m$ is not, in general, equal to the
separation between neighbouring abscissa points.
The \texttt{LauIntegrals} class handles the calculation of the general
weights and abscissas for the integration range $(-1, 1)$, while the
\texttt{LauDPPartialIntegralInfo} class scales these using the half-ranges
and mean values of $m_{13}$ and $m_{23}$, following the numerical recipe
given in Ref.~\cite{NRinC}.
This information is then used within the \texttt{LauIsobarDynamics} class
to find the normalisation integral for the total amplitude, as well as the
integrals for the fit fractions.

If the DP contains narrow resonances with widths below a threshold
value, which defaults to 20\mevcc but can be changed using the
\texttt{setNarrowResonanceThreshold} function in
\texttt{LauIsobarDynamics}, then the quadrature grid is split up into
smaller regions to ensure that the narrow lineshapes are integrated
correctly.
The range, in the invariant mass of the resonance daughters, for these sub-regions is taken to be $m_0 \pm 5\Gamma_0$,
where $m_0$ and $\Gamma_0$ are the nominal mass and width of the resonance.
The number of quadrature points ($\sim 1000$) along each sub-grid axis is
set by dividing the resonance mass range (ensuring the limits stay within
the DP boundary) by a $\delta m$ value which defaults to 1\% of $\Gamma_0$ and is configurable using the \texttt{setIntegralBinningFactor} function in \texttt{LauIsobarDynamics}.
Grid regions that are outside the narrow resonance bands use the default
bin width.
When there are narrow resonances along the diagonal axis $m_{12}$, the
integration scheme switches to use the SDP defined in
Section~\ref{sec:sqdp}.
The number of points on the integration grid defaults to 1000 for each of
the \mpr and \thpr axes.
This can be tuned using the \texttt{setIntegralBinWidths} function.

The fact that resonance parameters can float in the fit means that the
integrals will need to be recalculated if and when those values change.
In order to minimise the amount of information that needs to be
recalculated at each fit iteration, a caching and bookkeeping system is
employed that stores the amplitudes of each component of the signal model
as well as the Gauss--Legendre weights and the efficiency for every point
on the integration grid.
At each fit iteration it then checks which, if any, resonance parameter
values have changed and with which amplitudes those parameters are
associated.
Only those affected amplitudes are recalculated (for each integration grid
point and for each event in the data sample).
While greatly improving the speed of fits, this comes at some cost in terms
of memory usage, in particular if the integration grid is very fine.
If the number of grid points is extremely large a warning message is
printed, which recommends that the integration scheme be tuned using the
\texttt{setNarrowResonanceThreshold} and \texttt{setIntegralBinningFactor}
functions of \texttt{LauIsobarDynamics} to reduce the number of points.

\subsection{Signal model and amplitude coefficients}
\label{sec:sigmodel}

Having defined the signal PDF for $P \rightarrow d_1 d_2 d_3$, as in Eq.~(\ref{eq:SigDPLikeEff}), it is necessary to define the parameterisation of the complex coefficients $c_j$ defined in Eq.~(\ref{eq:amp}).
Several different parametrisations have been used in the literature and are available in \laura --- a complete list is given in Table~\ref{tab:coeff-sets}.
These can be separated into two categories: cases in which it is assumed that there is no difference between the decay of $P$ and its \CP-conjugate $\Pbar$ and cases in which such \CP-violating differences are accommodated.

In the former case, the signal model is constructed by passing the corresponding \texttt{LauIsobarDynamics} instance to a \texttt{LauSimpleFitModel} object, which inherits from the abstract base class \texttt{LauAbsFitModel}.
The fit model classes implement the functions needed to generate events according to the DP model as well as to perform fits to data.
In the latter case, the decays of each of $P$ and $\Pbar$ should be represented by their own \texttt{LauDaughters} instance.
These are used to construct two instances of \texttt{LauIsobarDynamics}, which in turn are used to construct an instance of the \texttt{LauCPFitModel} class that, like \texttt{LauSimpleFitModel}, also inherits from \texttt{LauAbsFitModel}.

The $F_j$ terms in Eq.~(\ref{eq:SigDPLike}) are calculated using the amplitudes that make up the \texttt{LauIsobarDynamics} model.
The complex coefficients $c_j$ are each represented by an object inheriting
from the \texttt{LauAbsCoeffSet} base class, which provides an abstract
interface for combining a set of real parameters to form the complex $c_j$.
Each coefficient is constructed by providing the name of the resonance $j$
and a series of parameter values that will be used to form the complex number $c_j$.
They are then applied to the model using the \texttt{setAmpCoeffSet}
function of the \texttt{LauSimpleFitModel} or \texttt{LauCPFitModel} class.
Checks are made to ensure that the coefficient name matches that of one
of the components of the isobar model.
The coefficient is then assigned to that component.
As such the various coefficients are reordered to match the ordering in the
\texttt{LauIsobarDynamics} model(s).

\begin{table}[!htb]
\centering
\caption{List of coefficient sets to represent $c_j$ in Eq.~(\ref{eq:amp}), separated into cases where \CP conservation is assumed and those where \CP violation is accommodated in the model.
  Where parameters are preceded by $\pm$ signs in the expressions for $c_j$, the $+$ ($-$) sign corresponds to the usage for $P$ ($\Pbar$) decays.
  The corresponding class for each set is \texttt{Lau\emph{Label}CoeffSet}, where \texttt{\emph{Label}} is the set label given below.
}
\label{tab:coeff-sets}
\begin{tabular}{lll}
\hline
Set label                 & Parameters & $c_j$\\
\hline \\ [-2.5ex]
\texttt{MagPhase}         & $r$, $\phi$ & $r e^{i\phi}$ \\
\texttt{RealImag}         & $x$, $y$ & $x + iy$ \\
\hline \\ [-2.5ex]
\texttt{BelleCP}          & $a$, $b$, $\delta$, $\phi$ 
                          & $ae^{i\delta}(1 \pm be^{i\phi})$ \\
\texttt{CartesianCP}      & $x$, $y$, $\delta_x$, $\delta_y$ 
                          & $x \pm \delta_x + i(y \pm \delta_y)$ \\
\texttt{CartesianGammaCP} & $x$, $y$, $x_{\CP}$, $y_{\CP}$, $\Delta x_{\CP}$, $\Delta y_{\CP}$ 
                          & $(x + iy)[1 + x_{\CP} \pm \Delta x_{\CP} + $ \\
                          & & \hspace{1.85cm} $i(y_{\CP} \pm \Delta y_{\CP})]$ \\
\texttt{CleoCP}           & $a$, $b$, $\delta$, $\phi$ 
                          & $(a \pm b)e^{i(\delta \pm \phi)}$ \\
\texttt{MagPhaseCP}       & $r$, $\phi$, $\bar{r}$, $\bar{\phi}$ 
                          & $r e^{i\phi}$ for $P$ \\
                          & & $\bar{r} e^{i\bar{\phi}}$ for $\Pbar$ \\
\texttt{PolarGammaCP}     & $x$, $y$, $r$, $\delta$, $\gamma$ 
                          & $(x + iy)(1 + re^{i(\delta \pm \gamma)})$ \\
\texttt{RealImagCP}       & $x$, $y$, $\bar{x}$, $\bar{y}$ 
                          & $x + iy$ for $P$ \\
                          & & $\bar{x} + i\bar{y}$ for $\Pbar$ \\
\texttt{RealImagGammaCP}  & $x$, $y$, $x_{\CP}$, $y_{\CP}$, $\bar{x}_{\CP}$, $\bar{y}_{\CP}$ 
                          & $(x + iy)(1 + x_{\CP} + iy_{\CP})$ for $P$ \\
                          & & $(x + iy)(1 + \bar{x}_{\CP} + i\bar{y}_{\CP})$ for $\Pbar$ \\
\hline
\end{tabular}
\end{table}

\section{Implementation of efficiency and resolution effects}
\label{sec:eff-resol}

As discussed in Sec.~\ref{sec:expt-effects}, in an experimental analysis it is usually necessary to modify the signal PDF in order to account for effects such as the variation of the reconstruction and selection efficiency over the DP and detector resolution or misreconstruction.
In this section we describe the classes and methods in the \laura\ package that are used to implement these modifications to the pure physics PDF described previously.

All experimental effects are handled within \laura\ through histograms.
Optional spline interpolation is available to smooth effects due to the finite size of samples used to obtain the histograms.
In some analyses it may be possible to obtain reliable parametric descriptions of, for example, the efficiency variation over the DP.
Where this is a possible and desirable approach the user can simply generate a histogram from the parametric shape and use that as input.
The granularity of the histogram can be as fine as necessary to describe local variations; often the limiting factor on the binning will be the size of the sample used to obtain the shape.  

\subsection{Efficiency}
\label{sec:eff}

The variation of the signal efficiency over the DP,
$\epsilon(m^2_{13}, m^2_{23})$, is implemented by the \texttt{LauEffModel}
class.
Its constructor requires a \texttt{LauDaughters} object, which defines the kinematic boundary, as well as a \texttt{LauVetoes} object, which is used to specify any region in the DP that has been excluded from the analysis (perhaps to remove particular sources of background or to exclude a region of phase space where the efficiency variation is poorly understood). 
The signal efficiency is set to zero inside a vetoed region; the resulting discontinuity at the boundary motivates different treatment of vetoes to other sources of inefficiency that vary smoothly across the DP.
Vetoes can be added using the \texttt{addMassVeto} or \texttt{addMassSqVeto} functions of the \texttt{LauVetoes} class, which require the bachelor daughter index as well as the lower and upper invariant mass (or mass-squared) values for each exclusion region.
Since version \texttt{v3r2} of \laura, vetoes are automatically symmetrised
as appropriate for DPs containing two or three identical
particles in the final state.

All other information on the efficiency variation over the DP needs to be supplied in the form of a uniformly binned two-dimensional \root histogram.
Alternatively, a set of histograms can be provided; in this case the total efficiency is obtained by multiplying the efficiencies at the appropriate position in phase space from each of the components.
This provides a convenient way to assess the impact of systematic uncertainties from different contributions to the total efficiency.  
Each of these component efficiency histograms can have different binning.

Efficiency histograms will usually be constructed by applying all selection requirements to a simulated sample of signal decays that has been passed through a full detector simulation.
A ratio is then formed of all decays that survive the reconstruction and selection to all those that were originally generated.
Since the effect of explicit vetoes in the phase space is separately accounted for, it is advised that these are not applied when constructing the numerator of the efficiency histogram.

As previously mentioned in Sec.~\ref{sec:sqdp}, signal events often occupy
regions close to the kinematic boundaries.
The SDP transformation defined in Eq.~(\ref{eq:sqdp-vars}) can be used to
spread out these regions so that the efficiency variation can be modelled
more accurately.
As such, the \texttt{LauEffModel} class will accept histograms that have
been created in either $m^2_{13}$--$m^2_{23}$ or \mpr--\thpr space.

The histogram can then be supplied to the \texttt{LauEffModel} class via the \texttt{setEffHisto} or \texttt{setEffSpline} function as appropriate.
Where the total efficiency is to be obtained from the product of several components, further histograms can be included with the \texttt{addEffHisto} and \texttt{addEffSpline} functions.
In each case, a boolean argument \texttt{squareDP} is used to indicate the space in which the histogram has been defined.
For symmetric DPs, there is also the option to specify that the
histogram provided has already been folded and hence only occupies the upper
half of the full DP (or the corresponding lower half of the SDP).
Internally, each histogram is stored as a \texttt{Lau2DHistDP} or
\texttt{Lau2DSplineDP} object, which implement (optional) bilinear or cubic
spline interpolation methods, respectively.

Functionality is also available to help estimate systematic uncertainties
due to imperfect knowledge of the efficiency variation by creating Gaussian
fluctuations in the bin entries.
The fluctuations are based on the uncertainties provided  by the user,
which, depending on the function used to provide the histogram, can be
asymmetric.
It is up to the user as to whether the provided uncertainties are simply due to the limited size of the simulated sample or whether they also account for effects such as possible disagreements between data and simulation.
The fluctuations are activated by providing optional boolean arguments to the function where the histogram is provided.

\subsection{Resolution}
\label{sec:resol}

The self cross-feed component of the signal likelihood, modelled as described in Eq.~(\ref{eq:scflike}), is implemented using information from the \texttt{LauScfMap} class, which stores all possible values of $w_{\rm scf}$ via its \texttt{setHistos} function. 
As for the description of the efficiency, the histograms used to describe $w_{\rm scf}$ and $f_{\rm scf}$ can be constructed in either $m^2_{13}$--$m^2_{23}$ or \mpr--\thpr space.
However, to simplify the implementation in \laura\ it is currently required that all histograms related to the description of resolution have the same binning.

The implementation proceeds as follows.
Consider a SDP describing the true position, $s_{\rm true}$, divided into a uniformly binned two-dimensional histogram.
Each bin will have associated with it another two-dimensional histogram, with identical binning, whose entries give the migration probability $w_{\rm scf}(s_{\rm reco},s_{\rm true})$ of the true position (given by the original bin centre) being reconstructed in the bin that contains $s_{\rm reco}$.
These can be constructed as follows:
\begin{itemize}
\setlength\itemsep{0mm}
\item
Each histogram contains only the events that were generated in a given true bin.
\item
The events are plotted at their reconstructed co-ordinates.
\item
The histogram is then normalised.
\end{itemize}
Some histograms may be empty if there were no events generated in that bin, although this is of course dependent on the size of the samples used and the size of the bins.
The order of the histograms in the vector supplied to the \texttt{setHistos} function should be in terms of the \root ``global bin number''.

The \texttt{splitSignalComponent} method of \texttt{LauSimpleFitModel} and \texttt{LauCPFitModel} takes a \texttt{LauScfMap} object as an argument, as well as a two-dimensional \root histogram whose entries give $f_{\rm{scf}}(s_{\rm{true}})$.
The fit models then use this information to evaluate the self cross-feed contribution to the likelihood.
In the ideal case of Eq.~(\ref{eq:scflike}) this is described by an integral; in practice this becomes a summation,
\begin{equation}
\begin{array}{c}
\int f_{\rm scf}(s_{\rm true}) \, w_{\rm scf}(s_{\rm reco},s_{\rm true}) \, {\cal P}_{\rm sig}(s_{\rm true})\,ds_{\rm true} \\
\hookrightarrow \\
\sum_i \left[ 
  \left< \hat{f}_{\rm scf}(\hat{s}_{\rm true\,i}) \right> \,
  \left< \hat{w}_{\rm scf}(\hat{s}_{\rm reco},\hat{s}_{\rm true\,i}) \right> \,
  {\cal P}_{\rm sig}(\hat{s}_{\rm true\,i}) \,
  \frac{\Delta \Omega(\hat{s}_{\rm true\,i})}{\Delta \Omega(\hat{s}_{\rm reco})}   
\right] \, ,
\end{array}
\end{equation}
where the hat ($\,\hat{}\,$) notation is used to indicate quantities evaluated in the SDP (as is the case in this example), and the bracket ($\,\left<\,\right>\,$) notes quantities obtained from histograms.
The pure signal PDF ${\cal P}_{\rm sig}$ is as in Eq.~(\ref{eq:SigDPLikeEff}), evaluated at the position corresponding to the SDP point at the centre of the $\hat{s}_{\rm true\,i}$ bin.
The phase space factors $\Delta \Omega$ are equal to $\Delta \mpr \Delta \thpr \left| J \right|$ where the SDP bin size is given by $\Delta \mpr \Delta \thpr$ and the Jacobian of the SDP transformation is that of Eq.~(\ref{eq:sqdp-jacobian}); since equal binning is required, the ratio of phase space factors reduces to a ratio of Jacobians.

\section{Implementation of background components}
\label{sec:backgrounds}

\subsection{Dalitz-plot distributions}

Backgrounds in the data sample can be taken into account by including them in the total likelihood defined in Eq.~(\ref{eq:likelihood}). 
This means that the DP distributions of all background categories need to be provided.
In an analogous way to the implementation of the signal efficiency described in the previous section, the DP distribution of each background category is represented with a uniformly binned two-dimensional \root histogram.

Background contributions are handled in \laura\ as follows.
Firstly, the names of all background categories need to be provided to the fit model via the \texttt{setBkgndClassNames} function of the appropriate \texttt{LauAbsFitModel} class.
Then, each named category needs to have its DP distribution defined and added to the fit model.
This is achieved by supplying one (or more if the model subdivides the
data to account for effects such as \CP violation) \texttt{LauBkgndDPModel}
object.
The constructor of the \texttt{LauBkgndDPModel} class requires pointers to
the usual \texttt{LauDaughters} and \texttt{LauVetoes} objects.
The histogram is supplied to the \texttt{LauBkgndDPModel} object via its
\texttt{setBkgndHisto} (\texttt{setBkgndSpline}) function, and is then
internally stored as a \texttt{Lau2DHistDPPdf} (\texttt{Lau2DSplineDPPdf})
object that implements bilinear (cubic spline) interpolation.
The PDF value is then calculated as the interpolated number of background
events $B(m^2_{13}, m^2_{23})$ divided by the total integrated area of the
histogram:
\begin{equation}
\label{eq:bkgndpdf}
{\cal P}_{\rm{bkgnd}} = 
\frac{B(m^2_{13}, m^2_{23})}{\int\!\!\int_{\rm DP}~B(m^2_{13}, m^2_{23})~dm^2_{13}\,dm^2_{23}} \,.
\end{equation}

Like signal events, backgrounds also tend to populate regions close to the
kinematic boundaries of the DP.
Therefore, the use of histograms in the SDP space can improve the modelling
of backgrounds.
This is achieved by providing a histogram in \mpr--\thpr space and setting
the \texttt{squareDP} boolean flag to true in the \texttt{setBkgndHisto} or
\texttt{setBkgndSpline} functions of \texttt{LauBkgndDPModel}.
The normalisation of these PDFs then automatically includes the Jacobian
for transforming from normal to square Dalitz-plot space.

Some special treatment is necessary for backgrounds modelled from sources that contain contributions that are vetoed in the DP fit.
Following the example of Sec.~\ref{sec:expt-effects}, the combinatorial background to $\Bm \rightarrow \Km \pip \pim$ decays may be modelled from a sideband in the $B$ candidate mass distribution that also contains some genuine $\Dz \rightarrow \Km \pip$ decays.  
The histogram binning will introduce some smearing of such contributions, so that once the veto is applied later some residual background may remain.
In principle this can be avoided with sufficiently fine histogram bins, but this will often be impractical due to finite sample sizes.  
Instead, and in contrast to the procedure for efficiency histograms described in Sec.~\ref{sec:eff}, the veto should be applied when the background histogram is made.
This will, however, lead to an underestimation of the background that is being modelled (in the example above, of the random combinations of three tracks) within bins that lie partially inside the vetoed regions.
To correct for this effect, each background histogram needs to be divided by another histogram (with the same binning) whose entries contain the fraction of events, generated from a high statistics sample that is uniform in phase space, that lie outside any veto region; bins that are completely outside (inside) a veto have a weight of unity (zero), and division by zero is interpreted as zero weight.
The histogram supplied to \texttt{LauBkgndDPModel} should already have had this correction applied to it.

\subsection{Other discriminating variables}
\label{sec:other-vars}

When fitting a data sample that contains (significant) backgrounds,
additional discriminating variables can be included in the total likelihood
function defined in Eq.~(\ref{eq:likelihood}), in order to provide improved
separation between signal and background categories.
Examples of such discriminating variables include the mass of the parent particle $P$ candidate and the output of a multivariate discriminant to separate signal from background.
Assuming that all of the variables $\vec{x} = (x_1, x_2, ..., x_n)$ are uncorrelated, the PDF ${\cal P}^j_k$ of the signal or background category $k$, for event $j$, is given by the product of the individual variable PDFs (including that of the DP distribution):
${\cal P}^j_k(\vec{x}) = {\cal P}^j_k(x_1) \times {\cal P}^j_k(x_2) \times ... \times {\cal P}^j_k(x_n)$.

Additional PDFs are represented by classes that inherit from
\texttt{LauAbsPdf}, whose constructor requires the variable name, a vector
of the PDF parameters, as well as minimum and maximum abscissas that
specify the variable range.
Each class implements the member function for evaluating the PDF
function at a given abscissa value and that for evaluating the maximum
value of the PDF in the fitted range.
A list of classes that can be used for additional PDFs is given in App.~\ref{sec:pdfs}.
Each PDF needs to be added to the fit model;
for \texttt{LauSimpleFitModel} (\texttt{LauCPFitModel}),
signal PDFs are added using the \texttt{setSignalPdf}
(\texttt{setSignalPdfs}) function,
self cross-feed PDFs are included via \texttt{setSCFPdf}
(\texttt{setSCFPdfs}) function,
and background PDFs are added with the \texttt{setBkgndPdf}
(\texttt{setBkgndPdfs}) function.

An alternative approach is to perform first a fit to some or all of the other discriminating variables to determine the yields of signal and background.
These values can then be fixed in the fit to the DP to extract the
amplitude parameters.
Requirements can also be imposed on those variables, before or after such a fit, in order to enhance the signal purity among the events entering the DP fit.
This approach will have slightly reduced sensitivity but avoids the need to model correlations between the DP and the other variables (although classes are provided that allow the modelling of such correlations, see App.~\ref{sec:pdfs-DPdep}).
It is possible to fit only other discriminating variables, \ie\ excluding the DP variables, in \laura\ through the \texttt{useDP} function of \texttt{LauAbsFitModel}, which takes a boolean argument.

\section{Work flows}
\label{sec:workflow}

The \laura package is designed to perform three main work flows:
\begin{itemize}
\setlength\itemsep{0mm}
\item
  Monte Carlo generation of toy datasets from a defined PDF,
\item
  fitting a defined PDF to an input data set,
\item
  calculating per-event weights for an input data sample (usually full
  detector simulation) from a defined DP model.
\end{itemize}
In this section we describe the setup stages that are common to all tasks, then each of the three work flows in turn, and finally some variations that are available.
Code taken from the various examples included with the package is used to illustrate many of these points.
These code snippets are in \cpp\ but it is also possible to use the python
bindings that are generated by \texttt{rootcling} when the \laura\ library is
compiled and to write the application code in python.
An example that demonstrates how to do this, \texttt{GenFit3pi.py}, is
included in the package since version \texttt{v3r3}.

\subsection{Common setup}
\label{sec:setup}

The first setup task is to define the DP with which we are
working.  As mentioned in Section~\ref{sec:daughters} this is achieved
through the declaration of a \texttt{LauDaughters} object:
\begin{lstlisting}
bool squareDP = true;
LauDaughters* daughters = new LauDaughters("B+", "pi+", "pi+", "pi-", squareDP);
\end{lstlisting}
specifying the types of the parent particle and the three daughters.
The final argument specifies whether or not we will be using the variables
of the SDP (defined in Section~\ref{sec:eff-resol}) and as
such whether these should be calculated --- if they are not required then it
is more efficient to switch off this calculation.

Next to be defined is the efficiency model, including any explicit vetoes
in the Dalitz plane:
\begin{lstlisting}
LauVetoes* vetoes = new LauVetoes();
LauEffModel* effModel = new LauEffModel(daughters, vetoes);
\end{lstlisting}
Without further specification, this will give a uniform efficiency (of
unity) over the DP.
This can be modified by specifying vetoes and/or supplying the efficiency
variation in the form of a histogram, \eg
\begin{lstlisting}
vetoes->addMassVeto(2, 1.7, 1.9); // D0 veto, covering 1.7 < m13 < 1.9 GeV/c^2
TH2* effHist = ...;
effModel->setEffHisto(effHist);
\end{lstlisting}

The next step is to define the signal amplitude model.
Before defining any resonances the first thing to be done is to specify the
Blatt--Weisskopf barrier radii to be used for the different resonances.
This is done by accessing the singleton \texttt{LauResonanceMaker} factory object:
\begin{lstlisting}
LauResonanceMaker& resMaker = LauResonanceMaker::get();
resMaker.setDefaultBWRadius( LauBlattWeisskopfFactor::Parent, 5.0 );
resMaker.setDefaultBWRadius( LauBlattWeisskopfFactor::Light,  4.0 );
\end{lstlisting}
It is possible also to specify whether any of these values should be
floated in the fit.
By default they are fixed to the specified value.
The isobar model can now be created and the various resonances added to it:
\begin{lstlisting}
LauIsobarDynamics* sigModel = new LauIsobarDynamics(daughters, effModel);
LauAbsResonance* reson(0);
reson = sigModel->addResonance("rho0(770)",  1, LauAbsResonance::GS);
reson = sigModel->addResonance("f_0(980)",   1, LauAbsResonance::Flatte);
reson->setResonanceParameter("g1",0.2);
reson->setResonanceParameter("g2",1.0);
reson = sigModel->addResonance("f_2(1270)",  1, LauAbsResonance::RelBW);
reson = sigModel->addResonance("rho0(1450)", 1, LauAbsResonance::RelBW);
reson = sigModel->addResonance("NonReson",   0, LauAbsResonance::FlatNR);
\end{lstlisting}
Note how the returned pointer to the added resonance can be used to modify
parameters of the resonance.
It is also possible to specify that resonance parameters should be floated
in the fit using the \texttt{LauAbsResonance::floatResonanceParameter} member function.

The fit model can now be created based on the isobar model that has just
been defined.
\begin{lstlisting}
LauSimpleFitModel* fitModel = new LauSimpleFitModel(sigModel);
\end{lstlisting}
And the complex coefficients for each resonance defined using one of the
various parametrisations defined in Table~\ref{tab:coeff-sets}, in this
case using a simple magnitude and phase form:
\begin{lstlisting}
std::vector<LauAbsCoeffSet*> coeffset;
coeffset.push_back( new LauMagPhaseCoeffSet("rho0(770)",  1.00,  0.00,  true,  true) );
coeffset.push_back( new LauMagPhaseCoeffSet("f_0(980)",   0.27, -1.59, false, false) );
coeffset.push_back( new LauMagPhaseCoeffSet("f_2(1270)",  0.53,  1.39, false, false) );
coeffset.push_back( new LauMagPhaseCoeffSet("rho0(1450)", 0.37,  1.99, false, false) );
coeffset.push_back( new LauMagPhaseCoeffSet("NonReson",   0.54, -0.84, false, false) );
for (std::vector<LauAbsCoeffSet*>::iterator iter=coeffset.begin(); iter!=coeffset.end(); ++iter) {
    fitModel->setAmpCoeffSet(*iter);
}
\end{lstlisting}
The boolean arguments to the constructor indicate whether the value should
be fixed in the fit.
Since \laura\ version \texttt{v3r1}, the ordering of the components is defined
by the isobar model (first all coherent resonances in order of addition to the isobar model, then all incoherent resonances in order of addition), and the coefficients are automatically rearranged to match that order.
Thus, the user does not need to worry about the order of the components given here.

To either generate toy datasets or to fit data, it is necessary to specify the number of events in each of the signal and (if appropriate) background categories. 
The \laura\ work flow requires that these values are specified also in the case of weighting data, although they are not used.
In the case of fitting data the given values may be floated, as is often appropriate when additional discriminating variables are included in the fit (see Sec.~\ref{sec:other-vars}), or fixed, as is usually necessary when only the DP is being fitted.
In the former case, if the \texttt{doEMLFit} option is specified the fitted values of the yields can be expected to have the correct Poisson uncertainties.
The user should ensure that the specified number of events in the signal and background categories correspond to the number of events in the data file that will be read in to that fit.
The number of signal events is set as follows:
\begin{lstlisting}
LauParameter * nSigEvents = new LauParameter("nSigEvents",500.0,-1000.0,1000.0,false);
fitModel->setNSigEvents(nSigEvents);
\end{lstlisting}
The number of experiments, as well as which experiment to start with, must
also be specified:
\begin{lstlisting}
const int nExpt(1);
const int firstExpt(0);
fitModel->setNExpts( nExpt, firstExpt );
\end{lstlisting}
The above example values are appropriate when fitting a single data sample.
If generating and/or fitting toy pseudoexperiments then a larger number of
experiments can be specified.
Optionally, models for one or more background categories can also be specified along with the corresponding event yields.

Finally there are various control and configuration options that can be
set, which are listed in Table~\ref{tab:config-options}.
Once all the configuration is completed, the required operation (toy
generation, fitting, or event weighting) can be run:
\begin{lstlisting}
fitModel->run( command, dataFile, treeName, rootFileName, tableFileName );
\end{lstlisting}
where all the arguments are strings, with the following purposes
\begin{itemize}
\setlength\itemsep{0mm}
\item
\texttt{command} is either ``gen'', ``fit'' or ``weight''; 
\item
\texttt{dataFile} and \texttt{treeName} specify the name of the \root file
and tree from which the data should be read or to which the generated data
should be written (depending on the mode of operation);
\item
\texttt{rootFileName} specifies the name of the \root file to which the
results of the fit (fit convergence status, likelihood, parameter values,
uncertainties, correlations, etc.) should be saved;
\item
\texttt{tableFileName} specifies the name of the \LaTeX~file to which the
results of the fit or generation should (optionally) be written.
\end{itemize}
The \texttt{run} function performs the initialisation of all the PDFs that
have been defined and checks the internal consistency of the model, \eg\ that all event categories have a PDF defined for each variable.
All the parameters of the model are gathered into a single list and the
bookkeeping of those parameters that affect the DP normalisation
is set up.
Any constraints on the model parameters, both on individual and on
particular specified combinations of parameters, are also recorded.
Following these initialisation steps, the particular routines for toy
generation, fitting, or weighting of events are called, depending on the specified command.
Each of these routines are described in the following sections.

\begin{table}[!htb]
\caption{Control and configuration options for all fit models.}
\label{tab:config-options}
\centering
\begin{tabular}{l L{0.65\textwidth}}
\hline
Function name  & Description \\
\hline
\texttt{useDP}                 & Toggle use of the DP PDF in the likelihood \\
\texttt{doSFit}                & Activate per-event weighting of events in likelihood fit \\
\texttt{doEMLFit}              & Toggle use of the extended maximum likelihood fit \\
\texttt{doTwoStageFit}         & Toggle use of the two-stage fit \\
\texttt{useAsymmFitErrors}     & Toggle determination of asymmetric uncertainties from the fit, \eg\ using MINOS routine in \minuit \\
\texttt{useRandomInitFitPars}  & Toggle randomisation of initial values of isobar parameters \\
\texttt{addConstraint}         & Add a Gaussian constraint on a specified combination of fit parameters \\
\texttt{doPoissonSmearing}     & Toggle Poisson smearing of event yields in toy generation \\
\texttt{writeLatexTable}       & Toggle generation of a \LaTeX-formatted summary table \\
\texttt{writeSPlotData}        & Activate calculation of event \sPlot\ weights \\
\texttt{compareFitData}        & Activate automatic generation of toy MC datasets based on fit results \\
\hline
\end{tabular}
\end{table}

\tikzstyle{decision} = [diamond, draw, fill=blue!20, text width=7.5em, text badly centered, inner sep=0pt]
\tikzstyle{block} =  [rectangle, draw, fill=blue!20, text width=11.0em, text centered, rounded corners, minimum height=5ex, node distance=3cm]
\tikzstyle{line} = [draw, -latex']
\tikzstyle{cloud} = [draw, ellipse,fill=red!20, node distance=3cm, minimum height=2ex]

\subsection{Toy generation}
\label{sec:toy-generation}

The generation of ensembles of simplified pseudoexperiments, or toys, is of great importance in \laura.
Not only is toy generation used to test the performance of a fitter, to test its stability and study potential biases and/or correlations between fitted parameters, it can also be used to determine uncertainties on parameters (see Sec.~\ref{sec:uncertainties}).
Large samples of toy experiments, generated according to the results of a fit, can also be used to display the results of the fit projected onto any variable in any region of the phase space.  
This is often a convenient way of drawing the result of the fit, including contributions from separate components.

The first action specific to the generation of toy pseudoexperiments is to
ensure that the value of each fit parameter is set to that to be used for
generation, \ie the \texttt{genValue} of the \texttt{LauParameter}.
Then an ntuple (specifically, a \root\ {\tt TTree}) is created in which the generated events will be stored.
The branches to be stored are determined based on the PDFs that have been
configured in the initialisation step, and will include:
\begin{itemize}
\setlength\itemsep{0mm}
\item
\texttt{iExpt} and \texttt{iEvtWithinExpt}: these indicate to which
pseudoexperiment this event belongs and the position of the event within
that experiment;
\item
\texttt{evtWeight}: a weight for the event that is normally unity but can
take other values\footnote{
Where toy events are being generated automatically based on the outcome of
a fit and it happens that one of the category yields has been determined to
be a negative number, the number of events generated for that category will
be the absolute value of the determined yield and the events will be given
a weight of $-1$. This allows the PDFs to be plotted correctly by simply
plotting the weighted events. Care must be taken when fitting such samples
since the fit will not take account of these weights.};
\item
a set of MC truth branches that indicate whether a given event was
generated as signal or one of the background categories, \eg\ \texttt{genSig};
\item
\texttt{efficiency}: the value of the signal efficiency at the generated
DP position (only stored for signal events);
\item
the DP position in terms of the invariant masses and helicity
angle cosines: \texttt{m12}, \texttt{m13}, \texttt{m23},
\texttt{m12Sq}, \texttt{m13Sq}, \texttt{m23Sq}, \texttt{cosHel12},
\texttt{cosHel13}, \texttt{cosHel23};
\item
optionally, the position in the SDP coordinates:
\texttt{mPrime} and \texttt{thPrime};
\item
the values of all other variables used by additional PDFs (see
Section~\ref{sec:other-vars}).
\end{itemize}

\begin{figure}
\centering
\begin{tikzpicture}[node distance=4cm, auto]
	\node [block]                                        (genpos)        {Generate random DP point};
	\node [block, right of=genpos, node distance=6cm]    (calcamps)      {Calculate amplitudes, $A_i$, for all components};
	\node [block, below of=calcamps]                     (calcinco)      {Calculate intensities, $I_j$, for all incoherent components};
	\node [block, below of=calcinco]                     (calcint)       {Calculate total intensity and multiply by efficiency: $I_\mathrm{tot} = \left( |\sum_i A_i|^2 + \sum_j I_j \right) \times \epsilon$};
	\node [block, below of=calcint]                      (throw)         {Generate random number $x \in \left[0,1\right)$};
	\node [decision, below of=throw]                     (while)         {\large $x < \frac{I_\mathrm{tot}}{I_\mathrm{max}}$};
	\node [decision, below of=while]                     (check)         {$I_\mathrm{tot} < I_\mathrm{max}$};
	\node [block, below of=check]                        (accept)        {Accept event};
	\node [block, left of=check, node distance=6cm]      (abort)         {Abort generation};
	\path [line] (genpos) -- (calcamps);
	\path [line] (calcamps) -- (calcinco);
	\path [line] (calcinco) -- (calcint);
	\path [line] (calcint) -- (throw);
	\path [line] (throw) -- (while);
	\path [line] (while) -- node [near start] {yes} (check);
	\path [line] (while) -| node [near start] {no}  (genpos);
	\path [line] (check) -- node [near start] {yes} (accept);
	\path [line] (check) -- node [near start] {no} (abort);
\end{tikzpicture}
\caption{
Schematic of the \texttt{LauIsobarDynamics::generate} function.
}
\label{fig:sig-gen-flowchart}
\end{figure}
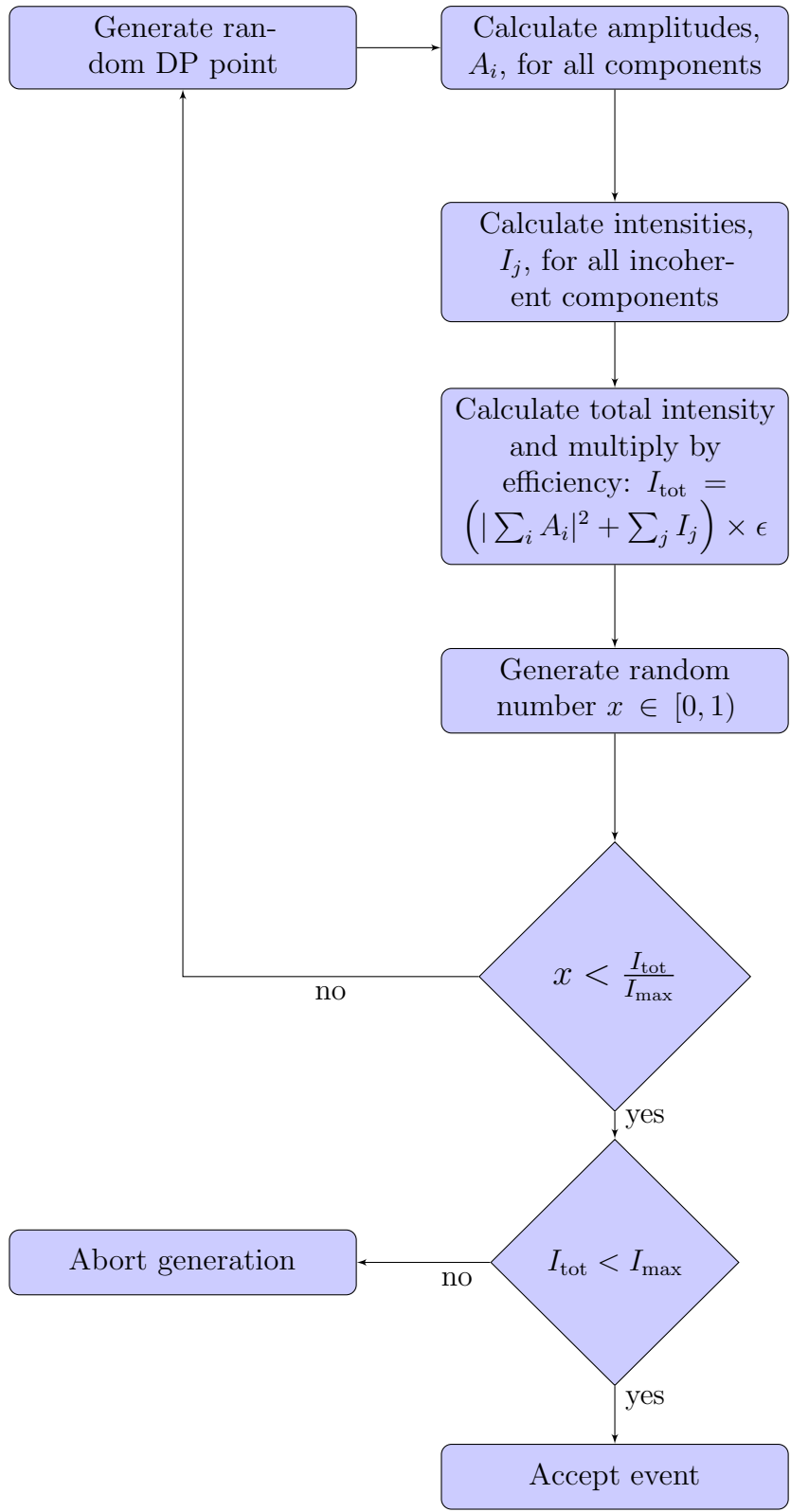

Each experiment is then generated in turn, with the following procedure
being followed for each:
\begin{enumerate}
\item
Determine the number of events to generate for each category.
This will either be the value of the yield parameter for the category or,
if \texttt{doPoissonSmearing} is specified (see Table~\ref{tab:config-options}), will be sampled from a Poisson distribution whose mean is equal to the yield parameter value.
\item
For each category, generate the number of requested events.
Truth information is stored as to which category the event belongs in
branches named ``\texttt{gen}CAT'', where ``CAT'' is replaced by the name of the
category.  For a given event, the variable corresponding to the category
being generated will have value 1, while all others will have value 0.
\item
For signal events, the generation of the DP position is performed
by the \texttt{LauIsobarDynamics::generate} function, the operation of
which is shown in Fig.~\ref{fig:sig-gen-flowchart}.
In essence, it is an ``accept/reject'' method in which the value of the
signal model intensity at a given point ($I_\mathrm{tot}$ in
Fig.~\ref{fig:sig-gen-flowchart}) is compared with a ``ceiling value''
($I_\mathrm{max}$ in Fig.~\ref{fig:sig-gen-flowchart}).
Note that the signal model intensity includes effects due to efficiency,
vetoes and resolution.
Since, in general, it is not computationally efficient to evaluate
analytically the maximum value of the intensity, an approximate value is
provided by the user via the \texttt{LauIsobarDynamics::setASqMaxValue}
function.
During the generation process it is checked that a larger value is not
encountered.
If it is, which would indicate that the generation is biased, the ceiling
value is increased and the generation processes starts again from the
beginning --- all previously generated experiments are discarded.
Similarly, too large a ceiling value can make the generation extremely
inefficient, so it is also attempted to detect this condition and correct it.
If a self cross-feed component has been included in the signal description,
once a DP point has been generated it may then be smeared
according to the procedure outlined in Section~\ref{sec:resol}.
Finally, the values of any other discriminating variables can be generated
from the corresponding PDFs.
\item
For background events, the DP position is generated by the
\texttt{LauBkgndDPModel::generate} function.
This performs a similar accept/reject routine to that used to generate
signal events but here the (optionally interpolated) histogram height is
compared with the maximum value of the histogram.
There is therefore no need to verify that the encountered value is less
than the maximum since this is true by construction.
Generated values within a veto region are automatically rejected.
Finally, the values of any other discriminating variables are generated
from the corresponding PDFs.
\end{enumerate}
Once all experiments have been generated successfully, all events are
written to the ntuple.

\subsubsection{Embedding events from a file}
\label{sec:embedding-events}

One possible variation of the scheme just outlined is that events for one
or more components of the model are loaded from a \root file rather than
being generated from the model PDFs.
These events could, for example, be taken from samples of full detector
simulation or from data control samples.
This procedure can be useful to check for biases that can result due to
certain experimental effects not being taken into account in the model 
(for example, experimental resolution or correlations between fit variables).
The events are sampled randomly from the provided file.
Depending on the number of events within the samples it may be necessary to
sample with replacement, \ie to allow events to be used more than once.

The exact procedure to enable the embedding of events from a \root file
varies slightly depending on the fit model class being used.
For \texttt{LauSimpleFitModel} there are two functions \texttt{embedSignal}
and \texttt{embedBkgnd}, which can be used to embed signal and background
events.
For \texttt{LauCPFitModel} there are four functions, to allow separate sources of events in the case that the parent candidate is the particle or antiparticle.
The arguments taken by the various functions are, however, essentially the
same:
\begin{itemize}
\item
\texttt{bkgndClass}: in the case of the functions for embedding background
events, the name of the background class must be specified as the first
argument;
\item
\texttt{fileName}: the name of the \root file from which the events should
be loaded;
\item
\texttt{treeName}: the name of the \texttt{TTree} object within the file in
which the events reside;
\item
the arguments \texttt{reuseEventsWithinEnsemble} and
\texttt{reuseEventsWithinExperiment} control whether or not the sampling of
events should allow replacement, firstly within the context of the entire
ensemble of experiments being generated (\ie replacement occurs once each
experiment has been generated) and secondly within each experiment (\ie
replacement occurs immediately);
\item
\texttt{useReweighting}: in the case of embedding signal events, this
argument controls whether an accept/reject routine should be used (based on
the configured amplitude model) when sampling the events.
As discussed in Sec.~\ref{sec:weighting-events}, reweighting allows an existing sample to be reused with an alternative model, which can avoid high computational costs associated with obtaining a new sample.
This option requires that the both the generated and reconstructed
DP coordinates are available in the provided sample since the
generated point is used to find the amplitude value to be used in the
accept/reject routine but the reconstructed point is the one saved to the
output ntuple.
This ensures that the effects of experimental resolution and
misreconstruction are preserved.
\end{itemize}
For example, to embed events for the signal component, sampling with
immediate replacement, and performing the accept/reject routine:
\begin{lstlisting}
fitModel->embedSignal( signalSampleFileName, signalSampleTreeName, kTRUE, kTRUE, kTRUE );
\end{lstlisting}
As for other control and configuration functions, such as those in
Table~\ref{tab:config-options}, these should be called prior to the
invocation of the \texttt{LauAbsFitModel::run} function.

\subsection{Fitting}
\label{sec:fitting}

While the primary goal of the \laura\ package is to facilitate fits to the DP distributions of various decays of interest in data, it is also essential to be able to fit simulated samples such as ensembles of toy experiments generated as discussed above.
The same work flow is used for both types of fits. 
Unbinned maximum likelihood fits are the default in \laura, with extended
maximum likelihood fits also an option as noted in Table~\ref{tab:config-options}.

The first action specific to the routine for fitting is to create the ntuple in which to store the results of the fit.
The branches to be stored are determined based on the PDFs that have been
configured in the initialisation step, and will include:
\begin{itemize}
\setlength\itemsep{0mm}
\item
\texttt{iExpt}: an integer that indicates to which pseudoexperiment the
results in this entry in the ntuple belong (zero for a fit to data);
\item
\texttt{fitStatus}: an integer that indicates whether or not the fit has
converged and the degree of accuracy of the resulting covariance matrix (0:
fit has failed to converge; 1: fit has converged but covariance matrix is
only approximate; 2: the covariance matrix is a full matrix but is forced
to be positive definite; 3: the covariance matrix is full and accurate);
\item
\texttt{EDM}: the estimated distance of the negative log likelihood to the true minimum of the function;
\item
\texttt{NLL}: the minimised value of the negative log likelihood;
\item
for each fit parameter: its fitted value, its value at initialisation (for toy data, this will be the true value) and, if it was floated in the fit, its
uncertainty (including asymmetric uncertainties, if calculated), pull and
global correlation coefficient;
\item
for each pair of floated fit parameters: their correlation coefficient;
\item
any extra parameters defined by the fit model, for example, the fit
fractions of each component and the interference fit fractions.
\end{itemize}
If information on the per-event likelihood values is to be stored in order
to allow the calculation of \sPlot\ weights~\cite{Pivk:2004ty},
this ntuple is also created.

The file containing the data to be fitted is opened and the ntuple retrieved.
An error will be returned if the file cannot be opened, if the ntuple cannot be found or if it does not have a flat format, \ie if it contains stored arrays.
Each experiment is then fitted in turn, with the following procedure being
followed for each:
\begin{enumerate}
\item
The data for the given experiment is read into memory.
The fit model then passes the data to each of the PDFs such that they can
store the values of any variables that they require and also calculate and
cache as much information as they can towards (and possibly including) the
likelihood value for each event.
\item
Optionally (see according to the specification of \texttt{useRandomInitFitPars}; see Table~\ref{tab:config-options}), the initial values of the isobar
coefficient parameters are randomised.
\item
The fitter is initialised by providing it with the list of fit parameters
and two boolean options to control whether a two-stage fit (see Sec.~\ref{sec:two-stage}) 
is to be performed and whether asymmetric uncertainties on the fit parameters should be determined.
\item
The minimisation of the negative log likelihood is then performed and the
uncertainties on the parameters and their correlations are determined.
The fit status information and the covariance matrix is stored, and the
final values and uncertainties are written back to the fit parameter
objects.
\item
Some final manipulation of the fit parameter values is performed, \eg such
that phases lie within a particular range, and their pull values are
calculated.
Quantities derived from the fitted parameters, such as fit fractions, are
also calculated.
All this information is written to the fit results ntuple.
\item
Optionally (see \texttt{writeSPlotData} in Table~\ref{tab:config-options}),
per-event likelihood information is stored in a separate ntuple.
\item
Optionally (see \texttt{compareFitData} in Table~\ref{tab:config-options}),
and only if the fit was successful, toy MC pseudoexperiments are generated
based on the fitted values of the parameters.
\end{enumerate}
Once all experiments have been fitted, information is printed on the
success rate of the fits and the fit results ntuple is written to file.
Optionally (see \texttt{writeSPlotData} in Table~\ref{tab:config-options}),
per-event weights are calculated from the likelihood information, using the
\sPlot\ method~\cite{Pivk:2004ty}, and written to an ntuple.

Due to the often complicated dependence of the likelihood on many
parameters in a typical DP analysis, generally there will appear local
minima in the likelihood as well as the global minimum.
Depending on the starting values of the fit parameters it is possible that
the fit will converge to one of the local minima.
As such it is highly advisable to perform multiple fits to each data sample
using randomised starting values for the parameters of the isobar
coefficients.
As mentioned in Table~\ref{tab:config-options}, the control function
\texttt{useRandomInitFitPars} can be used to toggle this behaviour.
It is then required only to re-run the executable to obtain a fit based on
a new randomised starting point.
The applications in the examples directory of the package all include a
compulsory command-line option to provide an integer to label the
particular fit (or set of fits, if fitting multiple experiments), which is
then incorporated into the name of the file containing the ntuple of
results for that fit or set of fits.
Once all fits have been performed, the fit that returns the best likelihood
value can then be selected as the likely global minimum for each experiment.
In the examples directory of the package a utility class
\texttt{ResultsExtractor} (with accompanying application code) is provided
to simplify the process of checking for multiple solutions and extracting
the results of the fit that is the candidate global minimum for each
experiment.
This application writes out a \root file that contains an ntuple that is
filled with the candidate global minimum results for each experiment, and a
histogram for each experiment that shows the frequency with which each
negative log likelihood value is obtained.

\subsubsection{Determination of uncertainties}
\label{sec:uncertainties}

In addition to central values, the fit will return estimates of the uncertainties of each fitted parameter.
However, it is typical to want to obtain results not only for the fitted parameters (\eg\ complex coefficients for each component of the model) but also for derived parameters (\eg\ fit fractions and interference fit fractions).
The central values of such derived parameters can be obtained algebraically, however non-linear effects typically render unreliable the determination of uncertainties by standard error propagation.
Instead, both central values and uncertainties can be obtained from ensembles of pseudoexperiments generated according to the result of the fit to data.  
In this approach the uncertainty is obtained from the spread of the distribution of the obtained values in the pseudoexperiments of the parameter of interest.
Since this procedure guarantees correct coverage of the uncertainties, and also allows study of effects such as biases and non-linear correlations in the results, it is often preferable to use it also for directly fitted parameters.
A further alternative to determining uncertainties on derived parameters is by bootstrapping~\cite{bootstrap,barlow}.

\subsubsection{Two-stage fit}
\label{sec:two-stage}

Fits with large numbers of free parameters can be slow to converge, and may be unstable.
These problems can be ameliorated with a two-stage fit, in which certain parameters are initially fixed to specified values while all other parameters are floated.
In the second stage, all parameters are floated ensuring that correlations between parameters are correctly accounted for.  
Note that parameters which are floated only in the second stage cannot have randomised initial values.

One situation in which this fit procedure has been found to be particularly useful is when resonance parameters such as masses and widths are to be floated.
Another is when \CP-violation parameters are to be determined.
It may be noted that in the \texttt{CartesianCP} coefficient set, one can simultaneously swap $x \leftrightarrow \delta_x$, $y \leftrightarrow \delta_y$ with an effect that is equivalent to rotating all amplitudes for $\Pbar$ decays by $\pi$.
Such a transformed set of parameters may be hard for the fit to distinguish from the untransformed case.
As such a two-stage fit, in which \CP-violation parameters are initially fixed to zero before being floated in the second stage, can help to resolve ambiguities and reject unrealistic multiple solutions.

\subsubsection{Relations between fit parameters}
\label{sec:formula-pars}

There are a number of situations where a given fit parameter can appear in
more than one PDF within the likelihood function.
Such cases must be flagged in the construction of the fit model in order to
ensure that such a parameter is provided only once to the minimiser.
Consider the following example that concerns a fit to the invariant
mass distribution of candidate $\Bp\to\Kp\pip\pim$ decays.
The signal is modelled as a Gaussian function, which has two parameters:
the mean $\mu$ (corresponding to the mass of the \Bp) and width $\sigma$
(corresponding to the experimental resolution).
A possible background arises from the decay
$\Bp\to\etapr\Kp;\etapr\to\pip\pim\gamma$, which is modelled using an ARGUS
threshold function (see App.~\ref{sec:argus-pdf}).
The threshold parameter of this function is the \Bp mass and as such it
needs to be encoded in the likelihood function that values of the mean of
the signal distribution and the threshold of this background are due to the
same parameter.
This can be achieved in \laura using the \texttt{LauParameter::createClone}
function, as follows:
\begin{lstlisting}
const Double_t mbMin(5.10);
const Double_t mbMax(5.60);

LauParameter* sig_mb_mean  = new LauParameter("sig_mb_mean",  5.28, 5.26, 5.30);
LauParameter* sig_mb_sigma = new LauParameter("sig_mb_sigma", 0.20, 0.10, 0.30);
std::vector<LauAbsRValue*> mbPars;
mbPars.push_back(sig_mb_mean);
mbPars.push_back(sig_mb_sigma);
LauAbsPdf* sig_mb_pdf = new LauGaussPdf("mB", mbPars, mbMin, mbMax);

LauParameter* bkg1_mb_m0 = sig_mb_mean->createClone();
LauParameter* bkg1_mb_xi = new LauParameter("bkg1_mb_xi", 20.0, 0.0, 50.0);
mbPars.clear();
mbPars.push_back(bkg1_mb_m0);
mbPars.push_back(bkg1_mb_xi);
LauAbsPdf* bkg1_mb_pdf = new LauArgusPdf("mB", mbPars, mbMin, mbMax);
\end{lstlisting}
For the isobar coefficients, it is possible to clone the parameters related
to a particular coefficient using the \texttt{LauAbsCoeffSet::createClone}
function.
One can specify precisely which parameters should be cloned;
for example, to allow only \CP-violating parameters to be
cloned while the \CP-conserving parameters are still free to vary
independently.

Extending the previous example to consider a second possible background
from $\Bp\to\Kp\pim\pip\piz$ decays, where the \piz is not reconstructed,
we encounter a case where the threshold parameter is now the difference
between the \Bp and \piz masses.
This scenario can be accommodated in \laura using a \texttt{LauFormulaPar}
object, the constructor of which takes as arguments two strings, which are the
name of the compound parameter and the formula to be used to combine the values
of the input parameters, and a \texttt{std::vector} of \texttt{LauParameter}
objects, which are the input parameters.
For the example in question this would be:
\begin{lstlisting}
std::vector<LauParameter*> inputPars;
LauParameter* mpiz = new LauParameter("mpiz", LauConstants::mPi0);
inputPars.push_back(sig_mb_mean);
inputPars.push_back(mpiz);
LauFormulaPar* bkg2_mb_m0 = new LauFormulaPar("bkg2_mb_m0", "[0] - [1]", inputPars);
\end{lstlisting}
This \texttt{LauFormulaPar} object can be used as one of the parameters for
the PDF:
\begin{lstlisting}
LauParameter* bkg2_mb_xi = new LauParameter("bkg2_mb_xi", 20.0, 0.0, 50.0);
mbPars.clear();
mbPars.push_back(bkg2_mb_m0);
mbPars.push_back(bkg2_mb_xi);
LauAbsPdf* bkg2_mb_pdf = new LauArgusPdf("mB", mbPars, mbMin, mbMax);
\end{lstlisting}
Note the syntax of the formula expression (from the \root \texttt{TFormula}
class) where the input parameters are referred to by their index in the
\texttt{std::vector}.
This syntax allows the usual arithmetic operators to be used as well as
functions such as those contained in the \texttt{TMath} namespace.

\subsubsection{Blinding of fit parameters}

In analyses of data when one is searching for a new signal or searching for
\CP violation it is quite a common practice to ``blind'' the value of the
observables of interest until the event selection and analysis procedures
have been finalised~\cite{Harrison:2002ip,Klein:2005di}.
This is commonly achieved by applying an offset, whose value is unknown to
the analysts but is determined uniquely from a so-called ``blinding
string'', when the parameter value is passed to the minimiser or is printed
out or otherwise saved at the end of the fit but the true, ``unblind'',
value is used when calculating the value of the likelihood function.
This same technique is used in \laura in the \texttt{LauBlind} class.
The hash of the blinding string is used as the seed for a random number
generator, using which a value is sampled from a normal distribution.
The offset is then calculated by multiplying this random number by a
scaling factor.
The blinding string and scaling factor are supplied by the user as follows:
\begin{lstlisting}
LauParameter* nSig = new LauParameter("signalEvents", 1000.0, -1000.0, 2000.0, kFALSE);
nSig->blindParameter("dalitzplot",1000.0);
\end{lstlisting}
Some care must be taken when defining the scaling factor and the allowed
range for the parameter to try and avoid situations where the parameter
hits one of the limits.
It is also advisable to use different blinding strings for each observable
that is to be blinded.

\subsubsection{Fitting with external constraints}
\label{sec:gauss-con}

When the value of a parameter is known from other measurements to within
some precision it can be useful to incorporate that information into the
likelihood function.
This is most commonly achieved by multiplying the likelihood by a Gaussian
term, where the mean and width of the Gaussian are the central value and
uncertainty from the external source and the abscissa is the value of the
parameter in the fit.
The addition of such terms to the likelihood expression can be carried out
in \laura using the function \texttt{LauParameter::addGaussianConstraint},
where the arguments are the central value and uncertainty from the external
measurement.

In the event that the constraint that needs to be applied is not just on
the value of a single parameter but on some combination of fit parameters,
this can be achieved via the \texttt{addConstraint} function that is
available in all fit model classes.
In addition to the mean and width of the Gaussian function, this function takes the following arguments:
\begin{itemize}
\item
a formula string that specifies how to combine the parameters, which should
use the syntax specified by \texttt{LauFormulaPar} (see Sec.~\ref{sec:formula-pars});
\item
a \texttt{std::vector} of strings that are the names of the fit parameters
(\ie\ the \texttt{LauParameter} objects in the list of fit parameters) whose
values are to be used in the formula to calculate the abscissa of the
Gaussian function.
Note that the order of the names within the vector must match the ordering
specified in the formula.
\end{itemize}
This feature can also be used to perform a scan of the likelihood as a function of any combination of fit parameters.
This can be achieved by setting the mean of the Gaussian constraint to one of a range of fixed values, in turn, and setting the width to a sufficiently small value that the quantity is effectively fixed.
The negative log-likelihood values obtained from independent fits at each of the scan points can be converted into whatever format is most convenient for the user.  

\subsubsection{Fitting with background subtraction via \sPlot\ weights}
\label{sec:sfit}

It is possible to perform fits with background subtraction via \sPlot\ weights as proposed in Ref.~\cite{Xie:2009rka}.
In this approach, the weights are obtained from a fit to a discriminating variable in which the signal and any background categories can be distinguished~\cite{Pivk:2004ty}; for the purposes of this discussion this will be considered to be the mass of a $B$ meson candidate for a particular decay to a three-body final state.
The advantage of this approach is that it becomes unnecessary to have models for the DP distributions of the background components, and therefore the discussion of Sec.~\ref{sec:backgrounds} becomes irrelevant.  
However, the formalism relies on the discriminating variable being uncorrelated with the DP variables.  
This is likely to be a good approximation for the signal, and also reasonable for combinatorial background.
However, additional sources of background such as those involving misidentified decays are likely to have significant correlations between the reconstructed mass and the DP position.  
Therefore this approach is only valid in the case that such backgrounds are negligible across the whole $B$ candidate mass range --- not only the signal region (in this approach to background subtraction, there is no concept of signal region).

Within \laura\ this approach to fitting can be implemented by calling the \texttt{doSFit} member function of \texttt{LauAbsFitModel}, which takes as argument the name of the branch that contains the relevant weights, as shown in Table~\ref{tab:config-options}.
It is also possible to pass an optional second argument, which is a scaling factor that is needed in order to obtain correct uncertainties from the likelihood; this is unnecessary if the uncertainties will be evaluated from pseudoexperiments.

\subsubsection{Simultaneous fitting of independent data samples}
\label{sec:jfit}

Simultaneous fits to independent data samples have a variety of applications.
For example, the samples may correspond to different experimental
conditions and so each may require a different model of the efficiency and
background distributions.
A simultaneous fit to the various sub-samples allows the statistical power
of the full sample to be used to determine the parameters of interest,
while accounting as accurately as possible for the changes in the
experimental environment.
Furthermore, it simplifies the evaluation of systematic uncertainties, in
particular the treatment of sources that are correlated among the samples.
The technique can also be employed in order to extract information that
would otherwise not be feasible, for example to perform a coupled-channel
analysis or to exploit flavour symmetries or other relations between decay
modes, in order to extract information on \CP violation observables, as
recently carried out in Ref.~\cite{Aaij:2016bqv}.

The implementation of simultaneous fitting in \laura is based on the \jfit
framework, an overview of which is given here and explained in more detail
in Ref.~\cite{Ben-Haim:2014afa}.
The framework is based around a master--worker architecture in which the
master drives the minimiser, combining the likelihood values for each
category that have been calculated by a number of workers.
The master and each of the workers run as separate processes that
communicate via network sockets.
This means that the calculation of the likelihood for each category is
performed in parallel, increasing the speed of the computations if
sufficient CPU cores are available.
Note that the framework even allows the master and various worker processes
to run on separate hosts with a modest performance penalty due to the increased latency in the communication between the processes, which depends on
the network speed between the hosts, as shown in Table~\ref{tab:simfit-speed}.

In order to modify an existing fit code to act as a worker process it is
sufficient to modify only the final line:
\begin{lstlisting}
fitModel->run( command, dataFile, treeName, rootFileName, tableFileName );
\end{lstlisting}
to become:
\begin{lstlisting}
if ( command == "fit" ) {
    fitModel->runSlave( dataFile, treeName, rootFileName, tableFileName, host, port );
} else {
    fitModel->run( command, dataFile, treeName, rootFileName, tableFileName );
}
\end{lstlisting}
where the additional arguments to the \texttt{runSlave} function are a
string and an unsigned integer that specify the hostname (\eg ``localhost''
when all processes are running on the same host) and port number on which
the master process is listening for connections.
The code to start a master process is extremely simple:
\begin{lstlisting}
LauSimFitMaster master( nSlaves, port );
master.runSimFit( ntupleName, nExpt, firstExpt, useAsymmErrors, twoStageFit );
\end{lstlisting}
The arguments to the constructor specify the number of worker processes
that are expected to connect and the port on which it should listen for
connections (a value of 0 indicates that it should use the first available port).
The arguments to the \texttt{runSimFit} function specify the name of the
file to which the fit results ntuple should be written, the number of
experiments to be run and the ID of the first experiment, whether or not
asymmetric uncertainties should be determined, and whether or not the fit
consists of two stages.
A shell script can then be used to launch a master process and the
appropriate number of worker processes.
An example script, \texttt{runMasterSlave.sh}, and source code for master
and worker executables, \texttt{Master.cc} and \texttt{Slave.cc}, are
included in the package examples.

During the set-up phase, each of the worker processes provides to the
master the list of fit parameters needed for that worker to calculate the
value of its likelihood function.
The master stores this information and configures the minimiser.
As part of this procedure, it must decide which parameters are common to
some or all workers, such that each of these are provided only once to the
minimiser.
This decision is based purely on the parameters' names, so it is vital that
these are correctly set when building the fit model in each of the workers.
In order to aid the user, a summary is printed by the master process at the
end of the set-up of the fit parameter lists.

\subsection{Weighting events}
\label{sec:weighting-events}

Typically when generating samples of full detector simulation for
three-body decays they are generated uniformly either in phase space or in
the SDP, such that they can be used to describe the
variation of the efficiency as a function of phase-space position.
While it would often be useful to also have samples that are generated
according to an amplitude model or indeed a selection of models, it can be
infeasibly expensive in terms of CPU time and disk space, particularly at
hadron collider experiments.
It is therefore convenient to be able to apply per-event weights to such
samples so that the weighted distributions reproduce those that would be
obtained if the samples had been generated according to a particular model.
The workflow to achieve this goal is described here.

The common parts of the workflow are the same as if preparing to generate
toy MC samples or to fit a data sample.
The only exception is that the argument to the \texttt{LauDaughters}
constructor that determines whether or not the SDP
coordinates are calculated takes on an additional significance --- it
indicates to the weighting procedure whether the provided sample was
originally generated uniform in the conventional or square Dalitz plot.
In case the original sample was not generated as a uniform distribution in either DP or SDP variables, this must be accounted for by applying a user-calculated  correction to the weighting factor.

The weights are calculated using the MC-truth coordinates of the events (so
as to preserve any resolution/migration effects from the full simulation).
Hence, it is required that these coordinates are available in the input file
with the following variable names: ``\texttt{m13Sq\_MC}'' and ``\texttt{m23Sq\_MC}''.
The weight is calculated as the total amplitude-squared divided by the
maximum value of the amplitude squared, which should be set by the user (as
described in Sec.~\ref{sec:toy-generation}).
If the input sample was generated uniformly in the SDP then the
weight is multiplied by the Jacobian of Eq.~(\ref{eq:sqdp-jacobian}).
The weights are written to an ntuple in a new file, which has the same name
as the input data file with ``\_DPweights'' appended.
The variables \texttt{iExpt} and \texttt{iEvtWithinExpt} are also written
to the ntuple and an event index is built from their values.
This allows the weights ntuple to be used in conjunction with the input
data ntuple via the \root ``friend tree'' mechanism.

\section{Performance}
\label{sec:performance}

A DP fitting package such as \laura\ must be highly performing in several different ways.  
First, it must allow the user to obtain a good fit to data, and to evaluate the goodness of the fit.
Several different methods to quantise the goodness of a multidimensional fit have been proposed in the literature and those which are available in \laura\ are described below.
A selection of examples of results obtained using \laura\ is then given.
Another important performance metric is the speed of execution; this is discussed for a number of example uses of the package.

\subsection{Goodness of fit}
\label{sec:performance:gof}

Evaluating the level of agreement between an amplitude fit and the data can be difficult. 
Three methods to perform this task are discussed further; a two-dimensional binned $\chi^{2}$ test, a mixed-sample test and a point-to-point dissimilarity test.
These methods are described in detail in Ref.~\cite{Williams:2010vh}.
It should be noted that all methods test whether the data and the model are consistent to within the statistical uncertainty of the data sample; in some cases it may be necessary to consider in addition whether systematic effects could lead to differences between the data and the model.

\subsubsection{Binned $\chi^{2}$ method}
The SDP distribution of the data is divided into bins with approximately equal bin content using an adaptive binning technique. 
The same binning distribution is applied to a sample of toy events that are generated from the amplitude fit model. 
A standard $\chi^{2}$ test is then performed to compare the data and toy distributions within the chosen binning scheme. 
The relevant test statistic is 
\begin{equation}
\chi^{2} = \sum^{n_{\rm bins}}_{i=1}\frac{\left(d_{i} - t_{i}\right)^{2}}{t_{i}},
\end{equation}
where $d_{i}$ and $t_{i}$ are the number of events in the $i^{\rm th}$ bin from data and toy, respectively, and $n_{\rm bins}$ is the number of bins.
Note that the generated toy sample can be much larger than the data, with the $t_i$ values obtained by scaling appropriately to correspond to the expectation for the data in each bin from the result of the fit.

A drawback of this method is that the minimum number of events in each bin should not be too small, in order to have a reliable test statistic.
But it should also not be too large, as this will cause the bins sizes to increase, leading to a loss of sensitivity to the variation of the amplitude over small scales. 
Typically a minimum number of events per bin of around 20 can be used, but the user should verify for themselves if this is appropriate in their case.

\subsubsection{Mixed-sample method}
The mixed-sample method tests how likely it is that the data and toy samples, produced from the fit model, come from the same parent distribution by evaluating
\begin{equation}
T_H = \frac{1}{n_{k}\left( n_{\rm data} + n_{\rm toy}\right)} \sum^{n_{\rm data} + n_{\rm toy}}_{i=1} \sum^{n_k}_{k=1} I\left(i,k\right),
\end{equation}
where $n_{\rm data}$ and $n_{\rm toy}$ are the number of data and toy events, respectively. 
The number of nearest-neighbours to each data or toy data point considered by the test is given by $n_{k}$. 
The term $I\left(i,k\right)$ is equal to 1 if the $i^{\rm th}$ event and its $k^{\rm th}$ neighbour belong to the same sample and is 0 otherwise. 
Reference~\cite{Williams:2010vh} suggests that $n_{k} = 10$ and $n_{\rm toy} = 10 n_{\rm data}$ are sensible values. 

The statistic $T_H$ can be calculated many times, by using subsamples of the data and toy events, to build up a distribution of values.
The quantity used to evaluate goodness-of-fit is $\left(T_H - \mu_{T}\right)/\sigma_{T}$, where $\mu_{T}$ and $\sigma_{T}$ are the mean and standard deviation of $T_H$, respectively. 
Thus, by definition, the distribution of $\left(T_H - \mu_{T}\right)/\sigma_{T}$ has a mean of 0 and a width of 1 in the case that the data and toy samples are identical.

\subsubsection{Point-to-point dissimilarity method}
The consistency of a data sample and a toy sample generated from a model obtained by fitting the data can also be assessed using the following test statistic,
\begin{equation}
  \label{eq:ptpdm}
T_h = \frac{1}{n^{2}_{\rm data}} \sum^{n_{\rm data}}_{i,j>i} \psi\left(|\vec{x}^{\rm data}_{i} - \vec{x}^{\rm data}_{j}|\right)
- \frac{1}{n_{\rm data}n_{\rm toy}} \sum^{n_{\rm data},n_{\rm toy}}_{i,j} \psi\left(|\vec{x}^{\rm data}_{i} - \vec{x}^{\rm toy}_{j}|\right).
\end{equation}  
Here, $\vec{x}$ denotes DP position, and $\psi(|\vec{x}_{i} - \vec{x}_{j}|)$ is a weighting function.
It can be shown that, in the limit of infinite statistics, with the choice $\psi(|\vec{x}_{i} - \vec{x}_{j}|) = \delta(|\vec{x}_{i} - \vec{x}_{j}|)$, the expression of Eq.~(\ref{eq:ptpdm}) is equivalent to a $\chi^2$ statistic~\cite{Williams:2010vh}.
In realistic scenarios, a form for the weighting function must be chosen, and the most appropriate choice may depend on the specific use case.
The choice 
\begin{equation}
\psi(|\vec{x}_{i} - \vec{x}_{j}|)
= e^{-{|\vec{x}_{i} - \vec{x}_{j}|}^{2}/2\sigma(\vec{x}_{i})\sigma(\vec{x}_{j})}\,
\end{equation}
has been shown to work well in DP analysis~\cite{Williams:2010vh}. 
The term $\sigma(\vec{x}) = \bar{\sigma} / (f(\vec{x}) \int dx^\prime)$, where $f(\vec{x})$ is the value of the model at position $\vec{x}$ and $\int dx^\prime$ is the area of the DP (which is included so that the mean value of the denominator becomes 1). 
The optimal value of the nuisance parameter $\bar{\sigma}$ is expected to be around the square of the typical width of the resonances in the DP in question, and thus usually $\sim 0.01 \gevgevcccc$. 
Unlike the mixed-sample test, the number of toy events should be large $(n_{\rm toy}\gg n_{\rm data})$ to avoid statistical fluctuations.

To compute a $p$-value using this test statistic, first the test statistic $T_h$ is calculated using the full available statistics. 
Then, a permutation test is performed as follows. 
The data and toy samples are pooled together and a new sample of size $n_{\rm data}$ is randomly selected from the pooled sample. 
The new sample is then treated as the data sample, while the remaining events become the toy sample, and a new value of $T_h = T_{\rm perm}$ is calculated. 
This is repeated many times, and the $p$-value of the test is obtained from the fraction of times that $T_{\rm perm} > T_h$.
This can then be repeated with additional toy samples to build up a distribution of $p$-values.

\subsection{Examples}
\label{sec:performance:examples}

The \laura\ package has been used for numerous publications by several experimental collaborations and groups of phenomenologists.
Below, several examples that demonstrate the features of the package are discussed.
In addition, \laura\ has also been used for various other studies of three-body charmless \B meson decays by the \babar\ collaboration~\cite{Aubert:2007xb,Aubert:2008rr,Aubert:2008aw,delAmoSanchez:2010ur,BABAR:2011aaa}, studies of charm decays by the LHCb collaboration~\cite{Aaij:2014afa}, unpublished studies by several collaborations (for example Refs.~\cite{delAmoSanchez:2010ad,Kohl:48803}), and investigations into the phenomenology of different three-body decays~\cite{Latham:2008zs,Gershon:2014yma,Nogueira:2016mqf}.

\begin{figure}[!htb]
\centering
\includegraphics[width=0.48\textwidth]{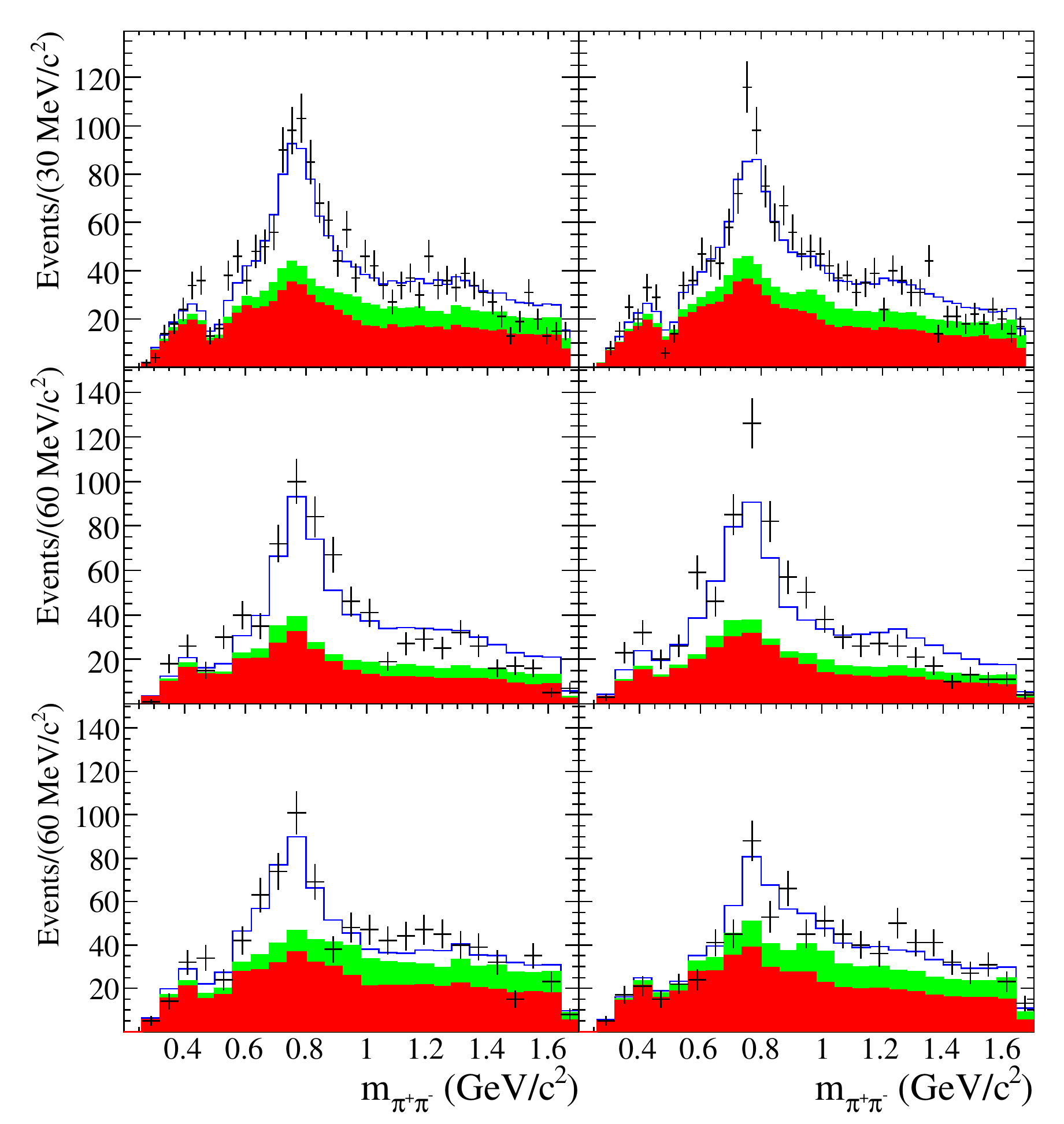}
\includegraphics[width=0.50\textwidth]{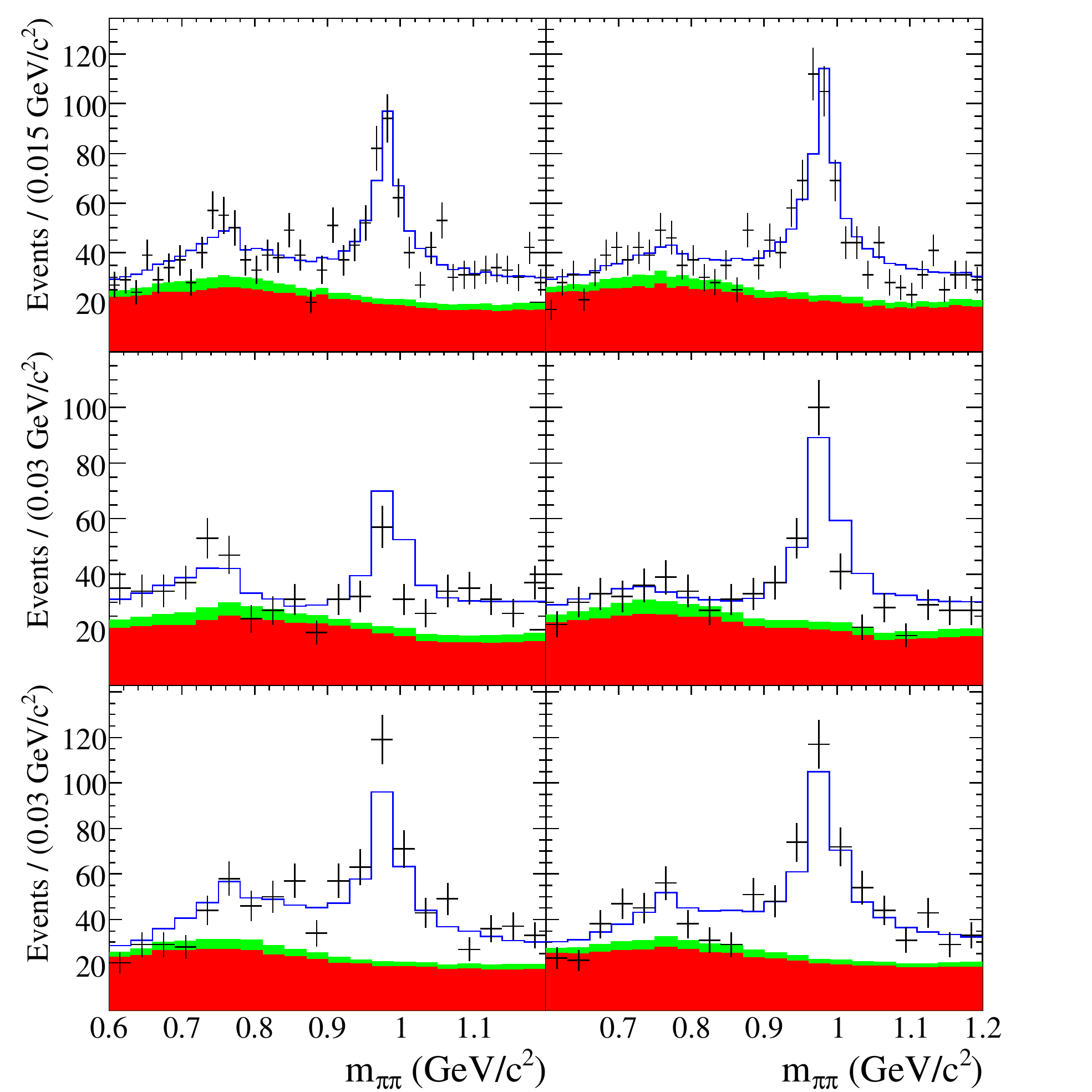}
\caption{
  Projections of the data and fit results onto the $\pip\pim$ invariant mass in the $\rho(770^0)$ region, for (left) $\Bp \to \pip\pip\pim$~\cite{Aubert:2009av} and (right) $\Bp \to \Kp\pip\pim$~\cite{Aubert:2008bj} candidates observed by the \babar\ collaboration.
  In both cases the top row shows all candidates, the middle row shows those with $\cos \theta_{\pip\pim} > 0$ and the bottom row shows those with $\cos \theta_{\pip\pim} < 0$, while in each row the left (right) plot is for $B^-$ ($B^+$) candidates.
  The data are the points with error bars, the red/dark filled histogram shows the continuum background component, the green/light filled histogram shows the background from other \B\ meson decays, and the blue unfilled histogram shows the total fit result.
}
\label{fig:3pi-Kpipi-examples}
\end{figure}

Studies of charmless three-body \B\ meson decays provide interesting opportunities to investigate the dynamics of hadronic \B\ decays including potential \CP\ violation effects.
The $\Bp \to \pip\pip\pim$~\cite{Aubert:2005sk,Aubert:2009av} and $\Bp \to \Kp\pip\pim$~\cite{Aubert:2005ce,Aubert:2008bj} decays have been investigated by the \babar\ collaboration using the \laura\ package.
In the most recent amplitude analysis of $\Bp \to \pip\pip\pim$ decays~\cite{Aubert:2009av}, the amplitude model includes contributions from the $\rho(770)^0$, $\rho(1450)^0$, $f_2(1270)$, $f_0(1370)$ resonances and a nonresonant component.
In the most recent amplitude analysis of $\Bp \to \Kp\pip\pim$ decays~\cite{Aubert:2008bj}, the amplitude model includes the $K^*(892)^0$, $K_2^*(1430)^0$, $\rho(770)^0$, $\omega(782)$, $f_0(980)$, $f_2(1270)$, $f_X(1300)$ and $\chi_{c0}$ resonances together with $K\pi$ S-wave and nonresonant components.
In both cases, \CP violation is allowed in the amplitudes.
Projections of the fit results around the $\rho(770)^0$ resonance are shown in Fig.~\ref{fig:3pi-Kpipi-examples}.
The $\Bp \to \pip\pip\pim$ data are consistent with \CP\ conservation while there is evidence for \CP\ violation in $\Bp \to \rho(770)^0\Kp$ decays, which becomes more evident when inspecting the data in different regions of the $\pip\pim$ helicity angle, $\theta_{\pip\pim}$.
As model-independent analyses of larger data samples of these decays by the LHCb collaboration~\cite{Aaij:2013sfa,Aaij:2013bla,Aaij:2014iva} have revealed large \CP\ violation effects that vary significantly across the DP, there is strong motivation for updated amplitude analyses.

\begin{figure}[!htb]
  \centering
  \includegraphics[width=0.34\textwidth]{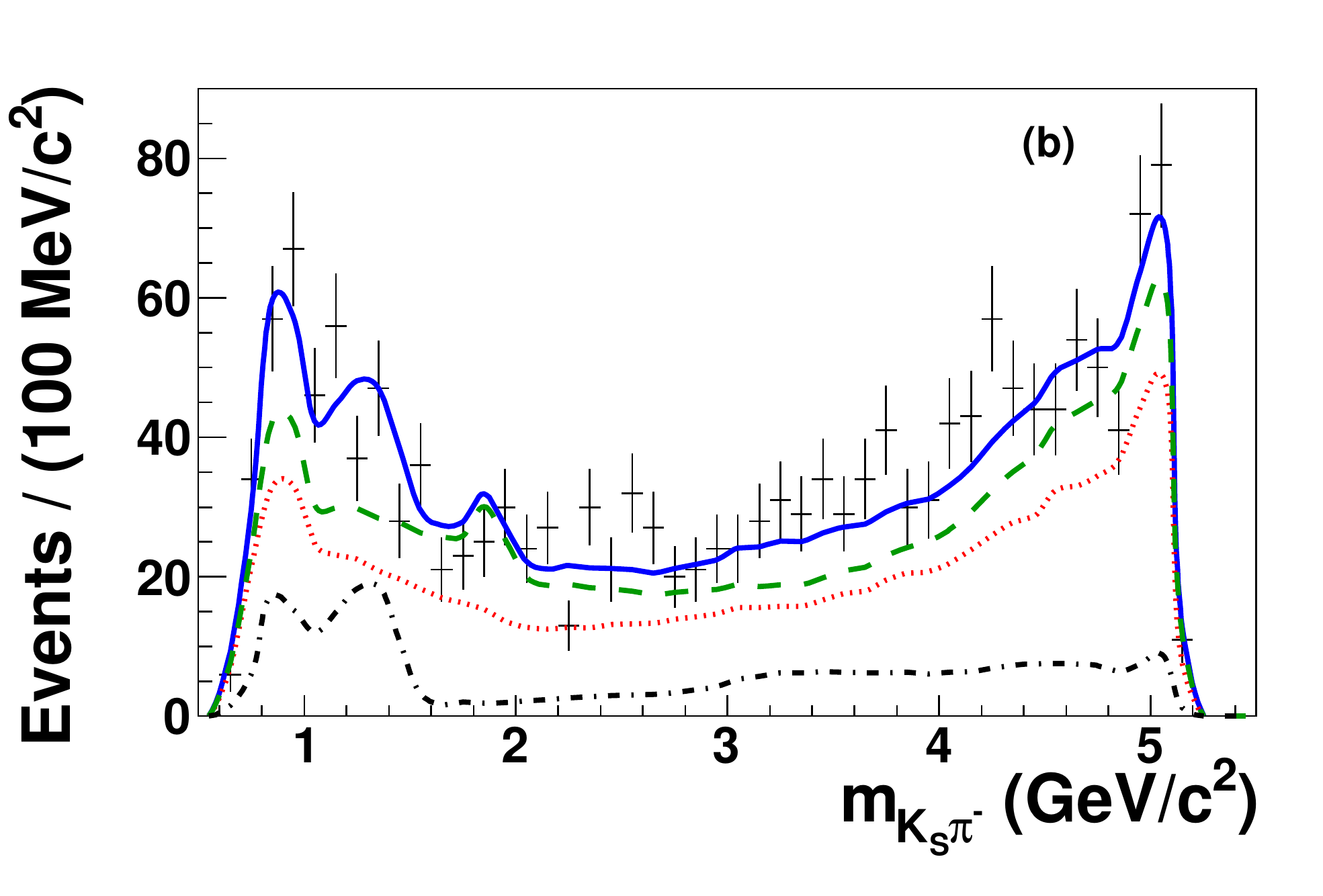}
  \includegraphics[width=0.34\textwidth]{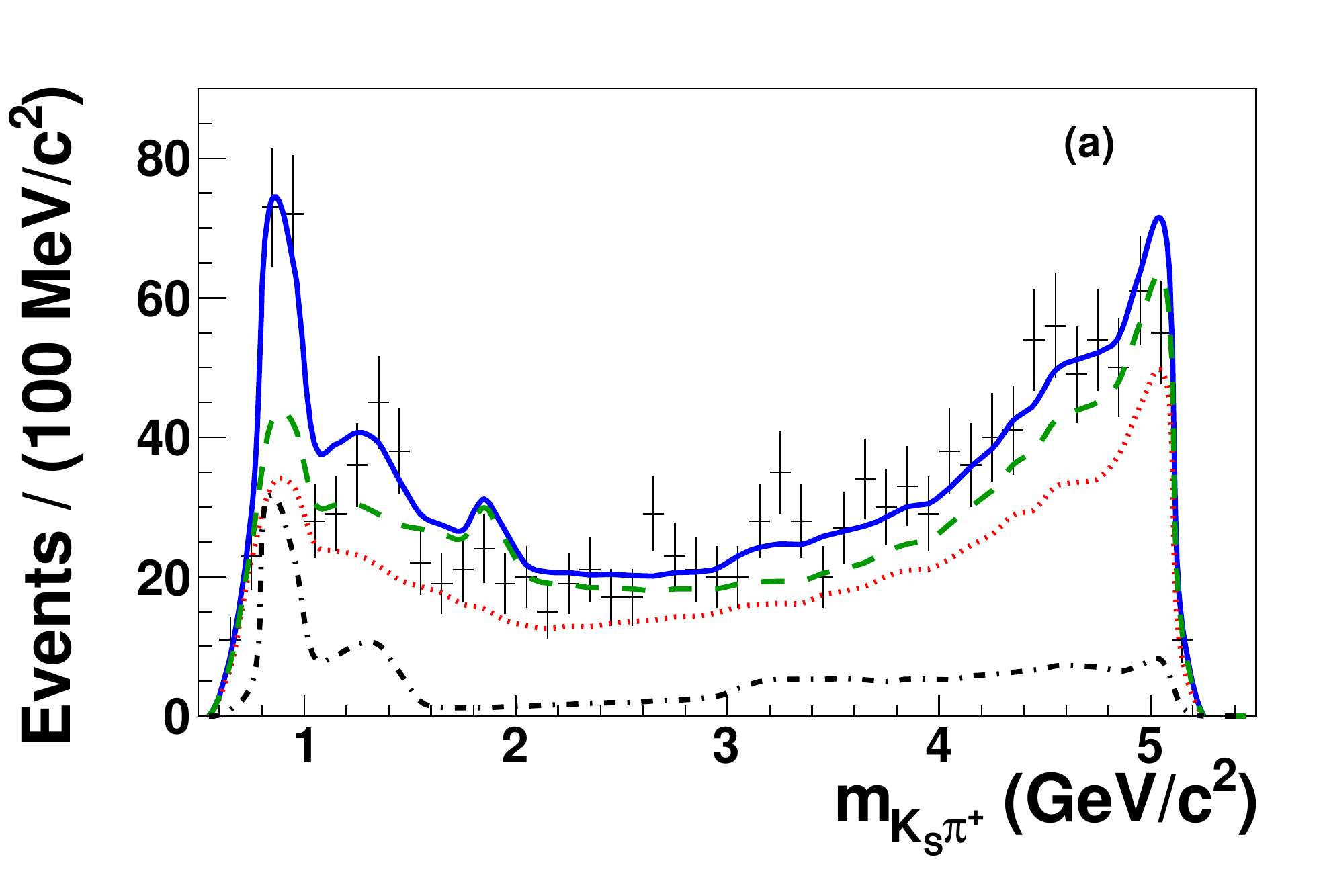}
  \includegraphics[width=0.34\textwidth]{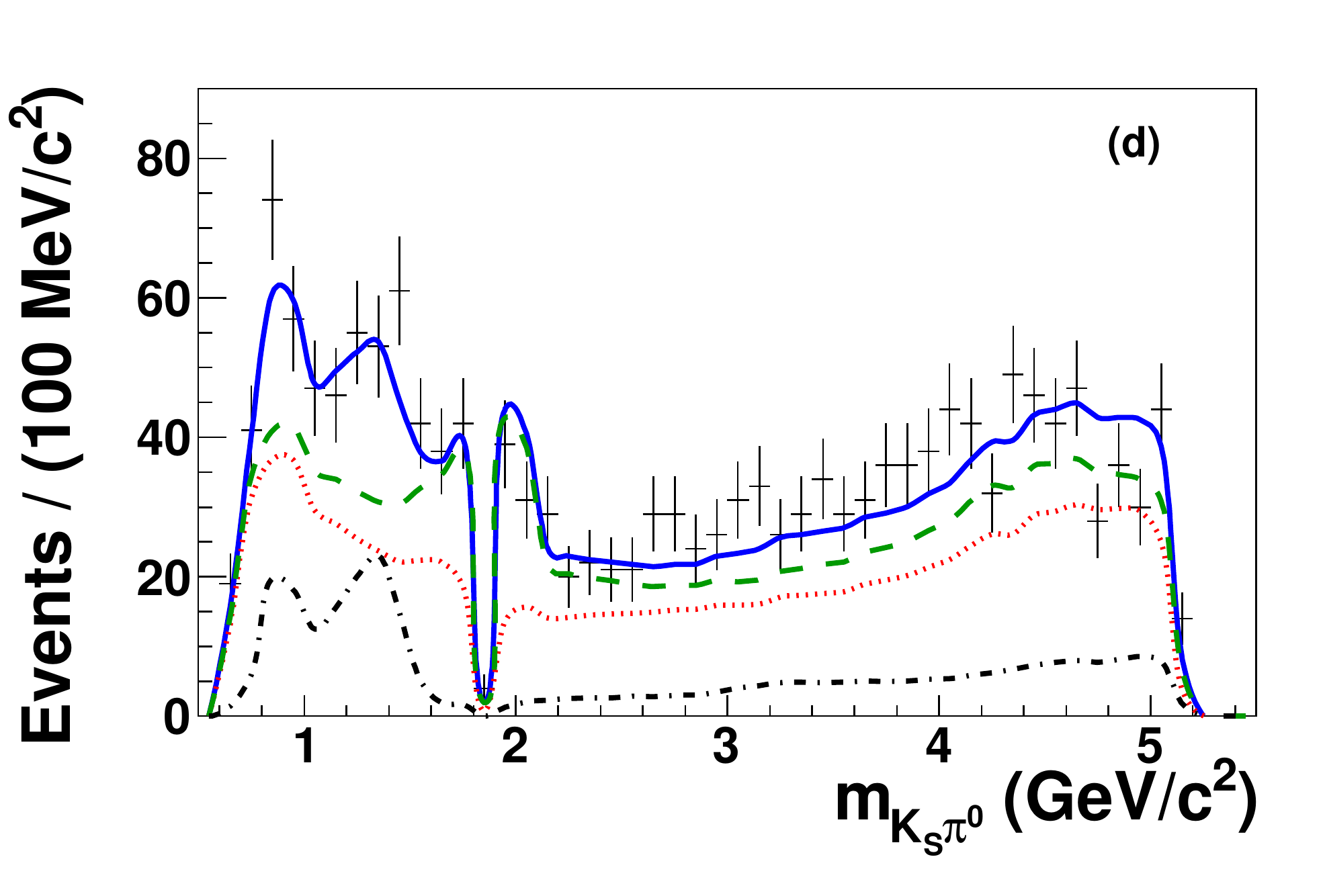}
  \includegraphics[width=0.34\textwidth]{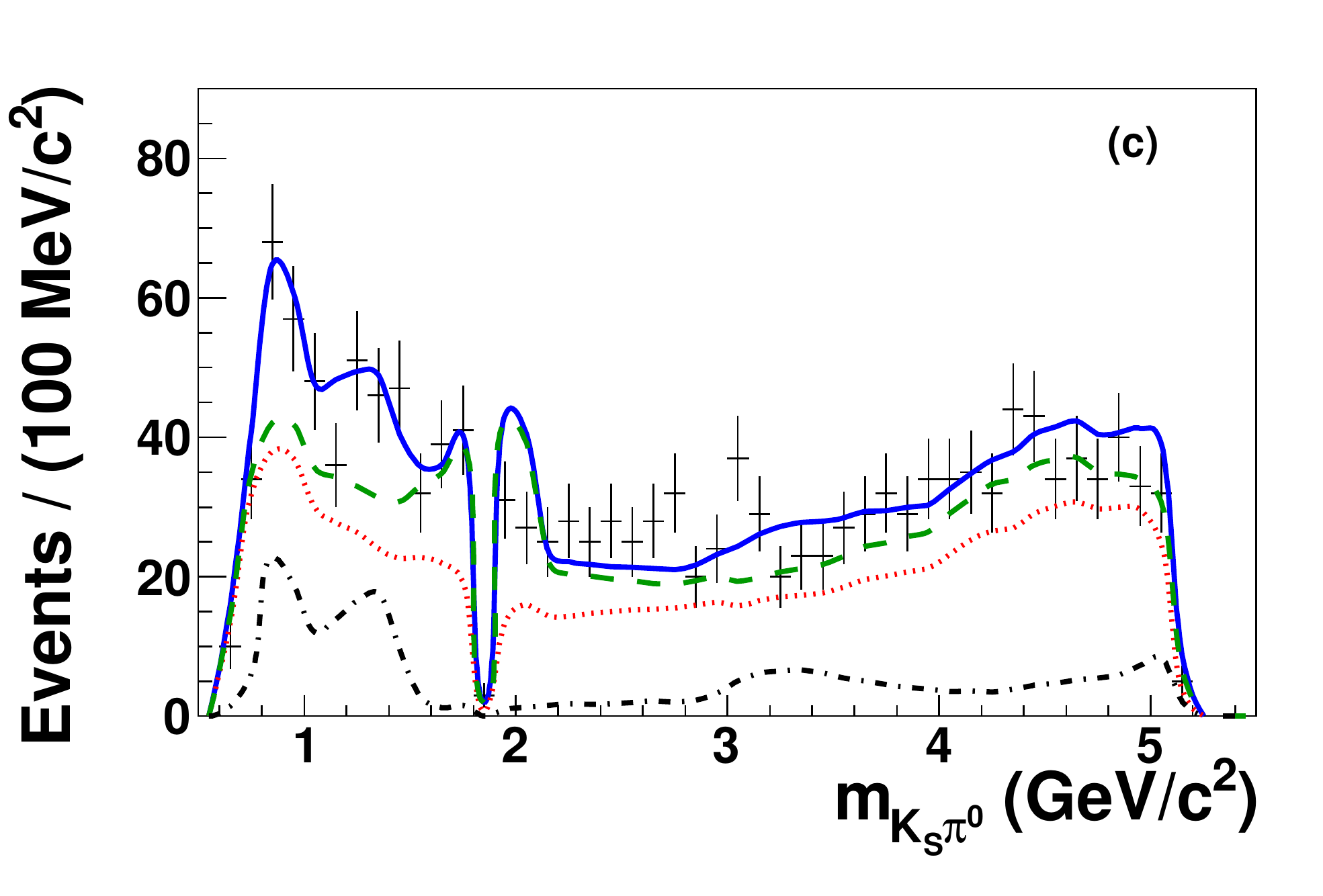}
  \includegraphics[width=0.34\textwidth]{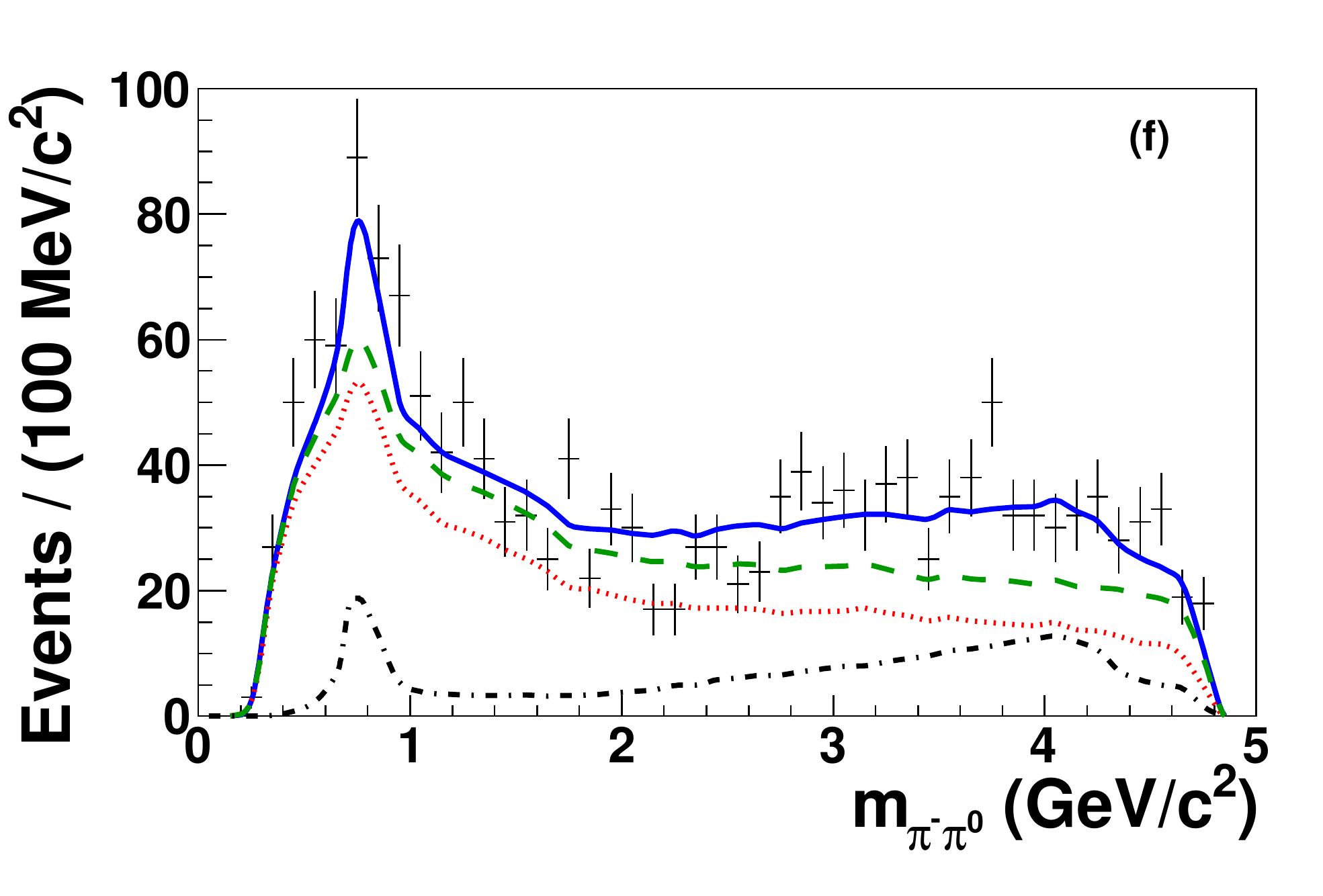}
  \includegraphics[width=0.34\textwidth]{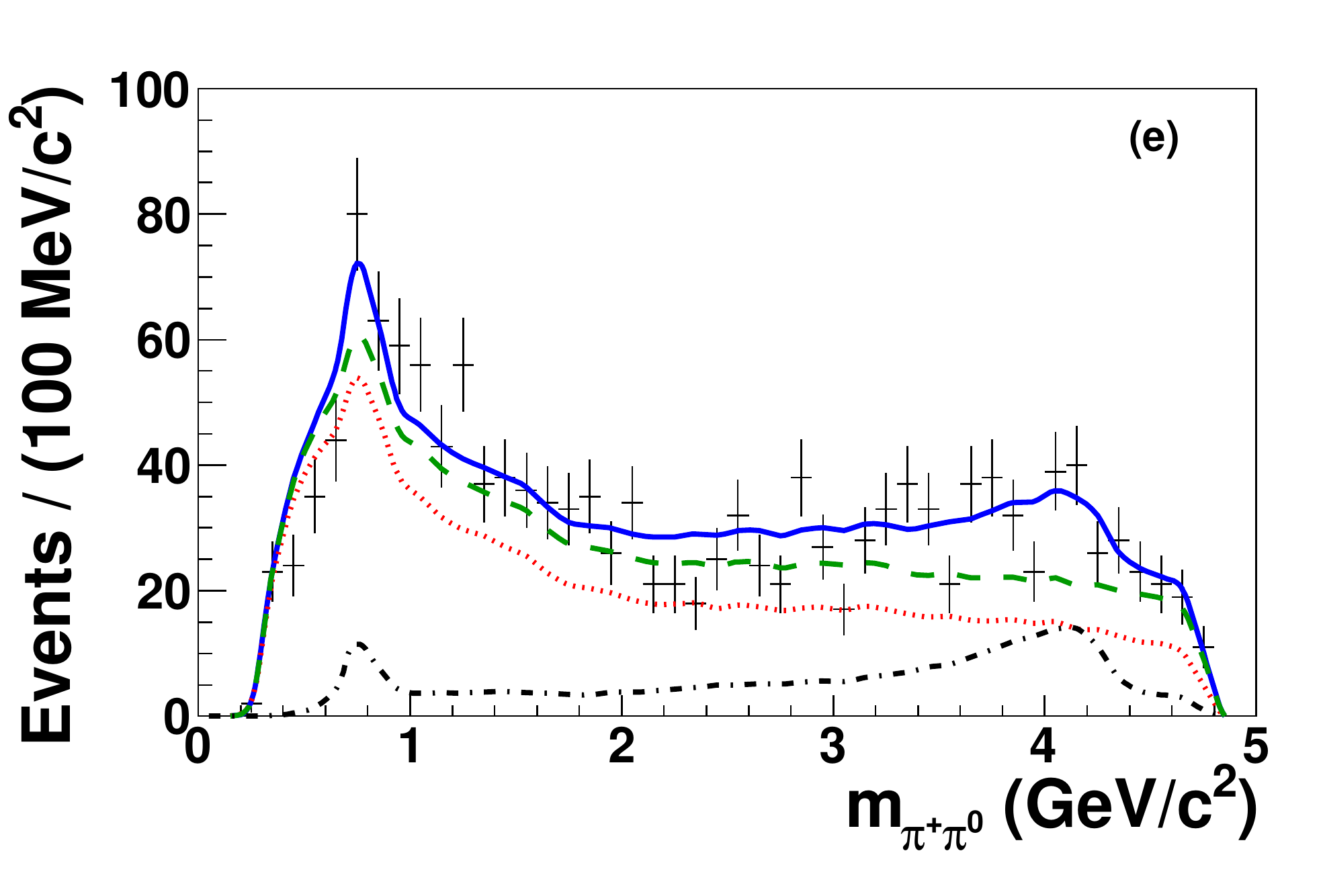}
\caption{
  Projections of the data and fit results onto (top) $\KS\pimp$, (middle) $\KS\piz$ and (bottom) $\pimp\piz$ invariant mass distributions for $\Bp \to \KS\pip\piz$ candidates observed by the \babar\ collaboration~\cite{Lees:2015uun}.
  Background from $\Dz \to \KS\piz$ has been vetoed.
  In each row the left (right) plot is for $B^-$ ($B^+$) candidates.
  The data are the points with error bars, the (black) dash-dotted curves represent the signal contribution, the dotted (red) curves to the continuum background component, the dashed (green) curves to the total background contribution and the solid (blue) curves the total fit result. 
}
\label{fig:KSpipi0-examples}
\end{figure}

Understanding the origin of these \CP violation effects requires related modes to also be studied.
The \laura\ package has also been used by the \babar\ collaboration for a time-dependent DP analysis of $\Bz \to \KS\pip\pim$ decays~\cite{Aubert:2009me}, as well as for an amplitude analysis of $B^{+} \to \KS \pi^{+} \pi^{0}$ decays~\cite{Lees:2015uun}.
In the latter, the modelling of the large background contribution, as well as of the smearing of the DP position due to the limited resolution of the neutral pion momentum (self cross-feed), is particularly important.   
In addition, correlations between the DP position and the variables that are used to discriminate signal decays from background contributions are taken into account as described in App.~\ref{sec:pdfs-DPdep}.
The amplitude model includes components from the $K^*(892)$ resonance and $K\pi$ S-wave (both appearing in both charged and neutral channels) as well as the $\rho(770)^+$ resonance.
Projections of the fit results are shown in Fig.~\ref{fig:KSpipi0-examples}.
The analysis reveals evidence for a \CP\ asymmetry in $\Bp \to K^*(892)^+\piz$ decays.

\begin{figure}[!htb]
  \centering
  \includegraphics[height=0.27\textwidth]{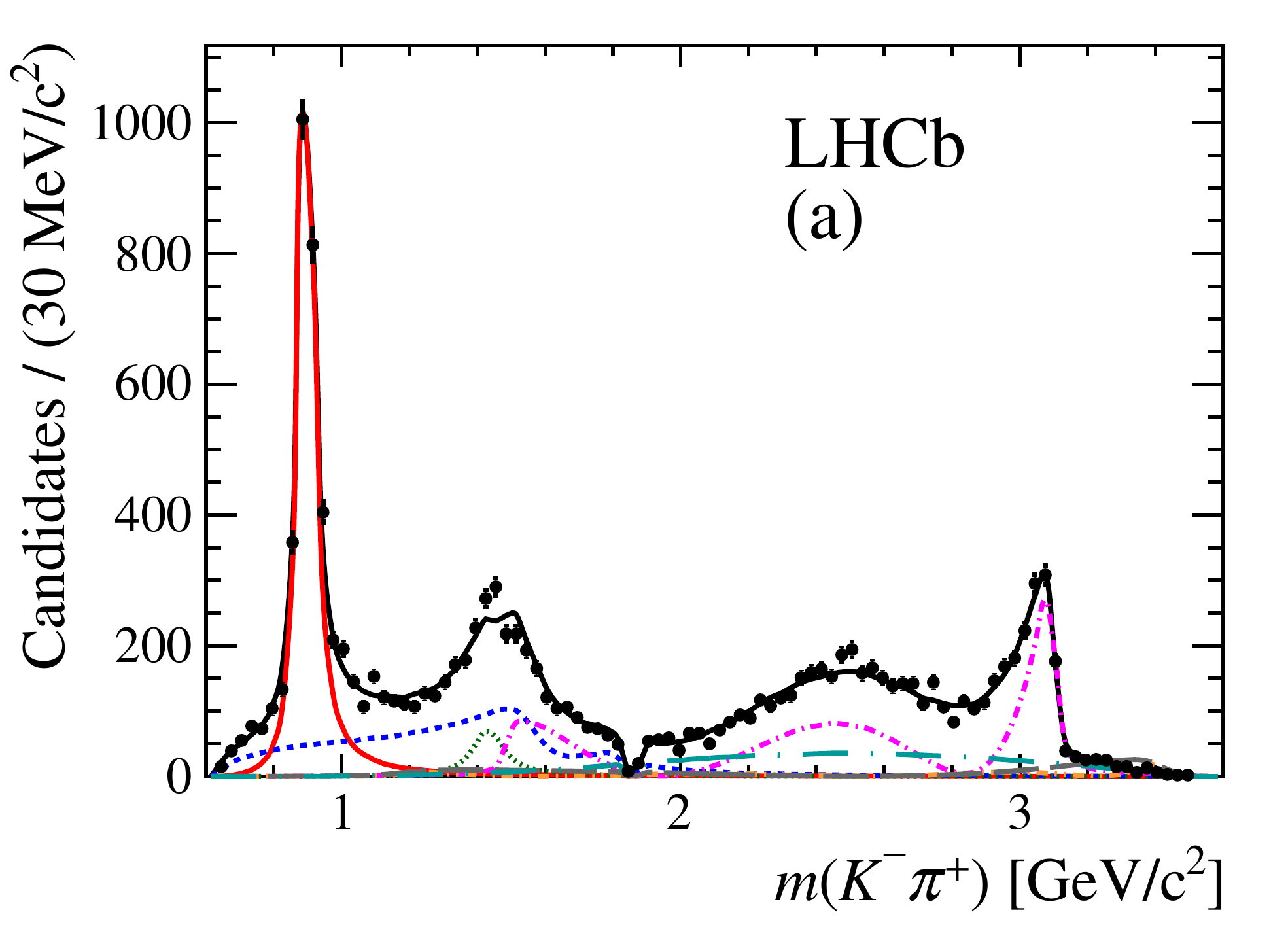}
  \includegraphics[height=0.27\textwidth,viewport=150 0 425 380,clip]{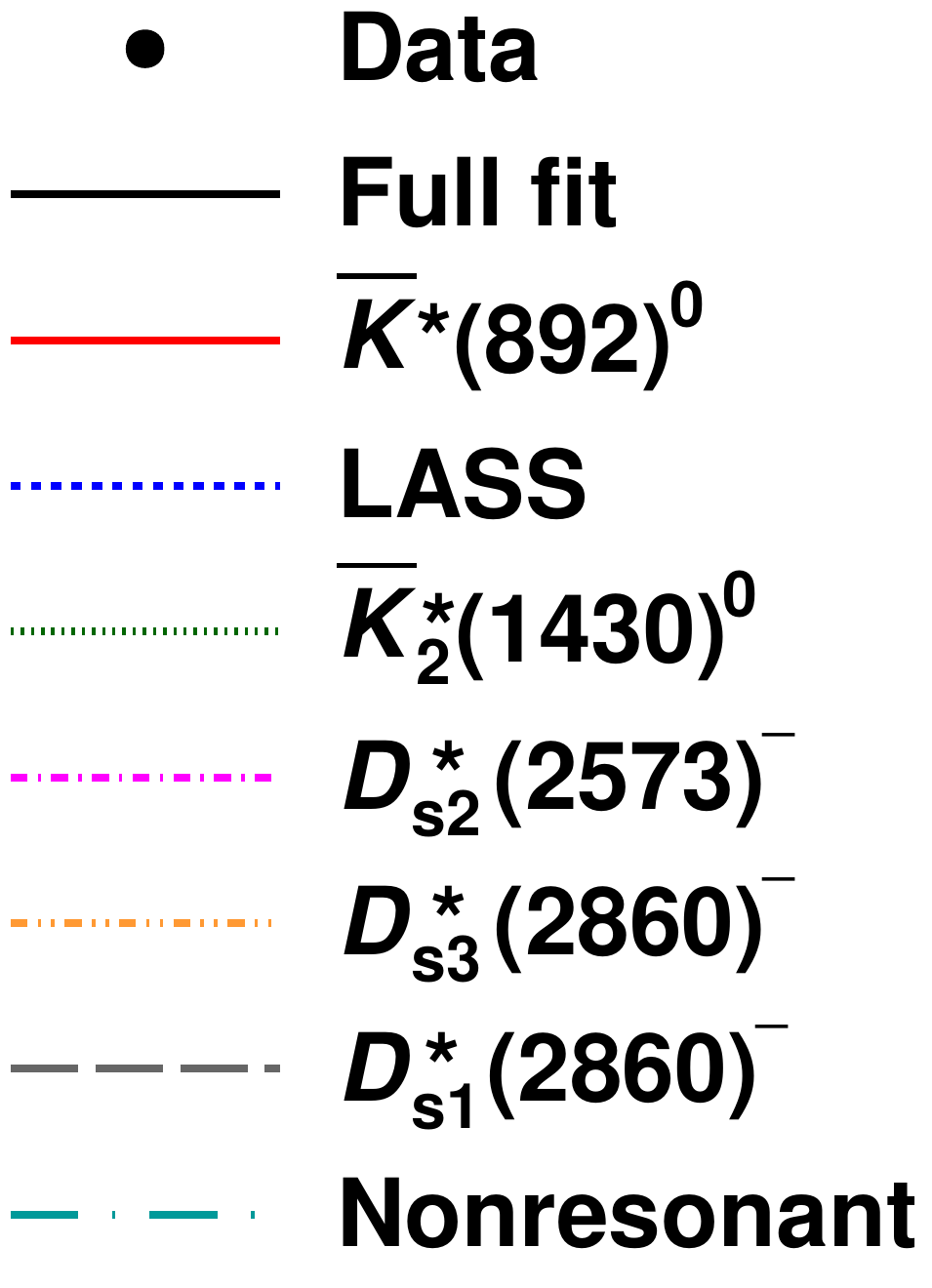}
  \includegraphics[height=0.27\textwidth]{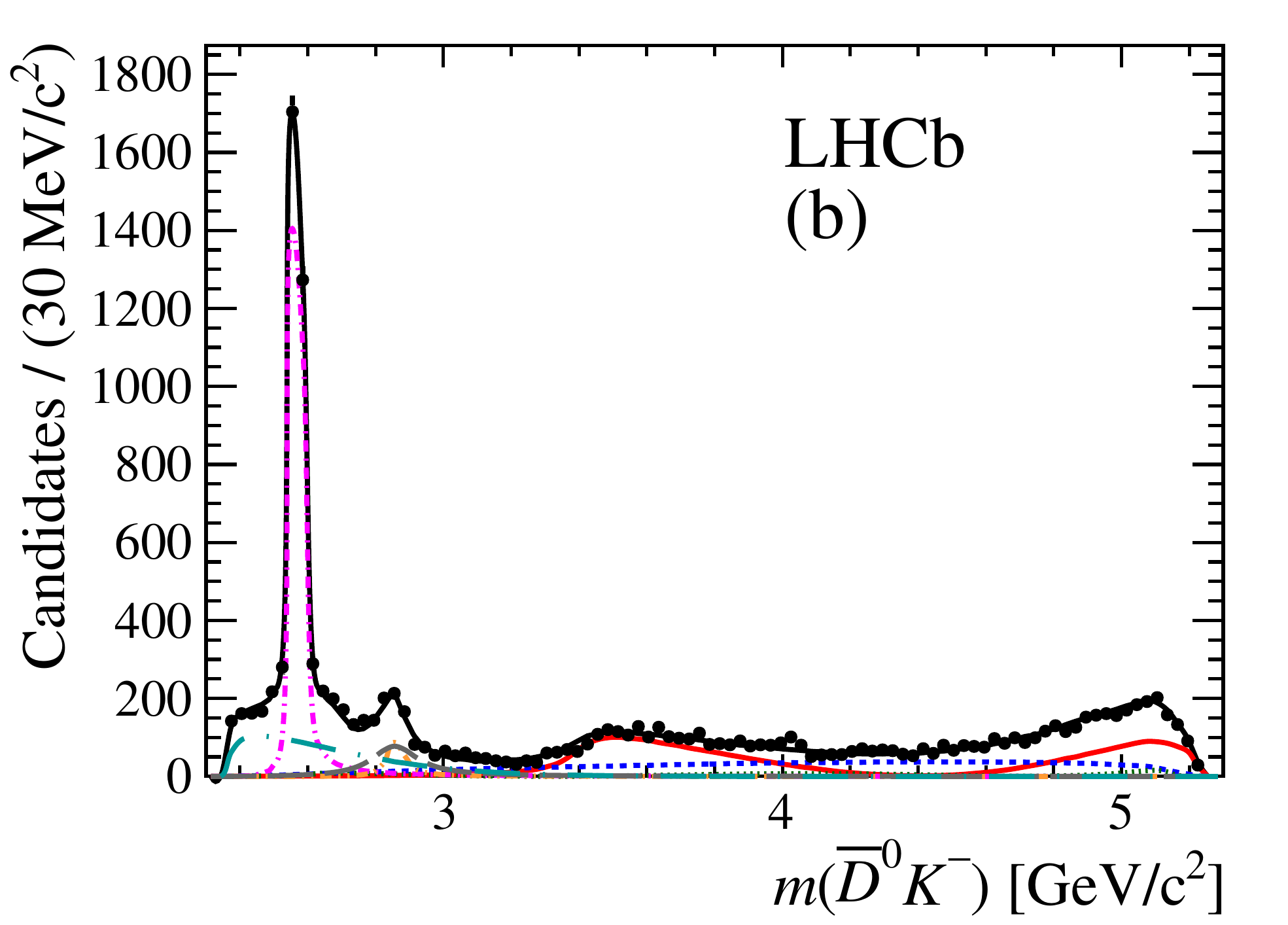}
  \includegraphics[width=0.35\textwidth]{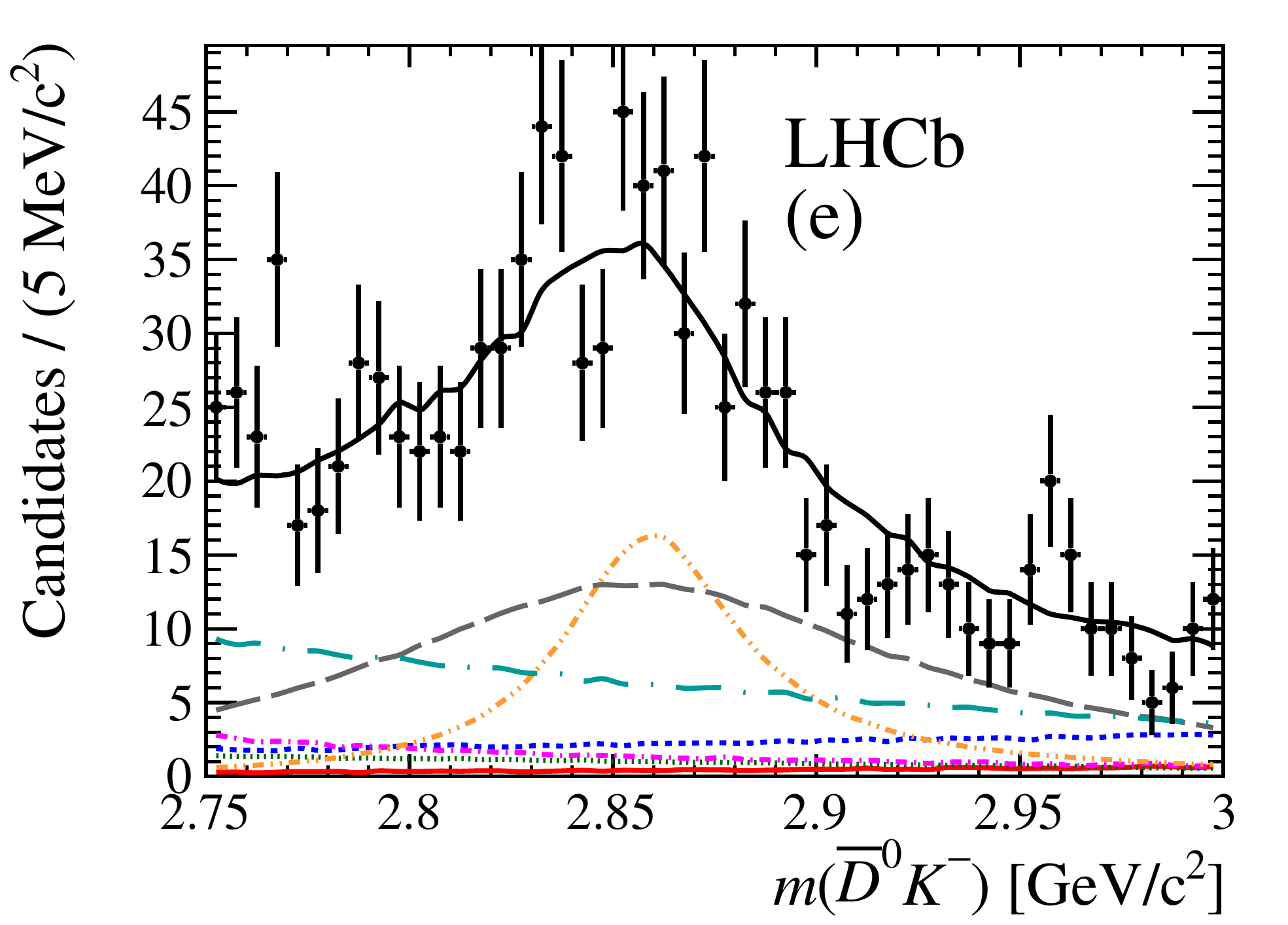}
  \includegraphics[width=0.38\textwidth]{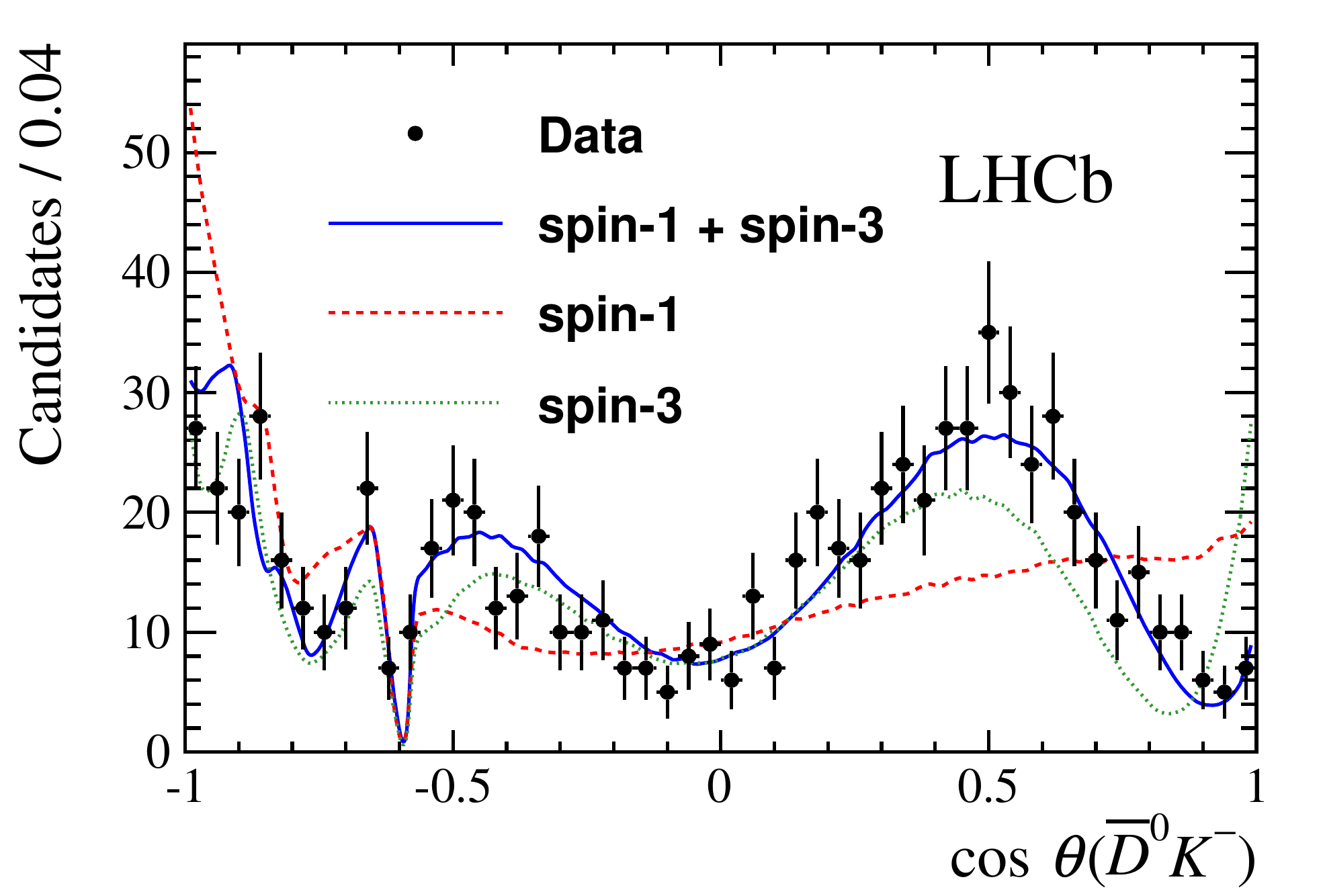}
\caption{
  Projections of the data and fit results onto (top left) $\Km\pip$ and (top right) $\Dzb\Kp$ invariant mass distributions for $\Bs \to \Dzb\Km\pip$ candidates observed by the LHCb collaboration~\cite{Aaij:2014xza,Aaij:2014baa}.
  Background from $\Dz \to \Km\pip$ has been vetoed.
  A legend describing the various contributions is also given, together with (bottom left) a zoom around $m(\Dzb\Km) \sim 2.86 \gevcc$ and (bottom right) a projection onto the cosine of the helicity angle $\theta(\Dzb\Km)$ for candidates in that region.
  In the last case, projections of the results of fits with models containing either or both of the $D_{s1}^*(2860)^-$ and $D_{s3}^*(2860)^-$ resonances are shown, demonstrating the need for both to obtain a good fit to the data.
}
\label{fig:BsDKpi-examples}
\end{figure}

The LHCb collaboration has used the \laura\ package for several studies of multibody $B$ meson decays to charmed final states, with important results for charm spectroscopy and \CP\ violation measurements.  
For example, the $\Bs \to \Dzb\Km\pip$ decay was found to have a DP structure that contains effects from overlapping spin-1 and spin-3 resonances with masses around $m(\Dzb\Km) \sim 2.86 \gevcc$~\cite{Aaij:2014xza,Aaij:2014baa}.
The neutral charm meson is reconstructed through its $\Dzb \to \Kp\pim$ decay.
The model contains contributions from the $\Kstarb(892)^0$, $\Kstarb(1410)^0$, $\Kbar{}^*_2(1430)^0$, $\Kstarb(1680)^0$ resonances as well as a $\Km\pip$ S-wave component, and $D_{s2}^*(2573)^-$, $D_{s1}^*(2700)^-$, $D_{s1}^*(2860)^-$, $D_{s3}^*(2860)^-$ resonances together with a nonresonant $\Dzb\Km$ S-wave amplitude and virtual contributions from the $D_{s\,v}^{*-}$, $D_{s0\,v}^*(2317)^-$ and $B_{v}^{*+}$ states.
The results of the analysis include the first experimental proof of the spin-2 nature of the $D_{s2}^*(2573)^-$ state, as well as world-leading measurements of the masses and widths of many of the resonances.
Projections of the DP fit results onto the data are shown in Fig.~\ref{fig:BsDKpi-examples}.

\begin{figure}[!htb]
  \centering
  \includegraphics[width=0.45\textwidth]{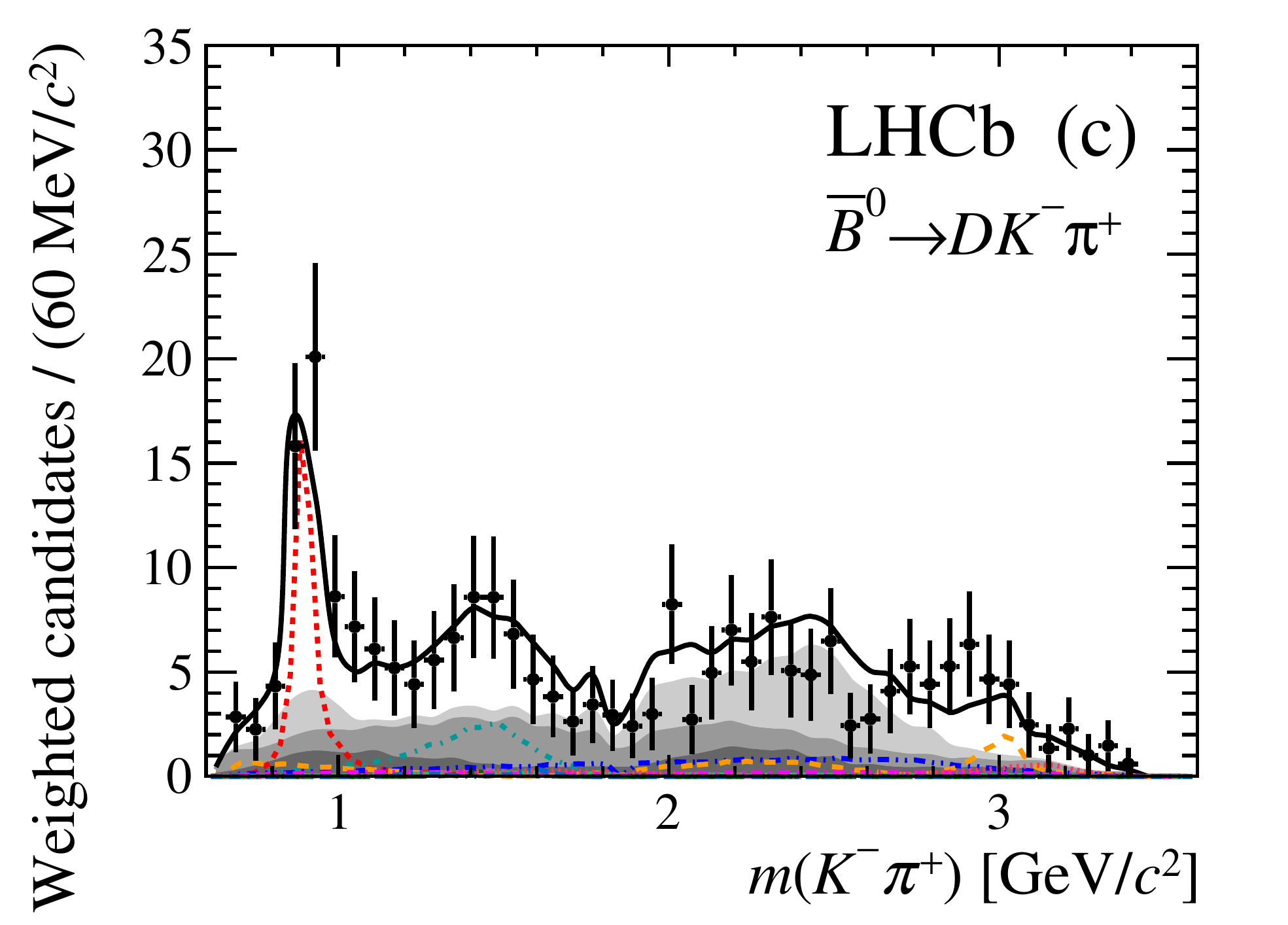}
  \includegraphics[width=0.45\textwidth]{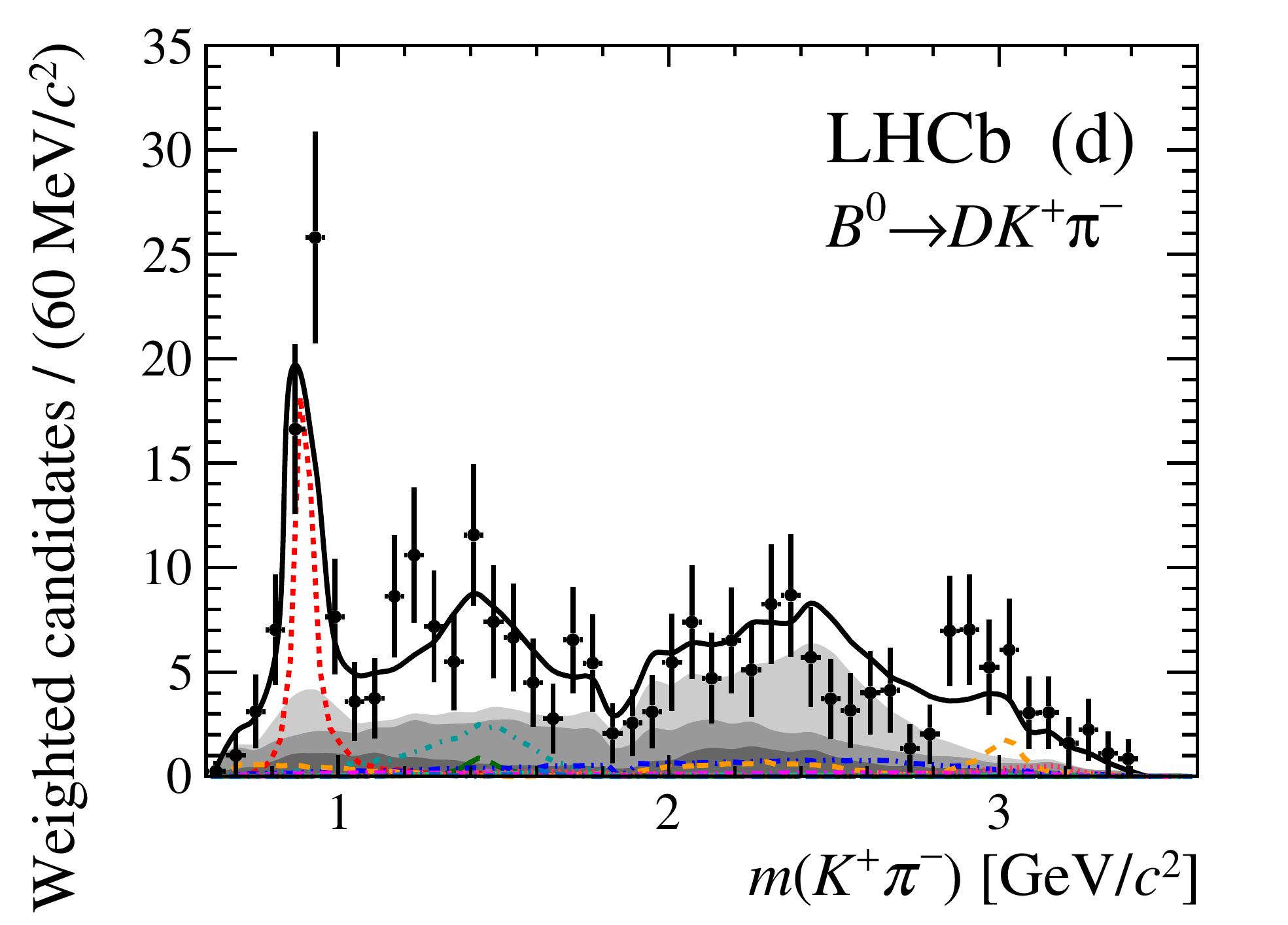}
  \includegraphics[width=0.65\textwidth]{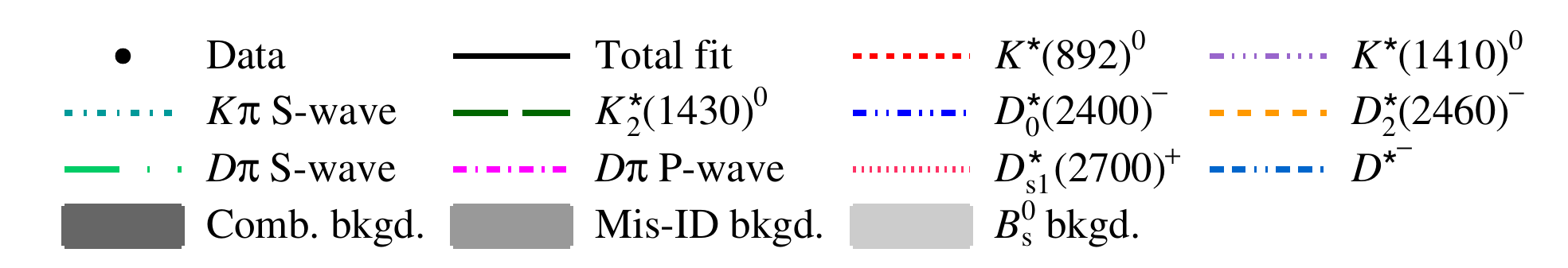}
\caption{
  Projections of the data and fit results onto $m(\Kmp\pipm)$ for (left) $\Bzb \to D\Km\pip$ and (right) $\Bz \to D\Kp\pim$ candidates observed by the LHCb collaboration~\cite{Aaij:2016bqv}.
  A legend describing the various contributions is also given.
}
\label{fig:B0DKpi-examples}
\end{figure}

A similar DP analysis with the \laura\ package has been performed by the LHCb collaboration for the $\Bz \to \Dzb\Kp\pim$ mode~\cite{Aaij:2015kqa}.
The model obtained is an essential input into a subsequent analysis of the same decay with the neutral charm meson reconstructed through $D$ decays to the \CP\ eigenstates $\Kp\Km$ and $\pip\pim$~\cite{Aaij:2016bqv}.
In the latter case, contributions from both $\Bz \to \Dz\Kp\pim$ and $\Dzb\Kp\pim$ amplitudes can interfere, giving sensitivity to the angle $\gamma$ of the CKM unitarity triangle.
A DP analysis allowing for \CP\ violation effects provides a powerful way to determine $\gamma$ without ambiguities~\cite{Gershon:2008pe,Gershon:2009qc}.
In this analysis, a simultaneous fit to the final states with different $D$ decays is implemented using the \jfit method described in Sec.~\ref{sec:jfit} and Ref.~\cite{Ben-Haim:2014afa}.
Projections of the fit result onto the data (weighted by the signal purity) are shown in Fig.~\ref{fig:B0DKpi-examples}.
No significant \CP\ violation effect is observed, and the resulting limits on $\gamma$ are not strongly constraining.  
The method is, however, expected to give competitive constraints on $\gamma$ as larger data samples become available and as additional $D$ meson decay modes are included in the analysis.
Moreover, the analysis also gives results for hadronic parameters that must be known in order to interpret results from quasi-two-body analyses of $\Bz \to D\Kstar(892)^0$ decays in terms of $\gamma$.
As such, the results have an important impact in combinations of results to obtain the best knowledge of $\gamma$~\cite{Aaij:2016kjh,Amhis:2016xyh}.

\begin{figure}[!htb]
  \centering
  \includegraphics[height=0.30\textwidth]{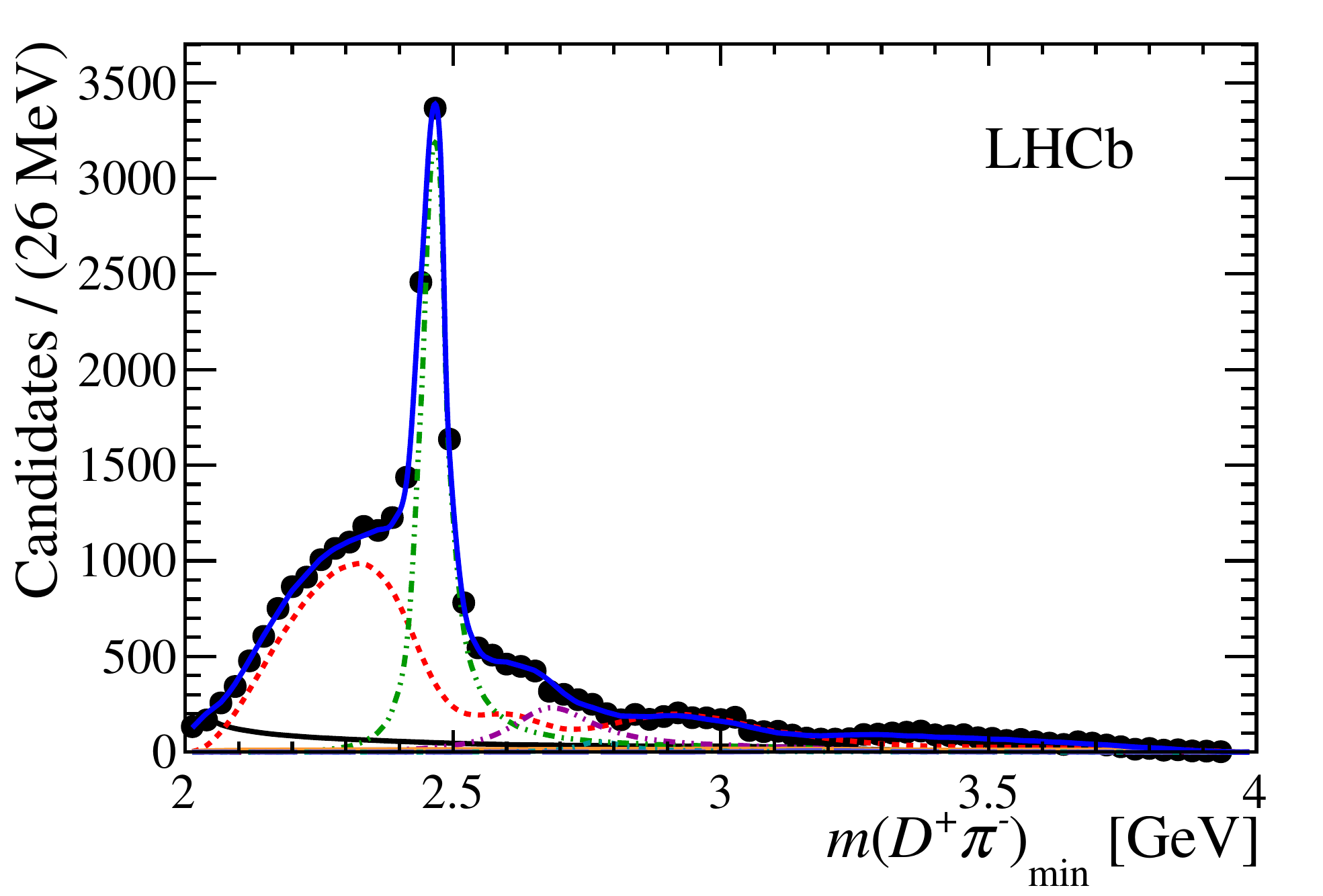}
  \includegraphics[height=0.30\textwidth]{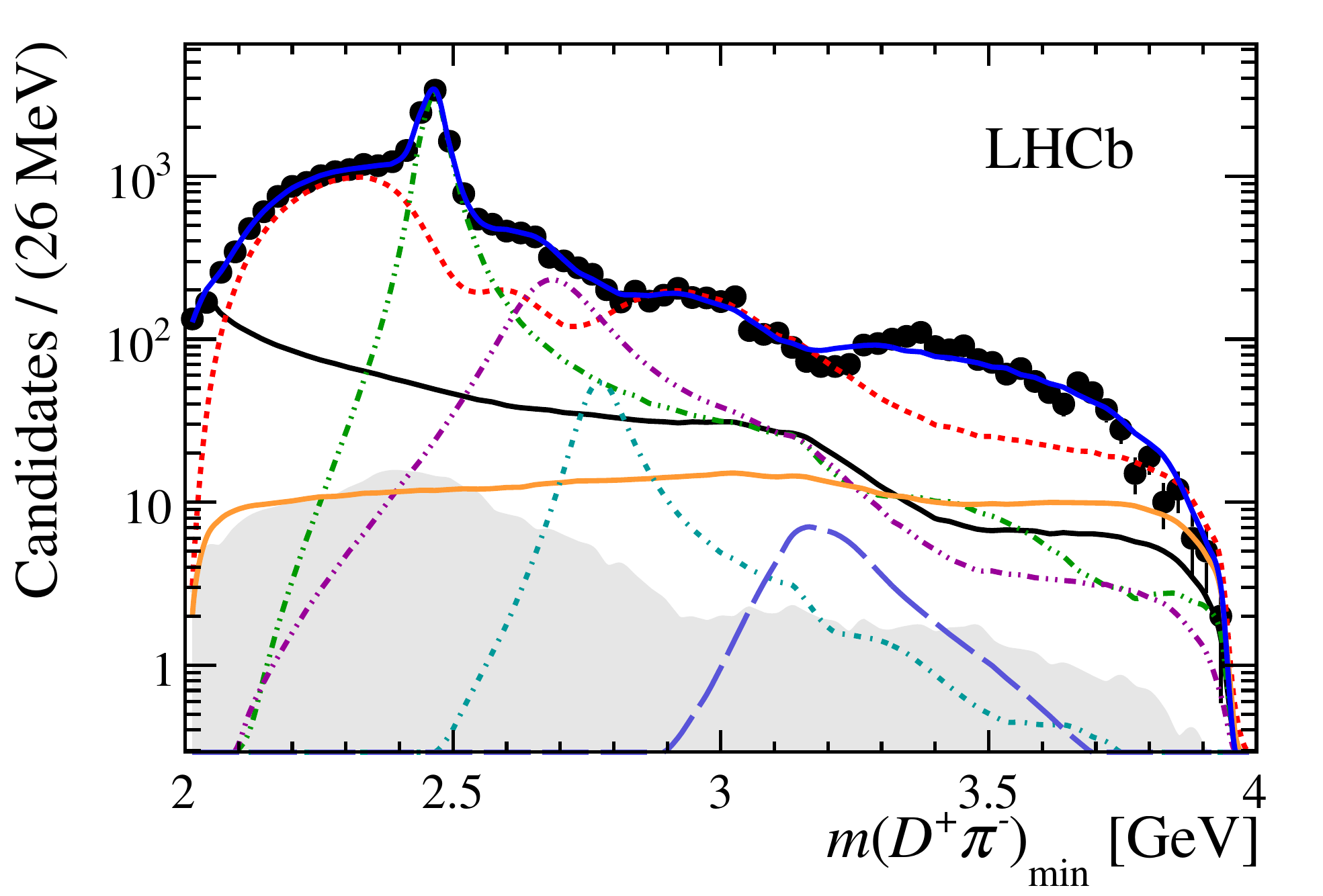}
  \raisebox{0.5\height}{\includegraphics[height=0.16\textwidth,viewport=45 20 425 160,clip]{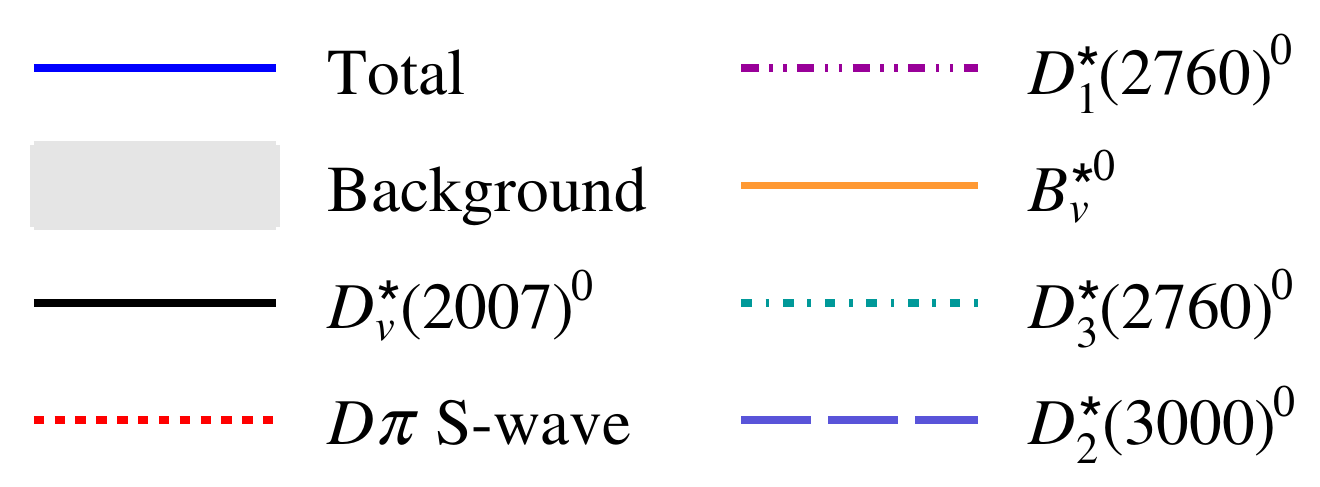}}
  \includegraphics[height=0.33\textwidth]{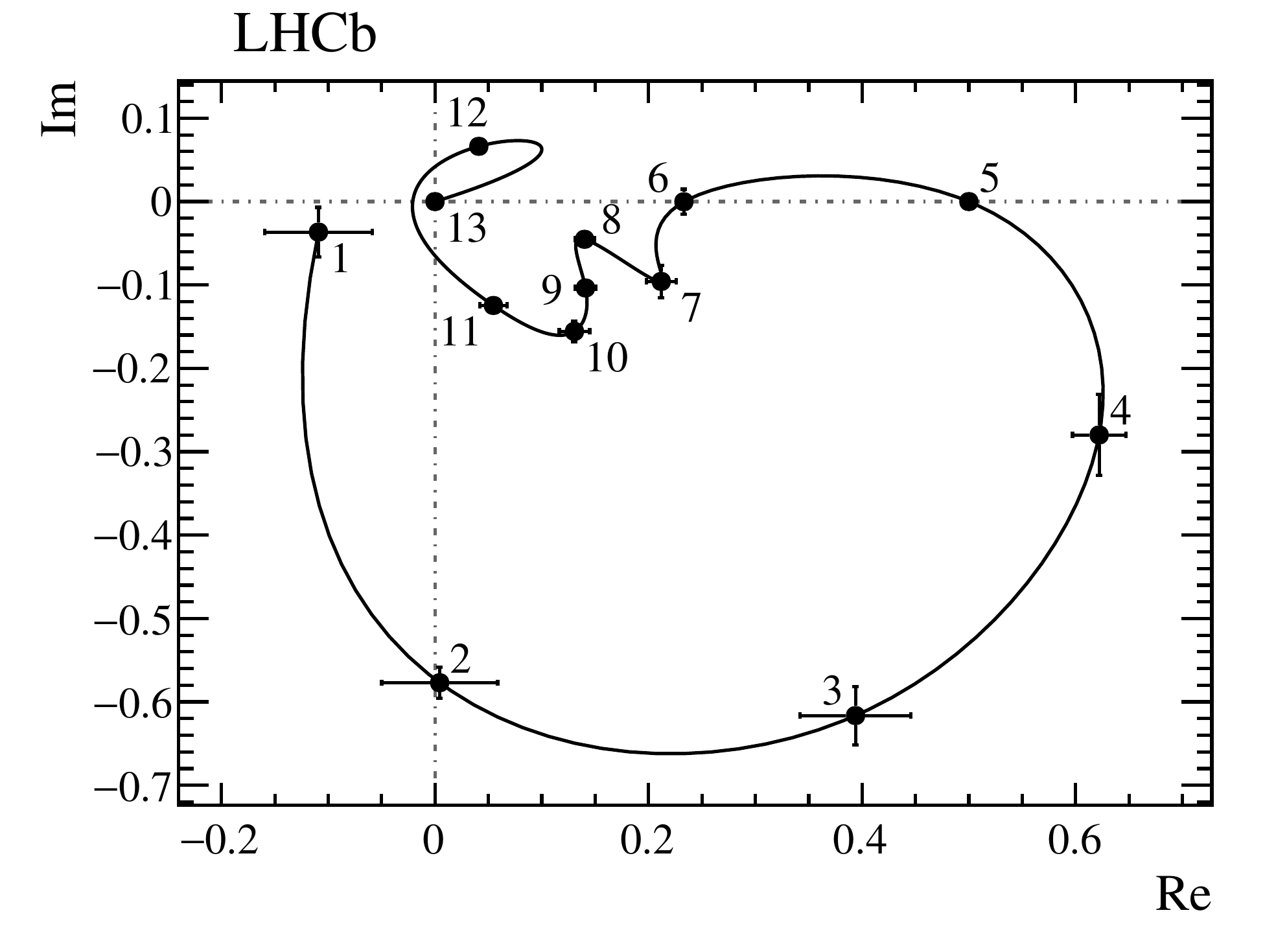}
\caption{
  Projections of the data and fit results onto $m(\Dm\pip)_{\rm min}$ for $\Bp \to \Dm\pip\pip$ candidates observed by the LHCb collaboration~\cite{Aaij:2016fma} on (top left) linear and (bottom right) logarithmic $y$-axis scales.
  Here, $m(\Dm\pip)_{\rm min}$ is the smaller of the two values of $m(\Dm\pip)$ for each $\Bp \to \Dm\pip\pip$ candidate.
  A legend describing the various contributions is also given.
  The (bottom right) Argand diagram of the $\Dm\pip$ S-wave amplitude shows the expected phase motion corresponding to the $\Dbar{}_0^*(2400)^0$ resonance.
  The numbered points correspond to the spline knots.
}
\label{fig:BDpipi-examples}
\end{figure}

Other decays of the type $B \to D^{(*)}K\pi$ have sensitivity to $\gamma$, and are also important to study as possible background contributions to the two-body $B \to D^{(*)}K$ type decays that are more conventionally used for this purpose.
The LHCb collaboration has also published results on the $\Bp \to \Dm\Kp\pip$~\cite{Aaij:2015vea} and $\Dp\Kp\pim$~\cite{Aaij:2015dwl} decays, which were obtained from analyses using the \laura\ package.
The higher-yield $B \to D^{(*)}\pi\pi$ channels provide some of the most interesting possibilities to explore charm spectroscopy.
An amplitude analysis of $\Bp \to \Dm\pip\pip$~\cite{Aaij:2016fma} has been performed by the LHCb collaboration, using the \laura\ package, in which the model contains contributions from the $\Dbar{}_2^*(2460)^0$, $\Dbar{}_1^*(2760)^0$, $\Dbar{}_3^*(2760)^0$ and $\Dbar{}_2^*(3000)^0$ resonances (the last three of which are either confirmed or observed for the first time), as well as virtual contributions from the $\Dbar{}_{v}^*(2007)^-$ and $B_{v}^{*0}$ states.
In the absence of sufficiently detailed models for the $\Dm\pip$ S-wave, a quasi-model-independent description based on spline interpolation is used.
Projections of the results of the fit onto the data are shown in Fig.~\ref{fig:BDpipi-examples}.

\subsection{Speed}
\label{sec:performance:speed}

It is essential for the \laura\ amplitude analysis package to run quickly,
since otherwise the large data samples available in modern experiments can
lead to unmanageably long execution time. 
In this subsection some performance benchmarks are provided.
More specifically, a selection of the examples that are provided with the
package (several of which are based on analyses presented in the previous
subsection) are run out of the box on the same machine (an Intel Core i5-3570
3.4\ghz quad-core CPU with 8\gbytes of RAM).
In each case, timings for both generation of 50 toy datasets and for
fitting those same 50 datasets are provided in
Table~\ref{tab:examples-speed}.
The fitting times are averaged over 20 fits with randomised starting parameters.
The scenario demonstrated in each example is as follows:
\begin{itemize}
\item
\texttt{GenFit3pi.cc}\\
Example analysis of the symmetric final state $\Bp\to\pip\pip\pim$, using
\texttt{LauSimpleFitModel} (\ie\ not including effects of \CP violation).
By default there are 1500 signal events per experiment and the signal
isobar model contains five components: $\rho(770)^0$, $f_0(980)$,
$f_2(1270)$, $\rho(1450)^0$, and a nonresonant component.
All of the resonance parameters are fixed in the fit.
There is also a background component, which defaults to being uniformly
distributed in the Dalitz plot and consists of 1250 background events.
A version of this example implemented in python, \texttt{GenFit3pi.py}, is
also included in the \laura\ package.
\item
\texttt{GenFit3K.cc}\\
Example analysis of the symmetric final state $\Bp\to\Kp\Kp\Km$, using
\texttt{LauSimpleFitModel} (\ie\ not including effects of \CP violation).
By default there are 5000 signal events per experiment and the signal
isobar model contains three components: $\phi(1020)$, $f_2^{\prime}(1525)$
and a nonresonant component.
The mass and width of both the $\phi(1020)$ and the $f_2^{\prime}(1525)$
are floating parameters by default and a two-stage fit is employed.
No background contributions are included.
To demonstrate further the impact on the performance of floating resonance parameters, this example is run with:
\begin{itemize}
\item no floating resonance parameters,
\item only the mass of the $\phi(1020)$ floating,
\item the mass and width of the $\phi(1020)$ floating,
\item the mass and width of both the $\phi(1020)$ and $f^{\prime}_2(1525)$ floating,
\end{itemize}
and the timings for each of those scenarios are provided in Table~\ref{tab:respars-speed}.
\item
\texttt{GenFit3KS.cc}\\
Example analysis of the fully-symmetric final state $\Bz\to\KS\KS\KS$, using
\texttt{LauSimpleFitModel} (\ie\ not including effects of \CP violation).
By default there are 10000 signal events per experiment and the signal
isobar model contains four components: $f_0(980)$, $f_0(1710)$, $f_2(2010)$
and $\chiczero$.
All of the resonance parameters are fixed in the fit.
No background contributions are included.
\item
\texttt{GenFitDs2KKpi.cc}\\
Example analysis of the decay $\Ds\to\pip\Kp\Km$, using
\texttt{LauSimpleFitModel} (\ie\ not including effects of \CP violation).
By default there are 10000 signal events and the signal isobar model
contains three components: $\phi(1020)$, $\Kstar(892)^0$ and a nonresonant
component.
All of the resonance parameters are fixed in the fit.
No background contributions are included.
\item
\texttt{GenFitEFKLLM.cc}\\
Example analysis of the decay $\Bz\to\Dzb\Kp\pim$, using
\texttt{LauSimpleFitModel} (\ie\ not including effects of \CP violation).
By default there are 5000 signal events per experiment and the signal
isobar model contains two components, which are the two parts of the EFKLLM
model for the $\Kp\pim$ S-wave (see Eq.~(\ref{eq:efkllm}) in
App.~\ref{sec:res-formulae}).
This example mainly serves to demonstrate how to use this particular model.
\item
\texttt{GenFitBelleCPKpipi.cc}\\
Example analysis of the decay $\Bp\to\Kp\pip\pim$, using
\texttt{LauCPFitModel}, which includes effects from \CP violation.
By default there are 5000 signal events per experiment and the signal
isobar model contains seven components: $\Kstar(892)^0$,
$\kaon^*_0(1430)^0$, $\rho(770)^0$, $f_0(980)$, \chiczero, and two
nonresonant components.
The slope of the $\pip\pim$ exponential nonresonant model and the effective
range and scattering length of the $\Kp\pim$ LASS nonresonant component are
floating parameters by default.
The isobar coefficients use the \texttt{LauBelleCPCoeffSet} form to
parameterise the \CP violation (see Table~\ref{tab:coeff-sets}) and a
two-stage fit is employed.
No background contributions are included.
\item
\texttt{GenFitKpipi.cc}\\
Example analysis of the decay $\Bp\to\Kp\pip\pim$, using
\texttt{LauCPFitModel}, which includes effects from \CP violation.
This example is based closely on the BaBar analysis from
Ref.~\cite{Aubert:2008bj}.
By default there are 4585 signal events per experiment and the signal
isobar model contains seven components: $\Kstar(892)^0$,
$\kaon^*_0(1430)^0$, $\kaon^*_2(1430)^0$, $\rho(770)^0$, $\omega(782)$,
$f_0(980)$, $f_2(1270)$, $f^{\prime}_0(1300)$, \chiczero, and a nonresonant
component.
All of the resonance parameters are fixed in the fit.
The isobar coefficients use the \texttt{LauCartesianCPCoeffSet} form to
parameterise the \CP violation (see Table~\ref{tab:coeff-sets}) and a
two-stage fit is employed.
Background contributions for combinatorial candidates and those from other
\B decays are included.
\item
\texttt{KMatrixExample.cc} and \texttt{KMatrixDto3pi.cc}\\
These examples demonstrate how to use the K-matrix description of the S-wave.
The first scenario is for the $\Bz\to\pip\pim\KS$ DP but the only terms
included in the signal model are $\rho(770)^0$, $\Kstar(892)^+$ and a
K-matrix component for the $\pip\pim$ S-wave.
By default there are 5000 signal events per experiment.
The second scenario is for the symmetric $\Dp\to\pip\pip\pim$ DP and includes in the signal model the $\rho(770)^0$, $f_2(1270)$, and the K-matrix.
By default there are 20000 signal events per experiment.
In both scenarios there are no background contributions and all resonance
parameters are fixed in the fit.
\item
\texttt{GenFitNoDP.cc} and \texttt{GenFitNoDPMultiDim.cc}\\
These examples demonstrate how to use the package to perform fits to
variables other than the DP.
The first case fits only a single variable (the invariant mass of the \Bp
candidate), while the second performs a 2D fit.
In both cases there are 5000 signal events.
In the first case there are two background components included, which have
yields of 7000 and 3000 events.
In the second case there is a single background component that has a yield
of 7000 events.
All yield parameters are floating, along with some of the shape parameters
of the PDFs.
Asymmetric uncertainties are evaluated in the first case.
\item
\texttt{runMasterSlave.sh}, \texttt{Master.cc} and \texttt{Slave.cc}\\
This example demonstrates how to perform a simultaneous fit to two categories
of data in the decay channel $\Bz\to\piz\piz\KS$.
The data are split based on the reconstruction category of the \KS candidate.
In one category there are 500 signal and 1200 background events,
while in the second there are 750 signal and 2500 background events.
The common signal DP model contains contributions from the $f_0(980)$,
$f_2(1270)$, $\Kstar(892)^0$, and $\kaon^*_0(1430)^0$ resonances.
The mass of the $\Kstar(892)^0$ resonance is a floating parameter of the
fit and a two-stage fit is employed.
The background component is distributed uniformly in the DP by default.
To demonstrate further the impact on the performance of simultaneous fitting, this example is run with:
\begin{itemize}
\item the Master and the two Slave processes all running on the same host,
\item the Master and the two Slave processes running on three separate hosts,
\end{itemize}
and the timings for each of these scenarios are provided in
Table~\ref{tab:simfit-speed}.
To run this particular example, the hosts have dual Intel Xeon
E5-2620v2 2.1\ghz 6-core CPUs and 64\gbytes of RAM and are connected via
10\gbsps ethernet.
\end{itemize}

\begin{table}[!htb]
\caption{
  Speed of execution of the examples provided with the \laura package.
  The times given are the total to generate/fit 50 toy experiments.
  The examples differ, as explained in detail in the text, not only through the complexity of the model but also through the number of signal and background events in each pseudoexperiment.
}
\label{tab:examples-speed}
\centering
\begin{tabular}{lrr}
\hline
Example                        & Execution time of  & Execution   \\
                               & toy generation     & time of fit \\
\hline
\texttt{GenFit3pi.cc}          & $  7\sec$          & $    9\sec$  \\
\texttt{GenFit3K.cc}           & $291\sec$          & $ 2513\sec$  \\
\texttt{GenFit3KS.cc}          & $ 78\sec$          & $   20\sec$  \\
\texttt{GenFitDs2KKpi.cc}      & $ 47\sec$          & $    9\sec$  \\
\texttt{GenFitEFKLLM.cc}       & $106\sec$          & $    3\sec$  \\
\texttt{GenFitBelleCPKpipi.cc} & $ 14\sec$          & $18909\sec$  \\
\texttt{GenFitKpipi.cc}        & $ 36\sec$          & $ 1073\sec$  \\
\texttt{KMatrixExample.cc}     & $194\sec$          & $  217\sec$  \\
\texttt{KMatrixDto3pi.cc}      & $ 46\sec$          & $  251\sec$  \\
\texttt{GenFitNoDP.cc}         & $  2\sec$          & $   11\sec$  \\
\texttt{GenFitNoDPMultiDim.cc} & $  3\sec$          & $    6\sec$  \\
\hline
\end{tabular}
\end{table}

\begin{table}[!htb]
\caption{
Speed of execution of the \texttt{GenFit3K.cc} example provided with the
\laura package with different sets of floating resonance parameters.
The times given are the total to fit 50 toy experiments.
}
\label{tab:respars-speed}
\centering
\begin{tabular}{lr}
\hline
Floated parameters & Execution time of fit \\
\hline
None                                                    & $   6\sec$ \\
Mass of $\phi(1020)$                                    & $ 702\sec$ \\
Mass and width of $\phi(1020)$                          & $1090\sec$ \\
Mass and width of $\phi(1020)$ and $f^{\prime}_2(1525)$ & $2513\sec$ \\
\hline
\end{tabular}
\end{table}

\begin{table}[!htb]
\caption{
Speed of execution of the fitting portion of the \texttt{runMasterSlave.sh}
example provided with the \laura package with the Master and two Slave
processes running either on the same or separate hosts.
The times given are the total to fit 50 toy experiments.
}
\label{tab:simfit-speed}
\centering
\begin{tabular}{lr}
\hline
Host setup                                             & Execution time of fit \\
\hline
Master and two Slave processes on single host          & $1972\sec$             \\
Master and two Slave processes on three separate hosts & $2151\sec$             \\
\hline
\end{tabular}
\end{table}

\section{Future developments}
\label{sec:future-devel}

There are several features that would help to improve and extend the functionality of \laura in the future.
Some of these potential future developments are described below.
In addition, it is anticipated that the \laura\ code will be continually
updated to take advantage of features in the latest \cpp\ standards.

\subsection{Plotting the amplitude}
\label{sec:amplitude-plots}

Currently the user can easily make plots of the DP distribution or its projections from the result of the fit (see Sec.~\ref{sec:toy-generation}).
It can also be of interest to draw amplitude-level quantities, for example to show the phase variation with two-body invariant mass of a particular partial wave. 
While this is currently possible in \laura\ (see Fig.~\ref{fig:BDpipi-examples} for an example), it would be desirable to provide an interface to simplify matters for the user.

\subsection{Decay-time-dependent fits}

As mentioned in Sec.~\ref{sec:introduction}, the evolution of DP structure with decay time of a $B$ or $D$ meson can be of interest to study \CP\ violation effects.
For example, studies of $\Bz \to \pip\pim\piz$ and $D\pip\pim$ decays are of interest to measure the angles $\alpha$ and $\beta$ of the CKM Unitarity Triangle with low theoretical uncertainty.
Studies of $\Bz \to \KS\pip\pim$ and $\KS\Kp\Km$ decays enable determinations of $\beta$ that are not as clean, but are potentially sensitive to effects of physics beyond the Standard Model.
Also, the $\Dz \to \KS\pip\pim$ channel appears the most sensitive to possible \CP\ violation effects associated with charm mixing.
Many other channels are potentially of interest.  

The \laura\ package has been used for decay-time-dependent DP analysis, for example, in Refs.~\cite{Aubert:2009me,Latham:2008zs}. 
However, this implementation was experiment-specific and therefore unsuitable for use in the more general case. 
Further development is necessary to establish a model that can deal generically with issues such as flavour tagging, decay time resolution and acceptance as well as production and detection asymmetries.

\subsection{Alternative handling of resolution effects}

The treatment of resolution discussed in Sec.~\ref{sec:expt-effects} and Sec.~\ref{sec:resol} is completely general, but is likely to be inefficient in certain cases.
For example, in the $\Bp \to \Kp\Km\Kp$ decay, there are narrow contributions from the $\phi(1020)$ and $\chi_{c0}$ resonances in specific regions of the DP, for which resolution effects may be important.
The rest of the DP is populated with broad or nonresonant states so that resolution can be safely neglected. 
An approach in which Gaussian (or non-Gaussian) smearing of Dalitz plot position can be used in only selected regions of the phase space may be useful.
In such a case it will be necessary to take care to avoid issues due to edge effects, including the possibility of an event being smeared to positions outside the kinematic boundary of the DP.

\subsection{Non-zero spins}

There are many interesting three-body decays that include particles with non-zero spin in the initial and final states, which cannot be fitted using the current version of \laura. 
Large samples of $b$-baryons are available in the LHCb data samples, which have baryons in the initial and final states, for example $\Lb\to\Dzb\proton\pim$ and 
$\Xibm\to\proton\Km\Km$ decays. Decays of $\B\to\jpsi hh^\prime$, where $h$ and $h^\prime$ are charged pions or kaons, are also interesting and contain the spin-1 $\jpsi$ particle. 
Another group of decay modes, $\B \to \Dstar hh^\prime$, which include the vector \Dstar meson, are interesting for spectroscopy of $D^{**}$ and $D_{s}^{**}$ states.

To enable the \laura package to cope with the decays above, several things would require updating or changing. Firstly, the phase space of the problem is expanded from two to five 
dimensions where additional degrees of freedom would be some angular variables. 
The helicity of particles of non-zero spin, like the $\jpsi$, must also be considered, requiring a sum over the helicity states. For the initial and final state this sum must be incoherent 
and for the intermediate states a coherent sum is required.
The Zemach spin terms currently implemented must also be changed for the
non-zero spin particles. 
One potential way to calculate the spin factors would be to interface \laura with the \texttt{qft++} package~\cite{Williams:2008wu}, which allows the spin terms to be calculated for any process.

\subsection{Genetic algorithms}

To ensure that the global minimum in the NLL has been found, \laura allows many fits to be performed with randomised starting values for the floated fit parameters. This method 
works well, but can become time consuming if the global minimum is not found in a high proportion of fit attempts. 
Genetic algorithms could provide a method to find sensible starting values for the floated parameters such that the global minimum is always found. This could be 
achieved by interfacing to existing software packages with implementations of genetic algorithms.

\subsection{Interface to \evtgen}

The \evtgen package~\cite{Lange:2001uf} is designed to predominantly simulate the decays of $b$- and $c$-hadrons. It is important in experimental particle physics to produce simulated data samples that are 
realistic and match the true data samples. Typically in three-body decays the simulated events are produced flat in the DP, without resonant contributions. 
It would be beneficial if \laura could be used directly to provide realistic DP distributions for simulated samples of three-body decays.

\section{Summary}
\label{sec:summary}

The \laura\ package provides a flexible and optimised framework for Dalitz-plot analysis.
While it can be used for the decay of any stable spin-zero particle to any final state containing three stable spin-zero particles, it has until now been most widely used for decays of $B$ or $D$ mesons to three pseudoscalars.  
Features included in \laura\ allow the physics of such decays to be probed in detail, including studies of the resonances appearing in the contributing partial waves, and investigations of \CP-violating effects.   
Use of the \laura\ software has resulted in numerous publications to date, with many more expected in future with the increasingly large data samples available at LHCb, Belle~II and other experiments.

\section{Acknowledgements}
\label{sec:acknowledgements}

The \laura\ package has been under development for many years, with support
principally from the Science and Technology Facilities Council (United
Kingdom) and by the European Research Council under FP7.  
Individual authors have received support from Marie Sk\l{}odowska-Curie
Actions and from the University of Warwick.  
We thank the \laura\ user community for feedback, bug reports, testing 
and other contributions to improve the package, 
in particular Louis Henry, Adlene Hicheur, Patricia Magalh\~{a}es,
Jussara Miranda, Sian Morgan and Charlotte Wallace.
We also acknowledge productive discussions regarding aspects of Dalitz plot
analysis with many members of the \babar, Belle and LHCb collaborations, notably 
Eli Ben-Haim, Alex Bondar, Jeremy Dalseno, Bill Dunwoodie, Brian Meadows and Anton Poluektov.
Similarly, instructive communications with members of the theory community have
been of great benefit; in particular we thank Vladimir Anisovich,
David Bugg, Leonard Lesniak, Benoit Loiseau, Mike Pennington,
Alessandro Pilloni, Andrey Sarantsev and Adam Szczepaniak. Furthermore, we thank 
Bertram Kopf and Matthias Steinke for providing access to the PAWIAN software 
(PANDA collaboration) for cross-checking the $K$-matrix implementation.
Finally, we acknowledge useful input on technical features of the package from
Ren\'{e} Brun and Bertrand Echenard.

\clearpage


{\noindent\bf\Large Appendices}

\appendix

\setcounter{table}{0}
\renewcommand{\thetable}{A\arabic{table}}

\section{Formulae for available lineshapes}
\label{sec:res-formulae}

This section presents the complete formulae for all resonance shapes implemented
in \laura. Table~\ref{tab:resForms} gives the list of shapes, together with
the corresponding \texttt{LauResonanceModel} enumeration integer that is required to
specify the resonance type for the \texttt{LauIsobarDynamics addResonance} function, as 
well as the equation number(s) that provide the expression for the resonance
mass term $R(m)$ used in Eq.~(\ref{eq:ResDynEqn}).
The K-matrix shape is particularly complicated and is therefore described in a dedicated subsection.

\begin{table}[!hbt]
\caption{
  List of the allowed resonance shape types.
  The \texttt{LauResonanceModel} case-sensitive enumeration in the \texttt{LauAbsResonance} 
abstract class specifies the integer that selects the resonance type for the \texttt{addResonance}
 function. For example, the simple Breit--Wigner integer type is \texttt{LauAbsResonance::BW}.}
\centering
\resizebox{\textwidth}{!}{
\begin{tabular}{lll}
\hline
Shape name                                                 & Enumeration                  & $R(m)$ Eq. \\
\hline   
Simple Breit--Wigner                                       & \texttt{BW}                  & (\ref{eq:SimpleBW}) \\
Relativistic Breit--Wigner (RBW)                           & \texttt{RelBW}               & (\ref{eq:RelBWEqn}) \\
Modified Breit--Wigner from Gounaris--Sakurai (GS)         & \texttt{GS}                  & (\ref{eq:GS}) \\
Flatt\'e or coupled-channel Breit--Wigner                  & \texttt{Flatte}              & (\ref{eq:Flatte}) \\
$\sigma$ or $f_0(500)$                                     & \texttt{Sigma}               & (\ref{eq:sigma}) \\
$\kappa$ or low-mass $K\pi$ scalar                         & \texttt{Kappa}               & (\ref{eq:sigma}) \\
Low-mass $D\pi$ scalar                                     & \texttt{Dabba}               & (\ref{eq:dabba}) \\
LASS $K\pi$ S-wave                                         & \texttt{LASS}                & (\ref{eq:LASSEqn}) \\
Resonant part of $K\pi$ LASS                               & \texttt{LASS\_BW}            & (\ref{eq:LASSEqn}) (2$^{\rm{nd}}$ term)\\
Non-resonant part of $K\pi$ LASS                           & \texttt{LASS\_NR}            & (\ref{eq:LASSEqn}) (1$^{\rm{st}}$ term)\\
Form-factor description of the $K\pi$ S-wave               & \texttt{EFKLLM}              & (\ref{eq:efkllm}) \\
S-wave using $K$-matrix and $P$-vector                     & \texttt{KMatrix}             & (\ref{eq:KMatProd})--(\ref{eq:prodPoleSVP}) \\
Uniform non-resonant (NR)                                  & \texttt{FlatNR}              & $R(m) \equiv 1$ \\
Theoretical NR model                                       & \texttt{NRModel}             & (\ref{eq:nrmodel}) \\
Empirical NR exponential                                   & \texttt{BelleNR}             & (\ref{eq:expnonres}) \\
Empirical NR power-law                                     & \texttt{PowerLawNR}          & (\ref{eq:nonrespower}) \\
Empirical NR exponential for symmetrised DPs               & \texttt{BelleSymNR}          & (\ref{eq:symnr}) \\
Empirical NR Taylor expansion for symmetrised DPs          & \texttt{TaylorNR}            & (\ref{eq:taylornr}) \\
Empirical NR polynomial                                    & \texttt{PolNR}               & (\ref{eq:polynr}) \\
Model-independent partial wave (magnitude \& phase)        & \texttt{MIPW\_MagPhase}      & (\ref{eq:mipw}) \\
Model-independent partial wave (real \& imaginary)         & \texttt{MIPW\_RealImag}      & (\ref{eq:mipw}) \\
Incoherent Gaussian shape                                  & \texttt{GaussIncoh}          & (\ref{eq:incohgauss}) \\
$\rho-\omega$ mixing: GS for $\rho$, RBW for $\omega$      & \texttt{RhoOmegaMix\_GS}     & (\ref{eq:rhoomega}) \\
  \hspace*{6em} neglecting $\Delta^2$ denominator term     & \texttt{RhoOmegaMix\_GS\_1}  & (\ref{eq:rhoomega}) \\
$\rho-\omega$ mixing: RBW for both $\rho$ and $\omega$     & \texttt{RhoOmegaMix\_RBW}    & (\ref{eq:rhoomega}) \\
  \hspace*{6em} neglecting $\Delta^2$ denominator term     & \texttt{RhoOmegaMix\_RBW\_1} & (\ref{eq:rhoomega}) \\
\hline
\end{tabular}
}
\label{tab:resForms}
\end{table}

The simple Breit--Wigner lineshape is given by
\begin{equation}
\label{eq:SimpleBW}
R(m) = \frac{1}{m - m_0 - \frac{i}{2}\Gamma_0} \equiv 
\frac{(m - m_0) + \frac{i}{2}\Gamma_0}{(m - m_0)^2 + \frac{\Gamma^2_0}{4}} \,,
\end{equation}
where $m_0$ is the pole mass and $\Gamma_0$ is the resonance width.
The more commonly used relativistic Breit--Wigner lineshape is described in Sec.~\ref{sec:lineshapes}.
We note here that the relativistic Breit--Wigner lineshape can also describe so-called virtual contributions, from resonances with masses outside the kinematically accessible region of the Dalitz plot, with one modification:
in the calculation of the momenta, the mass $m_0$ is set to a value $m_0^{\rm{eff}}$ within the kinematically allowed range. 
This is accomplished with the {\it ad-hoc} formula
\begin{equation}\label{eq:effmass}
 m_0^{\rm{eff}}(m_0) = m^{\rm{min}} + \frac{1}{2}(m^{\rm{max}} - m^{\rm{min}}) 
\left[ 1 + \tanh\left( \frac{m_0 - \frac{m^{\rm{min}}+m^{\rm{max}}}{2}}
{m^{\rm{max}}-m^{\rm{min}}} \right) \right]\, ,
\end{equation}
where $m^{\rm{max}}$ and $m^{\rm{min}}$ are the upper and lower limits of the kinematically allowed mass range.
For virtual contributions, only the tail of the RBW function enters the Dalitz plot.

The Gounaris--Sakurai form of the Breit--Wigner lineshape~\cite{GS} is usually used 
as an alternative model for the $\rho$ resonance. It is given by
\begin{equation}
\label{eq:GS}
R(m) = \frac{1+D\cdot\Gamma_0/m_0}
                {(m_0^2 - m^2) + f(m) - i\, m_0 \Gamma(m)} \,,
\end{equation}
where
\begin{equation}
\label{eq:GSfm}
f(m) = \Gamma_0 \,\frac{m_0^2}{q_0^3}\,
       \left[\;
             q^2 \left[h(m)-h(m_0)\right] +
             \left(\,m_0^2-m^2\,\right)\,q^2_0\,
             \frac{dh}{ds}\bigg|_{m_0}
       \;\right] \,,
\end{equation}
$q$ is the magnitude of the momentum of one of the daughter particles
in the resonance rest-frame,
\begin{equation}
\label{eq:GShm}
h(m) = \frac{2}{\pi}\,\frac{q}{m}\,
       \ln\left(\frac{m+2q}{2m_\pi}\right) \,,
\end{equation}
and
\begin{equation}
\label{eq:GSdh}
\frac{dh}{ds}\bigg|_{m_0} =
h(m_0)\left[(8q_0^2)^{-1}-(2m_0^2)^{-1}\right] \,+\, (2\pi m_0^2)^{-1} \,.
\end{equation}
%
The constant parameter $D$ is given by~\cite{GS}
\begin{equation}
\label{eq:GSd}
D = \frac{3}{\pi}\frac{m_\pi^2}{q_0^2}\,
    \ln\left(\frac{m_0+2q_0}{2m_\pi}\right)
    + \frac{m_0}{2\pi\,q_0}
    - \frac{m_\pi^2 m_0}{\pi\,q_0^3} \,.
\end{equation}

The Flatt\'e~\cite{Flatte:1976xu}, or coupled two-channel Breit--Wigner, lineshape
is usually used to model $f_0(980)$, $K^{*}_0(1430)$ and $a_0(980)$ states.
It was originally introduced in the form
\begin{equation}
\label{eq:Flatte}
R(m) = \frac{1}{(m_0^2 - m^2) - i m_0 [\Gamma_1(m) + \Gamma_2(m)]} \,.
\end{equation}
The decay widths in the two systems are usually represented by products of
couplings and dimensionless phase-space factors:
\begin{equation}
\label{eq:Gamma1}
\Gamma_1(m) = g_1 f_A \left(\frac{1}{3}\sqrt{1 - (m_{1,1} + m_{1,2})^2/m^2} + 
\frac{2}{3}\sqrt{1 - (m_{1,3} + m_{1,4})^2/m^2} \right) \,,
\end{equation}
\begin{equation}
\label{eq:Gamma2}
\Gamma_2(m) = g_2 f_A \left(\frac{1}{2}\sqrt{1 - (m_{2,1} + m_{2,2})^2/m^2} + 
\frac{1}{2}\sqrt{1 - (m_{2,3} + m_{2,4})^2/m^2} \right) \,.
\end{equation}
Here the fractional coefficients come from isospin conservation,
$m_{i,j}$ denotes the invariant mass of the daughter particle $j$ (1--4) in
channel $i$ (1--2),
and $g_1$ and $g_2$ are coupling constants whose values are assumed to be those
provided in Table~\ref{tab:flattepars}. 
The Clebsch-Gordan coefficients in Eqs.~\ref{eq:Gamma1} and~\ref{eq:Gamma2} are not guaranteed to be correct for every possible resonance that could be modelled with the Flatt\'e lineshape, but are appropriate for every case considered in Table~\ref{tab:flattepars}. 
The expressions for the widths are continued analytically ($\Gamma \ra i |\Gamma|$) when $m$ is below any of the specific channel thresholds,
contributing to the real part of the amplitude, while the Adler-zero 
term $f_A = (m^2 - s_A)/(m^2_0 - s_A)$ can be used to suppress false kinematic 
singularities when $m$ goes below threshold~\cite{Bugg:2003kj} 
(otherwise $f_A$ is set to unity).

Variants of the Flatt\'e lineshape have been used in the literature.
In some cases, \eg\ Refs.~\cite{Ablikim:2004wn,Abele:1998qd}, the constant
$m_0$ that multiplies the widths in the denominator of Eq.~(\ref{eq:Flatte})
is absorbed into the couplings.
As a consequence the couplings have dimensions of mass-squared, and are
sometimes denoted as $g_i$~\cite{Ablikim:2004wn} and sometimes as $g_i^2$~\cite{Abele:1998qd}.
In Table~\ref{tab:flattepars} all values have been converted to be consistent
with Eqs.~(\ref{eq:Flatte})--(\ref{eq:Gamma2}).
In \laura\ it is only possible to use the Flatt\'e lineshape for the systems
specified in Table~\ref{tab:flattepars}.
At construction time the resonance name is checked and the corresponding
parameter values are set; these can be modified by the user if desired.

\begin{table}[!htb]
\caption{
  The four daughter particles used for each channel term $m_{ij}$, as well as the coupling ($g_1$, $g_2$) and Adler-zero ($s_A$) constants for the Flatt\'e lineshapes.
  Units of \gev\ for $g_{1,2}$ (or $\gev^2$ for $m_0g_{1,2}$ when taken from Refs.~\cite{Ablikim:2004wn,Abele:1998qd}) and \gevgevcccc\ for $s_A$ are implied.
}
\centering
\resizebox{\textwidth}{!}{
\begin{tabular}{lllllll}
\hline
Resonance & Channel 1 & Channel 2 & $g_1$ or $m_0g_1$ & $g_2$ or $m_0g_2$ & $s_A$ & Reference \\
\hline
$f_0(980)$ & \piz,\piz,\pipm,\pipm & \Kpm,\Kpm,\Kz,\Kz & 0.165 & $4.21g_1$ & --- & \cite{Ablikim:2004wn} \\
$K^{*}_0(1430)^0$ & \Kz,\piz,\Kpm,\pipm & \Kz,\etapr,\Kz,\etapr & 0.304 & 0.380 & 0.234 & \cite{Bugg:2003kj} \\
$K^{*}_0(1430)^{\pm}$ & \Kpm,\piz,\Kz,\pipm & \Kpm,\etapr,\Kpm,\etapr & 0.304 & 0.380 & 0.234 & \cite{Bugg:2003kj} \\
$a_0(980)^0$ & \etaz,\piz,\etaz,\piz & \Kpm,\Kpm,\Kz,\Kz & 0.105 & $1.03g_1$ & --- & \cite{Abele:1998qd} \\
$a_0(980)^{\pm}$ & \etaz,\pipm,\etaz,\pipm & \Kpm,\Kz,\Kpm,\Kz & 0.105 & $1.03g_1$ & --- & \cite{Abele:1998qd} \\
\hline
\end{tabular}
}
\label{tab:flattepars}
\end{table}

The $\sigma$ or $f_0(500) \ra \pi\pi$ and $\kappa$ or $K^*_0(800) \ra K\pi$
low-mass scalar resonances can be described using the form
\begin{equation}
\label{eq:sigma}
R(m) = \frac{1}{M^2 - s -iM\Gamma(s)} \,,
\end{equation}
where $M$ is the mass where the phase shift goes through 90$^{\circ}$ for real $s \equiv m^2$,
and the width
\begin{equation}
\label{eq:sigmawidth}
\Gamma(s) = \sqrt{1 - (m_1 + m_2)^2/s}\left(\frac{s - s_A}{M^2 - s_A}\right)(b_1 + b_2s)e^{-(s - M^2)/A} \,,
\end{equation}
where the square-root term is the phase space factor, which requires
the invariant masses of the daughter particles $m_1$ and $m_2$, $s_A$ is the Adler-zero constant,
while $b_1$, $b_2$ and $A$ are additional constants~\cite{Bugg:2003kj}.
Table~\ref{tab:sigmakappa} gives the default values of the parameters.
\begin{table}[!htb]
\caption{Default values of the parameters for the $\sigma$ and $\kappa$ lineshapes
based on BES data~\cite{Bugg:2003kj}.}
\centering
\begin{tabular}{llllll}
\hline
Resonance & $M$ (\nbspgevcc) & $b_1$ (\nbspgevcc) & $b_2$ (\nbspgevcc) & $A$ (\nbspgevgevcccc) & $s_A$ \\
\hline
$\sigma$  & 0.9264 &  0.5843 & 1.6663 & 1.082 & $0.5m^2_{\pi}$ \\
$\kappa$  & 3.3 & 24.49 & 0.0 & 2.5 & $m^2_K - 0.5m^2_{\pi}$ \\
\hline
\end{tabular}
\label{tab:sigmakappa}
\end{table}

The $D\pi$ S-wave can be parameterised using the form provided by Bugg~\cite{Bugg:2009tu},
who labels the pole state as ``dabba'':
\begin{equation}
\label{eq:dabba}
R(m) = \frac{1}{1 - \beta(m^2-s_0) - ib\rho(m^2-s_A)e^{-\alpha(m^2-s_0)}} \,,
\end{equation}
where $\rho$ is the Lorentz invariant phase space factor $\sqrt{1 - s_0/m^2}$,
$s_0$ is the square of the sum of the invariant masses of the $D$ ($m_D$) and 
$\pi$ ($m_{\pi}$) daughters, $s_A$ is the Adler-zero term 
$m^2_D - 0.5m^2_{\pi}$ that comes from chiral symmetry breaking~\cite{Adler:1965ga}, while
$b$ = 24.49, $\alpha$ = 0.1 and $\beta$ = 0.1.

The RBW function is a very good approximation for
narrow resonances well separated from any other resonant or nonresonant
contribution in the same partial wave.  
This approximation is known to be invalid in the $K\pi$ S-wave, since the 
$\Kstarbsubz(1430)$ resonance interferes strongly with a slowly varying 
nonresonant term~\cite{Meadows:2007jm}.
The so-called LASS lineshape~\cite{lass} has been developed to combine these 
amplitudes,
\begin{eqnarray}
\label{eq:LASSEqn}
R(m) & = & \frac{m}{q \cot{\delta_B} - iq} + e^{2i \delta_B} 
\frac{m_0 \Gamma_0 \frac{m_0}{q_0}}
{(m_0^2 - m^2) - i m_0 \Gamma_0 \frac{q}{m} \frac{m_0}{q_0}}\, , \\
{\rm with} \ \cot{\delta_B} & = & \frac{1}{aq} + \frac{1}{2} r q \, ,
\end{eqnarray}
where $m_0$ and $\Gamma_0$ are now the pole mass and width of the $\Kstarbsubz(1430)$, 
and $a$ and $r$ are parameters that describe the shape.  
Most implementations of the LASS shape in amplitude analyses of \B meson
decays~\cite{Aubert:2004cp,Aubert:2005ce} apply a cut-off
to the slowly varying part close to the charm hadron mass ($\sim 1.7\gevcc$).

An alternative representation of the $K\pi$ S-wave amplitude can be made using the 
\texttt{EFKLLM} model described in Ref.~\cite{PhysRevD.79.094005} (the acronym comes
from the surnames of the authors of that paper), which uses a tabulated form-factor $f_0^{K\pi}(m^2)$ 
that is interpolated using two splines (one each for the magnitude and phase parts), multiplied by a 
scaling power-law mass-dependence $m^{\ell}$, leading to
\begin{equation}
\label{eq:efkllm}
R(m) = f_0^{K\pi}(m^2) \cdot m^{\ell} \, ,
\end{equation}
where suggested values for the exponent $\ell$ are zero for $\kappa$ (this is also the default value)
and $-2$ for $\Kstarbsubz(1430)$.

Because of the large phase-space available in three-body \B meson decays, it 
is possible to have nonresonant amplitudes (\ie\ contributions that are not associated 
with any known resonance, including virtual states) that are not constant
across the Dalitz plot. One possible parameterisation, based on theoretical
considerations of final-state interactions in $\Bpm \to \Kpm\pip\pim$ decays~\cite{Bediaga:2008zz}, uses the form
\begin{equation}
\label{eq:nrmodel}
R(m) = \left[ m_{13}m_{23} f_1(m^2_{13}) f_2(m^2_{23}) e^{-d_0 m^4_{13}m^4_{23}} \right]^{\frac{1}{2}} \,,
\end{equation}
where
\begin{equation}
\label{eq:nrmodelf}
f_j(m^2) = \frac{1}{1 + e^{a_j(m^2 - b_j)}} \,,
\end{equation}
with the constant parameters $d_0 = 1.3232\times10^{-3}$\,GeV$^{-8}$,
$a_1 = 0.65$\,GeV$^{-2}$, $b_1 = 18$\,GeV$^2$, $a_2 = 0.55$\,GeV$^{-2}$
and $b_2 = 15$\,GeV$^2$ in natural units.

There are several empirical methods that can be used to model the nonresonant
contributions. One is to use an exponential form factor~\cite{Garmash:2004wa}
\begin{equation}
\label{eq:expnonres}
R(m) = e^{-\alpha m^2} \, ,
\end{equation}
while another form is simply a power-law distribution
\begin{equation}
\label{eq:nonrespower}
R(m) = m^{-2\alpha} \,,
\end{equation}
where in both cases $\alpha$ is a parameter that must be determined from the data.
For symmetric DPs, the exponential form is modified to
\begin{equation}
\label{eq:symnr}
R(m) = e^{-\alpha m^2_{13}} + e^{-\alpha m^2_{23}} \,,
\end{equation}
while a Taylor expansion up to first order can also be used:
\begin{equation}
\label{eq:taylornr}
R(m) = 1 + \frac{\alpha(m^2_{13} + m^2_{23})}{m^2_P} \,,
\end{equation}
where $m_P$ is the invariant mass of the parent $P$.
Another possible description for non-symmetric DPs is based
on the polynomial function~\cite{Lees:2012kxa}
\begin{equation}
\label{eq:polynr}
R(m) = \left[m - \frac{1}{2}\left(m_P + \frac{1}{3}(m_1 + m_2 + m_3)\right)\right]^n \,,
\end{equation}
where $m_k$ is the invariant mass of daughter particle $k$ and $n$ is
the integer order equal to 0, 1 or 2; a quadratic dependence in $m$ can be constructed
by using up to three polynomial $R(m)$ terms, one for each order along with
their individual $c_j$ amplitude coefficients.

We next come to the model that implements the $\rho-\omega$ mass mixing amplitude
described in Ref.~\cite{Rensing:259802}
\begin{equation}
\label{eq:rhoomega}
A_{\rho-\omega} = A_{\rho} \left[ \frac{1 + A_{\omega} \Delta|B| e^{i \phi_{B}}}{1 - \Delta^2 A_{\rho} A_{\omega}} \right],
\end{equation}
where $A_{\rho}$ is the $\rho$ lineshape, $A_{\omega}$ is the $\omega$ lineshape, $|B|$ and 
$\phi_{B}$ are the relative magnitude and phase of the production amplitudes of
$\rho$ and $\omega$, and $\Delta \equiv \delta(m_{\rho} + m_{\omega})$, where $\delta$ governs 
the electromagnetic mixing of $\rho$ and $\omega$ (with pole masses $m_{\rho}$ and $m_{\omega}$).
Here, the amplitude $A_{\omega}$ is always given by the RBW form of Eq.~(\ref{eq:RelBWEqn}), while 
the amplitude $A_{\rho}$ can either be represented using
the Gounaris--Sakurai formula given in Eq.~(\ref{eq:GS}) or the RBW form; the required shape is 
selected using either the \texttt{RhoOmegaMix\_GS} or \texttt{RhoOmegaMix\_RBW} enumeration integer 
labels given in Table~\ref{tab:resForms}.
When ignoring the small $\Delta^2$ term in the denominator of Eq.~(\ref{eq:rhoomega}),
this is equivalent to the parameterisation described in Ref.~\cite{Akhmetshin:2001ig}; this option can be
chosen using either the \texttt{RhoOmegaMix\_GS\_1} or \texttt{RhoOmegaMix\_RBW\_1} enumeration labels, 
depending on what lineshape is needed for the $\rho$ resonance.
From SU(3) symmetry, the $\rho$ and $\omega$ are expected to be
produced coherently, which gives the prediction $|B|e^{i\phi_B} = 1$.
To avoid introducing any theoretical assumptions, however, it is advisable that these 
parameters are left floating in the fit. In general $\delta$ is complex, although the
imaginary part is small so this is neglected. The theory prediction for
$\delta$ is around $2 \mev$~\cite{PhysRev.134.B671}, and previous analyses
have found $|\delta|$ to be $2.15 \pm 0.35 \mev$~\cite{Rensing:259802} and
$1.57 \pm 0.16 \mev$, and $\rm{arg}~\delta$ to be $0.22 \pm 0.06$~\cite{Akhmetshin:2001ig}.
These parameters can be also be floated in the fit.

If the dynamical structure of the DP cannot be described by any of the
forms given above, then the \texttt{LauModIndPartWave} class can be used to define
a model-independent partial wave component, using splines
to produce an amplitude. It requires a series of mass points called ```knots'', in 
ascending order, which sets the magnitude $r(m)$ and phase $\phi(m)$ at each point $m$ that 
can be floated when fitted to data:
\begin{equation}
\label{eq:mipw}
R(m) = r(m)\left[\cos \phi(m) + i\sin \phi(m) \right] \,.
\end{equation}
The amplitude for points between knots is found using cubic spline interpolation, and
is fixed to zero at the kinematic boundary. There are two implementations for representing
the amplitudes: one uses magnitudes and phases (\texttt{MIPW\_MagPhase}), while the other 
uses real and imaginary parts (\texttt{MIPW\_RealImag}).

Finally, the incoherent Gaussian lineshape form is given by
\begin{equation}
\label{eq:incohgauss}
R(m) = e^{-(m - m_0)^2/2G^2_0} \,,
\end{equation}
where $m_0$ is the mass peak and $G_0$ is the width. This can be used
to parameterise the amplitude for a very narrow resonance, where the measurement
of the width is dominated by experimental resolution effects, producing a lineshape 
that is indistinguishable from a Gaussian distribution. The narrow width ensures that 
the resonance will not interfere significantly with other resonances in the DP,
\ie\ it will be incoherent. This form could also be used to
parameterise narrow background resonance contributions that would otherwise be 
excluded with mass vetoes, such as the $\Dz$ meson decay to $\Km \pip$ in the charmless mode
$\Bm \ra \Km \pip \pim$, when used with the \texttt{addIncoherentResonance} function of 
\texttt{LauIsobarDynamics}.

\subsection{$K$-matrix}
\label{sec:KMatrix}

The isobar model, described earlier in Sec.~\ref{sec:DalitzGeneralities}, can be used
to describe the dynamics of three-body decays when the quasi two-body 
resonances are relatively narrow and isolated. However, this model does not 
satisfy scattering ($S$-matrix) unitarity, thereby violating the conservation 
of quantum mechanical probability current, when there are broad, overlapping 
resonances (with the same spin-parity), such as the intermediate S-wave
states $\sigma$ for $\pi\pi$ and $\kappa$ for $K\pi$ channels. 
Assuming that the dynamics is dominated by two-body 
processes, meaning that the S-wave does not interact with the 
rest of the products in the final state, then unitarity is naturally conserved within 
the $K$-matrix approach~\cite{Chung:1995dx}, which was originally developed
for two-body scattering~\cite{Dalitz:1960du} and the study of resonances 
in nuclear reactions~\cite{Wigner:1946zz,Wigner:1947zz}, but was extended
to describe resonance production in a more general way~\cite{Aitchison:1972ay}.

The scattering matrix $S$, describing the general transformation of an initial state
to a final state, can be defined as
\begin{equation}
\label{eq:SMatrix}
S \equiv I + 2i T\,,
\end{equation}
where $I$ is the identity matrix, representing the case when the initial and final
states do not interact at all, and $T$ is the physical (observable) transition matrix. 
Conservation of scattering probability means that the $n \times n$ $S$ matrix, where
$n$ is the number of channels, is unitary ($S S^{\dagger}$ = $S^{\dagger}S = I$). 
The factor $2i$ is introduced so that the transition amplitude for a single resonance 
channel corresponds to a circle of unit diameter centred at (0, $i/2$) in the 
complex plane; physically allowed values of $T$ will be along the boundary
(elastic scattering) or inside (inelastic scattering) this unitarity circle.
Using Eq.~(\ref{eq:SMatrix}), it can be shown
that the $n \times n$ $K$ matrix operator, defined as
\begin{equation}
\label{eq:KMatrix}
K \equiv (T^{-1} + iI) ^{-1} \,,
\end{equation}
is Hermitian ($K$ = $K^{\dagger}$)~\cite{Chung:1995dx}.
Furthermore, $K$ is real and symmetric, owing to the time-reversal invariance
of the $S$ and $T$ matrices. Rearranging Eq.~(\ref{eq:KMatrix}) gives the following 
expression for the scattering transition operator in terms of the $K$ matrix:
\begin{equation}
\label{eq:Tmatrix}
T = (I - iK)^{-1} K.
\end{equation}
The normalisation of the two-body wave functions requires the inclusion of phase-space
factors in both the initial and final states~\cite{Chung:1971ri}. This then leads to 
the following definition of the Lorentz-invariant transition amplitude $\hat{T}$:
\begin{equation}
\label{eq:THat}
T_{uv} \equiv \{\rho_u^{\dagger}\}^{\frac{1}{2}} \, \hat{T}_{uv} \{\rho_v\}^{\frac{1}{2}} \,,
\end{equation}
where $u$ and $v$ indicate the channel indices (from 1 to $n$) and
$\rho$ is the normalised diagonal $n \times n$ phase space matrix, whose elements are
equal to $2q/m$, where $q$ is the magnitude of the momentum of either daughter
in the rest-frame of the two-body state that has invariant mass $m = \sqrt{s}$.
In general, the phase space element of channel $u$ is given by
\begin{equation}
\label{eq:rhoterm}
\rho_u = \sqrt{\left(1 - \frac{(m_{1u} + m_{2u})^2}{s}\right)
\left(1 - \frac{(m_{1u} - m_{2u})^2}{s}\right)} \,,
\end{equation}
where $m_{1u}$ and $m_{2u}$ are the rest masses of the two daughters~\cite{PDG2016},
and is continued analytically by setting $\rho_u$ to be $i|\rho_u|$ when the channel is below
its mass threshold, provided it does not cross into another channel.

The Lorentz-invariant form of the $K$ matrix, which will also be real and symmetric, 
can be written as
\begin{equation}
\label{eq:KHat}
\hat{K}^{-1} = \hat{T}^{-1} + i \rho \,,
\end{equation}
which then implies that the Lorentz-invariant transition amplitude is given by
\begin{equation}
\label{eq:THat2}
\hat{T} = (I - i \hat{K} \rho)^{-1} \cdot \hat{K} \,.
\end{equation}
We can now use this expression to give the general amplitude of the \emph{production} of 
overlapping resonance states. This model or ansatz~\cite{Aitchison:1972ay} describes the amplitude 
of channel $u$ in terms of the initial $\hat{P}$-vector preparation of channel states $v$,
that has the same form as $\hat{K}$, transforming (``scattering'') 
into the final state $u$ via the propagator term $(I - i \hat{K} \rho)^{-1}$:
\begin{equation}
\label{eq:KMatProd}
{\cal{F}}_u = \sum_{v=1}^{n} [I - i \hat{K} \rho]^{-1}_{uv} \cdot \hat{P}_{v} \,.
\end{equation}
The scattering $\hat{K}$ matrix can be parameterised as a combination of the summation 
of $N$ poles with real bare masses $m_{\alpha}$, together with nonresonant slowly-varying parts
(SVPs), so-called since they essentially have a $1/s$ dependence, with real (and symmetric) coupling 
constants $f^{\rm{scatt}}_{uv}$~\cite{Anisovich:2002ij}:
\begin{equation}
\label{eq:KMatTerms}
\hat{K}_{uv}(s) = \left( \sum_{\alpha=1}^N{\frac{g^{(\alpha)}_u g^{(\alpha)}_v}{m^2_{\alpha} - s}} + 
f^{\rm{scatt}}_{uv} \frac{m^2_0 - s^{\rm{scatt}}_0}{s - s^{\rm{scatt}}_0} \right) f_{A0}(s) \,,
\end{equation}
where $g_u^{(\alpha)}$ denotes the real coupling constant of the pole $m_{\alpha}$ to the channel $u$,
the factor
\begin{equation}
\label{eq:adler}
f_{A0}(s) = \left(\frac{1\,{\rm{\gevgevcccc}} - s_{A0}}{s - s_{A0}}\right)
\left(s - \frac{1}{2}s_A m_{\pi}^2\right)
\end{equation}
is the Adler zero term that suppresses the false kinematic singularity when $s$ goes 
below the $\pi\pi$ production threshold~\cite{Adler:1965ga},
while $m_0^2$, $s^{\rm{scatt}}_0$, $s_{A}$ and $s_{A0}$ are real constants of order unity;
typical values are $m_0^2 = 1\,\gevgevcccc$, $s^{\rm{scatt}}_0 = -5\,\gevgevcccc$, 
$s_A = 1$, and $s_{A0} = 0\,\gevgevcccc$.
Note that the real poles $m_{\alpha}$ are the masses of the so-called \emph{bare} states of the 
system, which do not correspond to the masses and widths of \emph{resonances} 
(mixtures of bare states) from the complex poles in the physical $T$ matrix.
The production vector $\hat{P}$ is parameterised in an analogous form to the $\hat{K}$ matrix:
\begin{equation}
\label{eq:PVect}
\hat{P}_v(s) = \sum_{\alpha=1}^N{\frac{\beta_{\alpha} g^{(\alpha)}_v}{m^2_{\alpha} - s} + 
f^{\rm{prod}}_{v}\frac{m_0^2 - s_0^{\rm{prod}}}{s - s_0^{\rm{prod}}}} \,,
\end{equation}
where $\beta_{\alpha}$ and $f^{\rm{prod}}_{v}$ (which both depend on the final state channel $u$)
are complex production constants for the poles and nonresonant SVPs, respectively, and 
$s_0^{\rm{prod}}$ is of order unity and is usually taken to be approximately equal to $s_0^{\rm{scatt}}$. 
It is important that the production and scattering processes use 
the same poles $m_{\alpha}$, otherwise the transition amplitude would vanish (diverge) 
at the $\hat{K}$-matrix ($\hat{P}$-vector) poles; the singularities need to cancel out
for the total amplitude. Also note that the Adler zero suppression
factor given in Eq.~(\ref{eq:adler}) is generally not needed for $\hat{P}$, since its inclusion 
does not improve the description of S-wave amplitudes found in experimental 
data~\cite{Link:2003gb,kmatrix,Aubert:2008bd,Link:2009ng}.

In order to clarify what amplitudes are used, we can separate out the production pole and SVP terms
shown above. The amplitude of each production pole $m_{\alpha}$ for the final state $u$ is given by
\begin{equation}
\label{eq:prodPole}
{\cal{A}}_{\alpha,u}(s) = \sum_{v=1}^{n} [I - i \hat{K} \rho]^{-1}_{uv}
\frac{\beta_{\alpha}g^{(\alpha)}_v}{m^2_{\alpha} - s}
\equiv \frac{\beta_{\alpha}}{m^2_{\alpha} - s} \sum_{v=1}^{n} [I - i \hat{K} \rho]^{-1}_{uv} g^{(\alpha)}_v \,,
\end{equation}
where we need to sum the propagator contributions over the channels $v$, while the SVP production
amplitudes are separated out for each individual $v \rightarrow u$ channel as
\begin{equation}
\label{eq:prodSVP}
{\cal{A}}_{{\rm{SVP}},uv}(s) = \frac{m_0^2 - s_0^{\rm{prod}}}{s - s_0^{\rm{prod}}}
[I - i \hat{K} \rho]^{-1}_{uv} f^{\rm{prod}}_{v} \,.
\end{equation}
We can then sum all of these contributions to give the total S-wave amplitude
\begin{equation}
\label{eq:prodPoleSVP}
{\cal{F}}_u = \sum_{\alpha=1}^{N} {\cal{A}}_{\alpha,u} + \sum_{v=1}^{n}{\cal{A}}_{{\rm{SVP}},uv} \,.
\end{equation}

The elements $\rho_{u}$ of the diagonal phase space matrix depend on the channel type $u$.
For $\pi\pi$ systems, the five available channels are $\pi\pi$, $K\bar{K}$, $4\pi$,
$\eta\eta$ and $\eta\etapr$ multimeson states~\cite{Anisovich:2002ij}; note that
$\etapr\etapr$ is above the open charm threshold and is not considered.
The phase space factor for $\pi\pi$ ($u=1$), $K\bar{K}$ ($u=2$) and $\eta\eta$ ($u=4$) is given by
Eq.~(\ref{eq:rhoterm}), with $m_{1u}$ and $m_{2u}$ equal to the rest masses of the
two pseudoscalars forming channel $u$ ($m_{1u}=m_{2u}$). For $\eta\etapr$ ($u=5$), the second 
multiplicative term involving $m_{\eta} - m_{\etapr}$ is ignored (set to unity), since below threshold
this crosses channels and we cannot continue this analytically in the usual way.
As given in Ref.~\cite{Anisovich:2002ij}, the phase space term for the $4\pi$ multimeson state ($u=3$) is
\begin{equation}
\label{eq:rhoterm2}
\rho_{3}(s) =
\begin{cases}
\rho_{31}(s) & \text{for } s < 1\, \gevgevcccc \\
\sqrt{1 - (16 m^2_{\pi}/s)}  & \text{for } s \geq 1\, \gevgevcccc \,;
\end{cases}
\end{equation}
\begin{equation}
\rho_{31}(s) = \rho_0 \int \frac{ds_1}{\pi} \int \frac{ds_2}{\pi} 
\frac{M^2_0 \Gamma(s_1) \Gamma(s_2) \sqrt{(s + s_1 - s_2)^2 - 4ss_1}}
{s[(M^2_0 - s_1)^2 + M^2_0\Gamma^2(s_1)][(M^2_0 - s_2)^2 + M^2_0\Gamma^2(s_2)]} \,. \nonumber
\end{equation}
Here, $s_1$ and $s_2$ refer to the invariant mass-squared of the two di-pion states
(which are simply considered as integration variables),
$M_0$ is the pole mass of the $\rho$ resonance (775\,\mev) and 
$\Gamma(s) = \Gamma_0 [1 - (4m^2_{\pi}/s)]^{3/2}$ is the energy-dependent width, where 
$\Gamma_0$ is taken to be 0.3\,\gev, which is approximately 75\% of the 
total width of the $f_0(1370) \ra 4\pi$ state~\cite{PDG2016}. The constant 
factor $\rho_0$ ensures continuity at $s = 1\,\gevgevcccc$, while the limits of 
integration are $4m^2_{\pi}$ to $(\sqrt{s} - 2m_{\pi})^2$ for $s_1$ and $4m^2_{\pi}$ to 
$(\sqrt{s} - \sqrt{s_1})^2$ for $s_2$ in order to satisfy kinematic constraints.
The $\rho_{31}$ term needs to be evaluated numerically and is approximated
very well by a 6$^{\rm{th}}$ order polynomial in $s$:
\begin{equation}
\rho_{31}(s) = 1.0789s^6 + 0.1366s^5 -0.2974s^4 - 0.2084s^3 + 0.1385s^2 - 0.0193s + 0.0005 .
\label{eq:rho4pi}
\end{equation}

For $K\pi$ systems, we have the three channels $K\pi$, $K\etapr$ and
$K\pi\pi\pi$ multimeson states~\cite{Anisovich:1997qp}. The phase space factors for the
first two channels ($u$ = 1,2) are again given by Eq.~(\ref{eq:rhoterm}),
while the multimeson phase space element is given by
\begin{equation}
\label{eq:rhoterm3}
\rho_{K\pi\pi\pi}(s) = 
\begin{cases}
r_0\left[1 - ((m_K - 3m_{\pi})/s)\right]^{5/2} & \; \text{for } s < 1.44\, \gevgevcccc \\
1 & \; \text{for } s \geq 1.44\, \gevgevcccc \,,
\end{cases}
\end{equation}
where $r_0$ is a constant of continuity at $s = 1.44\,\gevgevcccc$.

The $K$-matrix formalism is a way to describe the dynamics of a set of broad,
overlapping resonances with the same isospin $I_s$, spin $L$ and parity $P$.
Final states with different $I_sL^{P}$ values would require the appropriate
number of $K$-matrices. To avoid overcomplicating the Dalitz plot analysis,
the usual procedure is to parameterise only the S-wave ($L^{P}$ = 0$^{+}$) 
components with the $K$-matrix approach, and then combine the other 
(narrow) resonances with the isobar model. This means that the total amplitude 
would be given by
\begin{equation}
\label{eq:kmamp}
{\cal A}\left(m^2_{13}, m^2_{23}\right) = \sum_{I_s}{\cal{F}}_{u,I_s}(s) + 
\sum_{j=1, j\neq{\cal{F}}_u}^{N} c_j F_j\left(m^2_{13}, m^2_{23}\right) \,,
\end{equation}
where ${\cal{F}}_{u,I_s}(s)$ is the $K$-matrix amplitude defined in
Eq.~(\ref{eq:KMatProd}) for the channel $u$ and isospin state $I_s$.
The recommended procedure would then be to first use scattering data
to completely define the $K$-matrix elements in Eq.~(\ref{eq:KMatTerms}), such as
using the values quoted in Ref.~\cite{Aubert:2008bd} which are obtained from a 
global analysis of $\pi\pi$ scattering data~\cite{Anisovich:2002ij}.
Subsequently in the DP analysis the user can fit for the coefficients
$\beta_{\alpha}$ and $f^{\rm{prod}}_{v}$ of the $\hat{P}$-vector used in Eqs.~(\ref{eq:PVect}),~(\ref{eq:prodPole}) and~(\ref{eq:prodSVP}).

\subsubsection{Implementation details for $K$-matrix}
\label{sec:kmatrix-imp}

Special commands are required in order to use the $K$-matrix amplitude defined in 
Eqs.~(\ref{eq:KMatProd}) and~(\ref{eq:prodPoleSVP}), which is combined automatically with the other 
isobar resonances to produce the total dynamical amplitude given by Eq.~(\ref{eq:kmamp}).

First, the $(I - i\hat{K}\rho)^{-1}$ propagator term is formed using the \texttt{defineKMatrixPropagator}
function in \texttt{LauIsobarDynamics}, which requires a descriptive name, a text file containing a
keyword-defined list of the scattering and Adler zero coefficients, as well as an integer to 
specify which daughter is the bachelor particle. This function also requires the total 
number $n$ of $K$-matrix scattering channels (sum over $v = 1$ to $n$), the number of bare poles $N$
(sum of $m_{\alpha}$ terms), as well as the final channel index $u$. 
Note that the complete $\hat{K}$ matrix in Eq.~(\ref{eq:KMatTerms}),
which is real and symmetric, is found for all possible values of $u$ and $v$ in order 
to find the propagator; the specific $u$ index is only needed for finding
the final ${\cal{F}}_u$ amplitude. For $\pi\pi$ S-wave, we normally have five channels
($\pi\pi$, $K\bar{K}$, $4\pi$, $\eta\eta$ and $\eta\etapr$ multimeson states), 
five poles and the index $u$ is equal to 1 ($\pi\pi$). The production vector $\hat{P}$ defined in
Eq.~(\ref{eq:PVect}) is then formed using the \texttt{addKMatrixProdPole}
and \texttt{addKMatrixProdSVP} functions of \texttt{LauIsobarDynamics} that create
the $\beta_{\alpha}$ pole term given by Eq.~(\ref{eq:prodPole}) (which internally sums the propagator 
function over the initial channels $v$ owing to the $g_v$ coupling dependence) and the
slowly-varying $f_{v}^{\rm{prod}}$ term given by Eq.~(\ref{eq:prodSVP}), respectively. 
They each require a descriptive name, the name 
of the propagator term defined earlier and the pole or channel integer number (starting from 1).
These functions also accept a boolean \texttt{useProdAdler} to specify if the Adler zero suppression
factor given in Eq.~(\ref{eq:adler}) is also used for the production vector $\hat{P}$; 
by default \texttt{useProdAdler} is set to false.
Additional $K$-matrix amplitudes (e.g. for different isospin settings)
can be included by simply defining additional propagators with unique names together with
their required production terms.

Internally, the $K$-matrix propagator is defined by the \texttt{LauKMatrixPropagator} class,
in which each unique propagator is created using an instance of the \texttt{LauKMatrixPropFactory} factory 
method, while the \texttt{LauKMatrixProdPole} and \texttt{LauKMatrixProdSVP} classes represent the 
production pole and slowly-varying terms, respectively. Since \root\ does not implement complex 
matrices, the $K$-matrix propagator is expanded into real and imaginary parts using
the following method. If $A$, $B$, $C$ and $D$ are real matrices (\texttt{TMatrix} objects), then the 
propagator can be expressed as
\begin{equation}
\label{eq:KMatProp}
(I - i \hat{K} \rho)^{-1} \equiv C + iD \equiv (A + iB)^{-1}\, ,
\end{equation}
where $A$ is equal to $I + \hat{K}$Im($\rho$) and $B$ is $-\hat{K}$Re($\rho$). Both $A$ and $B$ are 
completely determined if we know the real, symmetric $\hat{K}$ matrix and the diagonal phase space 
matrix $\rho$ (which can have imaginary terms if the invariant mass is below threshold).
The real and imaginary propagator terms are then given by
\begin{equation}
C = (A + B A^{-1} B)^{-1}
\hspace{10mm}{\rm and}\hspace{10mm}
D = - A^{-1} B C \, .
\label{eq:KMatProp2}
\end{equation}

\subsubsection{Pedagogical  $K$-matrix plots}
\label{sec:kmatrix-plots}

\begin{figure}[!htb]
\centering
\includegraphics[width=\textwidth]{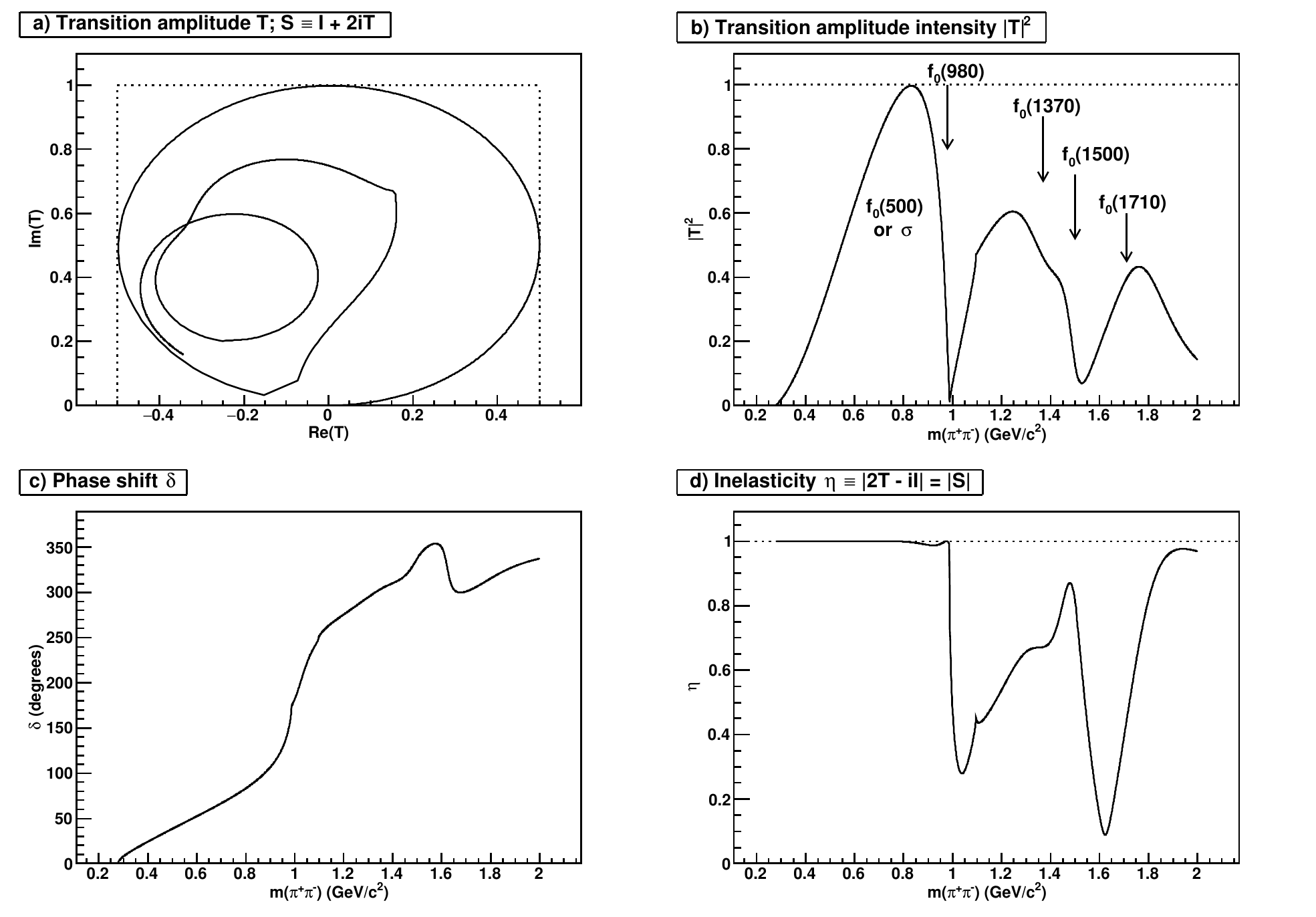}
\caption{\small Plots showing properties of the $\pi\pi \rightarrow \pi\pi$ $K$-matrix S-wave 
transition amplitude, corresponding to the $T_{11}$ matrix element: Argand diagram, intensity 
or amplitude squared (showing the location of various ``resonance structures''), 
phase shift $\delta$ and inelasticity $\eta$.}
\label{fig:KMatrixTAmpPlots}
\end{figure}
\begin{table}[!htb]
\centering
\caption{\small
$\hat{K}$-matrix parameters taken from Ref.~\cite{Aubert:2008bd}, which are obtained from a 
global analysis of $\pi\pi$ scattering data by Anisovich and Sarantsev~\cite{Anisovich:2002ij}. 
Only $f_{1v}$ parameters are listed here ($\pi\pi$ S-wave).
Masses $m_\alpha$ and couplings $g_u^{(\alpha)}$ are given in \gevcc, while units of \gevgevcccc 
for $s$-related quantities are implied; $s^{\rm{prod}}_0$ is taken from Ref.~\cite{Link:2003gb}.}
\begin{tabular}{ccccccc}
\label{table:kmatrixParameters}
$\alpha$ & $m_{\alpha}$ & $g^{(\alpha)}_1 [\pi\pi]$ & $g^{(\alpha)}_2 [K\bar{K}]$
& $g^{(\alpha)}_3 [4\pi]$ & $g^{(\alpha)}_4 [\eta\eta]$ & $g^{(\alpha)}_5 [\eta\eta']$ \\
\hline
$1$ & $0.65100$ & $0.22889$ &           $-0.55377$  &  \phantom{$-$}$0.00000$ &            $-0.39899$ &            $-0.34639$ \\
$2$ & $1.20360$ & $0.94128$ & \phantom{$-$}$0.55095$  &  \phantom{$-$}$0.00000$ &  \phantom{$-$}$0.39065$ &  \phantom{$-$}$0.31503$ \\
$3$ & $1.55817$ & $0.36856$ & \phantom{$-$}$0.23888$  &  \phantom{$-$}$0.55639$ &  \phantom{$-$}$0.18340$ &  \phantom{$-$}$0.18681$ \\
$4$ & $1.21000$ & $0.33650$ & \phantom{$-$}$0.40907$  &  \phantom{$-$}$0.85679$ &  \phantom{$-$}$0.19906$ &            $-0.00984$ \\
$5$ & $1.82206$ & $0.18171$ &           $-0.17558$  &            $-0.79658$ &            $-0.00355$ &  \phantom{$-$}$0.22358$ \\
\hline
\hline
& $s_0^{\rm{scatt}}$ & $f_{11}^{\rm{scatt}}$ & $f_{12}^{\rm{scatt}}$ & $f_{13}^{\rm{scatt}}$ & $f_{14}^{\rm{scatt}}$ & $f_{15}^{\rm{scatt}}$ \\
& $-3.92637$      & $0.23399$ & \phantom{$-$}$0.15044$ & $-0.20545$ & \phantom{$-$}$0.32825$ & \phantom{$-$}$0.35412$ \\
\hline
& $s_0^{\rm{prod}}$ & $m_0^2$ & $s_A$ & $s_{A0}$ & & \\
& $-3.0$          & $1.0$  & $1.0$ & $-0.15$ & &\\
\hline
\end{tabular}
\end{table}
In order to better understand the properties of the $K$-matrix description
we will now show a series of instructional plots. The first of these 
is Fig.~\ref{fig:KMatrixTAmpPlots} which 
shows the transition amplitude of the $\pi\pi \rightarrow \pi\pi$ S-wave, corresponding to the first 
element $T_{11}$ of the $T$ matrix defined in Eq.~(\ref{eq:THat}), using the parameters
given in Table~\ref{table:kmatrixParameters} and where we are not considering the effect of the 
production vector $\hat{P}$. 
Figure~\ref{fig:KMatrixTAmpPlots}a) shows the phase motion of the amplitude, which lies 
within a circle of unit diameter centred on $(0,i/2$), while Fig.~\ref{fig:KMatrixTAmpPlots}b) 
is the equivalent intensity or amplitude squared. First, we can see that the amplitude follows 
the unit circle anticlockwise, corresponding to the very broad $\sigma$ or $f_0(500)$ 
resonance structure, until we reach an invariant mass near to the threshold of the $f_0(980)$ resonance, 
where its interference with the $\sigma$ produces a striking dip in the intensity; the amplitude 
is purely elastic until we reach the $f_0(980)$. As we follow
the phase motion counterclockwise, new channels such as $K\bar{K}$ open up at higher energies, producing
other resonance structures that give extra interference terms and so the scattering process remains
inelastic. A more detailed discussion of these features is given in Ref.~\cite{Zou:1996en}, 
which has a slightly different amplitude intensity distribution at high invariant mass owing to 
different scattering data 
being considered. If we now imagine a vector $\ell$ starting from the centre of the unitarity circle 
$(0,i/2)$ and ending on the position of the complex amplitude $T_{11}$, then the phase shift 
$\delta$ is defined as half of the angle that $\ell$ subtends with the imaginary axis 
(anticlockwise is positive):
\begin{equation}
\label{eq:delta}
\delta \equiv \frac{1}{2}{\rm{tan}}^{-1}\left( \frac{{\rm{Im}}T_{11} - \frac{1}{2}}
{|{\rm{Re}}T_{11}|}\right) + \frac{\pi}{4} \hspace{0.5cm} \rm{radians} ,
\end{equation}
while the inelasticity $\eta$ is defined as twice the length of $\ell$
\begin{equation}
\label{eq:inelasticity}
\eta \equiv 2\left|T_{11} - {\frac{i}{2}}\right| = |S| = 
2\sqrt{({\rm{Im}}T_{11} - {\scriptstyle{\frac{1}{2}}})^2 + ({\rm{Re}}T_{11})^2}.
\end{equation}
Figure~\ref{fig:KMatrixTAmpPlots}c) shows the evolution of $\delta$ with invariant $\pi\pi$ mass,
while Fig.~\ref{fig:KMatrixTAmpPlots}d) shows the variation of the inelasticity $\eta$, where
a purely elastic (inelastic) process has $\eta = 1$ ($\eta = 0$); rapid changes in $\delta$ and $\eta$
are observed at the thresholds of various resonance structures.

\begin{figure}[!htb]
\centering
\includegraphics[width=\textwidth]{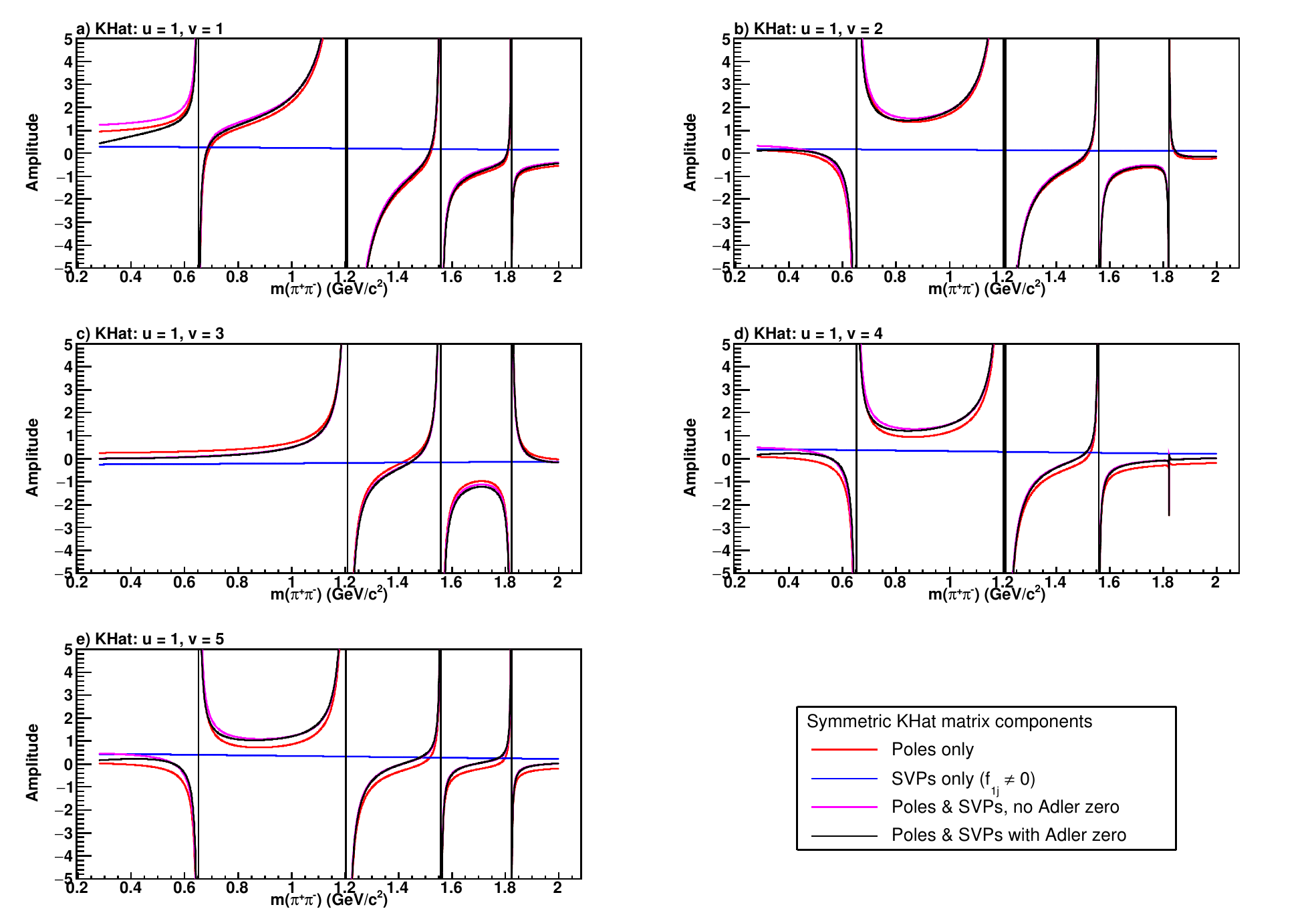}
\caption{\small First row of the $\hat{K}$ matrix, where red (blue) lines show only 
the pole (SVP) terms, while the black (magenta) lines show the full elements 
with (without) the Adler zero factor.}
\label{fig:K_1jPlots}
\end{figure}
We now move onto the $\hat{K}$ matrix itself, which is the main ingredient of the scattering propagator.
Figure~\ref{fig:K_1jPlots} shows the first row of the $\hat{K}$ matrix ($\hat{K}_{1j}$), 
where we have split up the various components that make up each matrix element. 
The red lines show only the bare pole $m_{\alpha}$ contributions, given by the first 
summation term on the right hand side of Eq.~(\ref{eq:KMatTerms}), where $u=1$, $v = 1-5$ 
and we sum over all five poles ($\alpha = 1-5$), and
the Adler zero suppression factor $f_{A0}(s)$ is set to unity.
All of the plots show the strong effect of the bare pole singularities.
The blue lines show the rather small SVP contributions, corresponding to the second term on the 
right hand side of Eq.~(\ref{eq:KMatTerms}) (with $f_{A0}(s)=1$). The summation of these 
various pole and SVP  contributions is given by the magenta lines, while the black lines show the 
inclusion of the Adler zero suppression factor of Eq.~(\ref{eq:adler}), which only significantly effects the 
overall shapes at low invariant mass ($m \lesssim 0.5\,\gevcc$). 
The absence of singularities for $\hat{K}_{13}$ below 
$1.2\,\gevcc$ is due to the fact that the coupling of the first two bare poles 
in the $4\pi$ channel is zero. Similar distributions are obtained for the other rows of the
$\hat{K}$ matrix.

\begin{figure}[!htb]
\centering
\includegraphics[width=\textwidth]{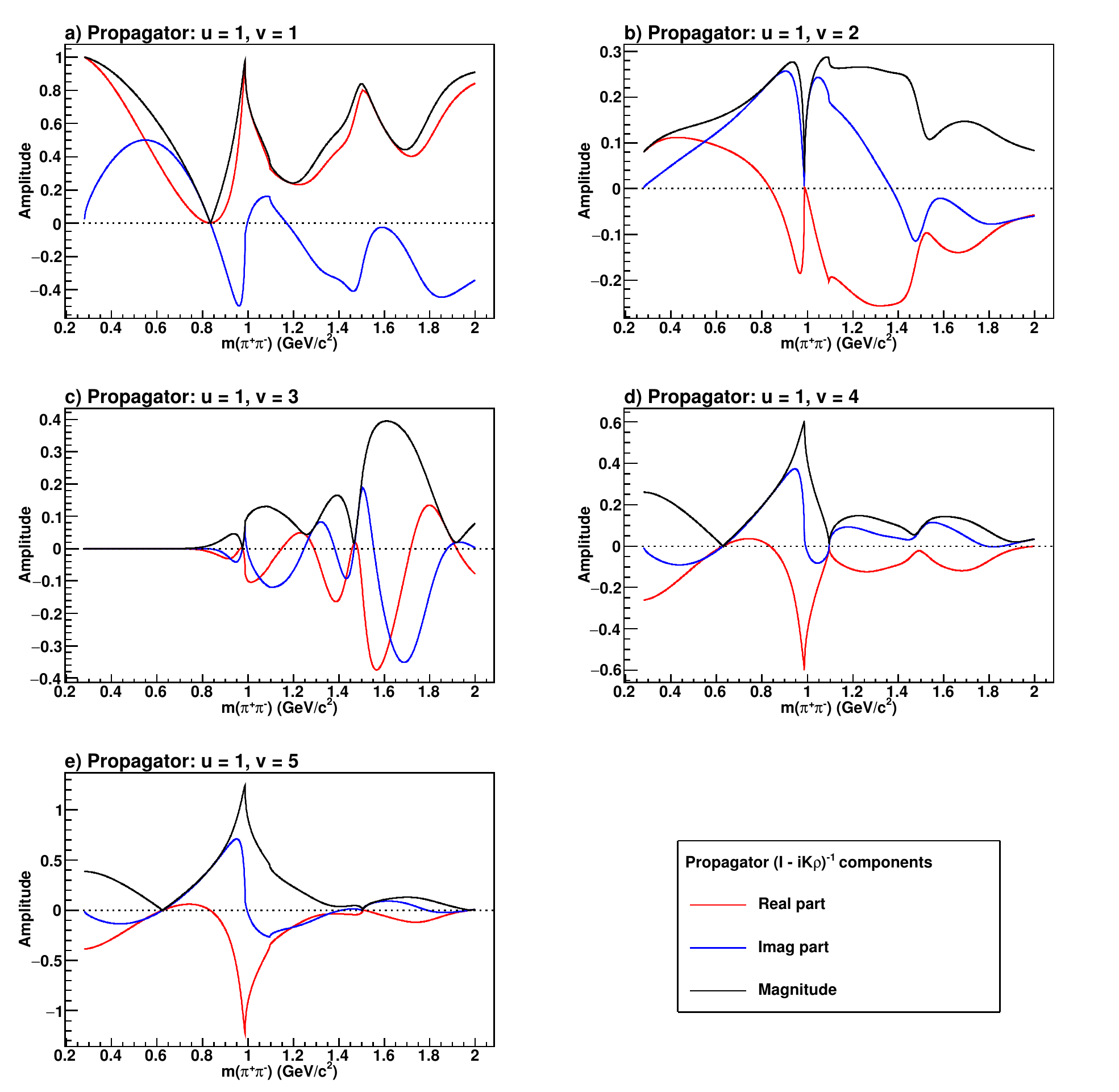}
\caption{\small Complex amplitude components of the first row of the 
propagator $[I - i\hat{K}\rho]^{-1}_{1v}$ ($v \rightarrow \pi\pi$ channels), where red (blue)
lines show the real (imaginary) components while the black curves show the magnitudes.
The dotted horizontal lines denote the zero amplitude level. Channels are a) $\pi\pi$, 
b) $K\bar{K}$, c) $4\pi$, d) $\eta\eta$ and e) $\eta\etapr$.}
\label{fig:PropAmp_0}
\end{figure}
Next, let us look at the $s$-dependence of the first row of the propagator matrix 
$[I - i\hat{K}\rho]^{-1}_{1v}$, since this will effectively modulate the individual
 production pole and SVP shapes that are combined to form the total $\pi\pi$ S-wave 
amplitude ${\cal{F}}_1(s)$ using Eq.~(\ref{eq:KMatProd}).
Figure~\ref{fig:PropAmp_0} shows the real and imaginary components, as well as the magnitude, of the 
propagator elements. The overall impression we get is that the propagator amplitudes have 
non-trivial variations as a function of $\sqrt{s}$, owing to the matrix inversion process 
mixing and transposing the superposition of the bare pole states $m_{\alpha}$. These poles
essentially produce the various cusps and peaks in the propagator amplitude, where
the channel couplings $g_{u,v}^{(\alpha)}$ and phase space $\rho_{u,v}$ (mass threshold) weighting factors 
shift and distort these features away from the original $m_{\alpha}$ values. 
The corresponding Argand 
diagrams show similar behaviour as the transition amplitude 
$T_{11}$ shown in Fig.~\ref{fig:KMatrixTAmpPlots}a), although in general they exhibit distortions due to the 
channel-dependent couplings and mass thresholds. In particular, the $\pi\pi \rightarrow \pi\pi$ propagator 
($u=1,v=1$) exactly matches the phase motion of $T_{11}$ if we first rotate $T_{11}$ by 90 degrees anticlockwise 
($\delta$ = 45 degrees) around the centre of the unitarity circle at $(0,\frac{i}{2})$ and then shift 
it by the translation $(\frac{1}{2}, -\frac{i}{2})$.

Discussing the features in Fig.~\ref{fig:PropAmp_0} in detail, 
the first pole ($m_1 = 0.651$\,\gevcc) produces the first cusps around 0.65--0.8\,\gevcc in all of the channels, 
except for $4\pi$ ($v=3$) which has a coupling of zero. 
The second pole ($m_2 = 1.2036$\,\gevcc) produces cusps at 1, 1.1 and 1.5\,\gevcc for the 
$K\bar{K}$ $(v=2)$, $\eta\eta$ $(v=4)$ and $\eta\eta'$ $(v=5)$ channels, respectively, and
also generates a very broad dip centred near 1.2\,\gevcc for the $\pi\pi$ $(v=1)$ channel. 
The third pole ($m_3 = 1.55817$\,\gevcc) generates broad peaks near 1.55\,\gevcc, where the
low mass end terminates in a cusp at the threshold of the given channel, except for the
$\eta\eta'$ mode where a narrow cusp at 1.5\,\gevcc is generated against a smoothly 
varying amplitude. The fourth pole ($m_4 = 1.21$\,\gevcc) creates structures very similar to the second 
pole owing to their almost degenerate mass values, with an additional broad peak around 1.1--1.3\,\gevcc 
present for the $4\pi$ channel, while the fifth pole ($m_5 = 1.82206$\,\gevcc) generates the broad peaks near 
1.8\,\gevcc for all channels, with additional cusps at 1 ($K\bar{K}$), 1.1 ($\eta\eta$) and 
1.5 \,\gevcc ($\eta\eta'$) that are very similar to those found for the second and fourth poles.

\begin{figure}[!htb]
\centering
\includegraphics[width=\textwidth]{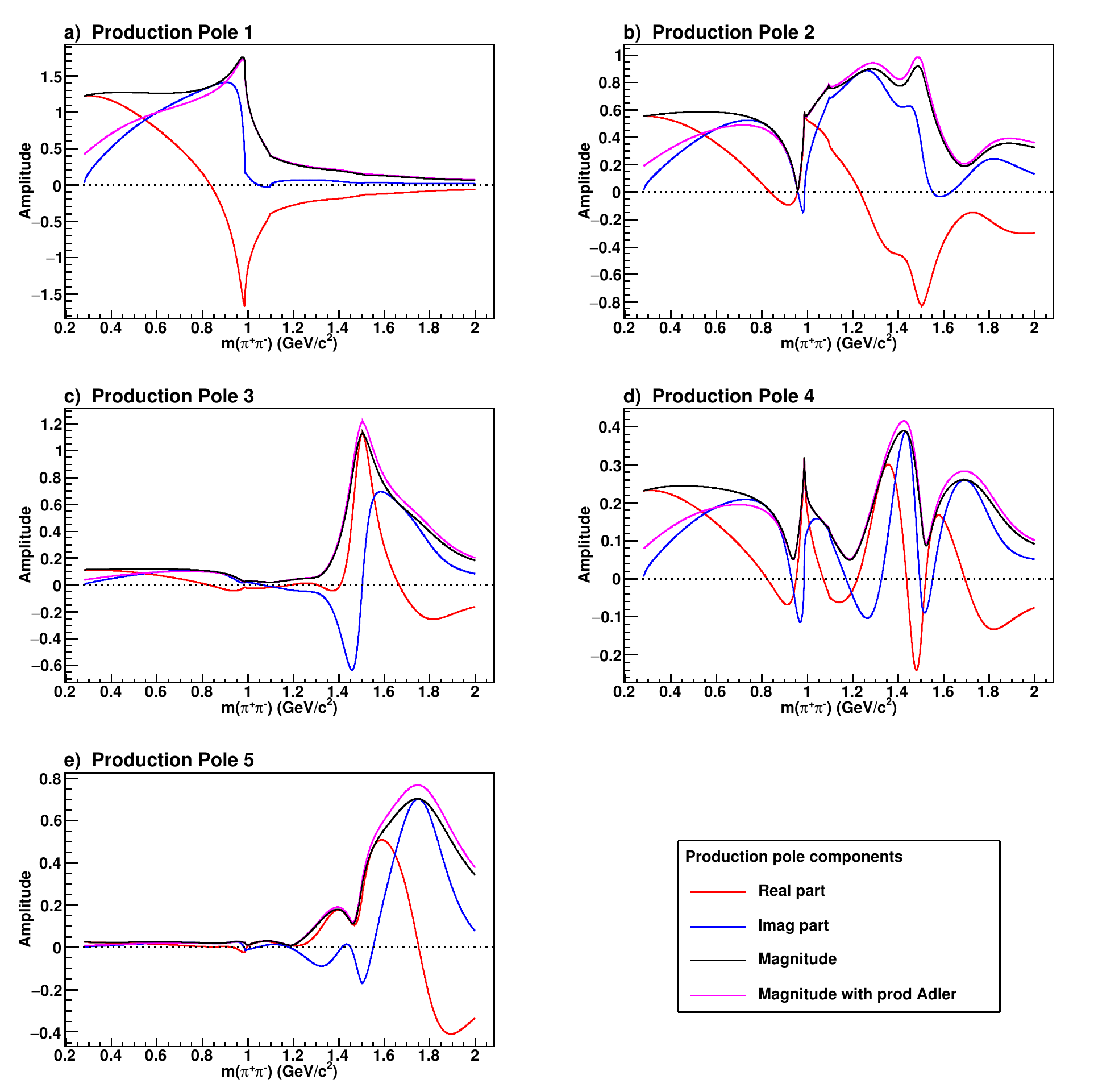}
\caption{\small Complex amplitude components for the production poles defined in Eq.~(\ref{eq:prodPole}), 
where the red (blue) lines show the real (imaginary) parts, while the black lines show the magnitude. 
The magenta lines show what happens to the magnitude when it is scaled by the Adler zero suppression factor
$f_{A0}(s)$. The dotted horizontal lines denote the zero amplitude level.}
\label{fig:ProdPoleAmp}
\end{figure}
Figure~\ref{fig:ProdPoleAmp} shows the pole production amplitudes ${\cal{A}}_{\alpha,u=1}(s)$ 
defined in Eq.~(\ref{eq:prodPole}), which are formed by modulating the pole singularity term
$1/(m^2_{\alpha} - s)$ with a weighted sum of the $s$-dependent propagator distributions for all channels $v=1$ to $5$ 
shown in Fig.~\ref{fig:PropAmp_0}, along with $\beta_{\alpha} \equiv 1$. Concentrating on the magnitudes, we can see that 
the first pole ($m_1 = 0.651$\,\gevcc) has an amplitude that begins rather flat from the $\pi\pi$ threshold
until it starts to peak near 1\,\gevcc before rapidly falling at higher invariant mass. Even though the
$\pi\pi$ propagator ($\pi\pi \rightarrow \pi\pi$) is slowly decreasing from unity at the $\pi\pi$ threshold down 
to zero near 0.8\,\gevcc, the presence of both the rapid rise of the amplitude from the pole 
singularity at 0.65\,\gevcc as well as the increasingly influential $K\bar{K}$ propagator 
($K\bar{K} \rightarrow \pi\pi$), owing to its larger coupling constant, ensures that the amplitude
remains fairly constant in the region below 1\,\gevcc. The propagators for all of the channels (except $4\pi$)
have a peak near 1\,\gevcc, and these combine to give the same local peak for the first production pole.
As we increase the invariant mass, the falling shape of the pole singularity starts to dominate the amplitude
modulation, and so we get a rapid reduction in the magnitude no matter what shapes the propagators have.
The amplitude for the second production pole ($m_2 = 1.2036$\,\gevcc) strongly depends upon the $\pi\pi$ 
and $K\bar{K}$ propagator shapes, where the former has a coupling constant almost double that of the latter. 
As we decrease the invariant mass from $1.2$\,\gevcc, the pole amplitude would be very small at 
the strong $K\bar{K}$ dip at 1\,\gevcc if not for the compensating sharp peak in the $\pi\pi$ propagator.
Likewise, the zero $\pi\pi$ propagator amplitude at 0.8\,\gevcc is nullified by the non-zero 
$K\bar{K}$ contribution. These two effects conspire to shift the location of the sharp dip in
the production pole amplitude by 50\,\mevcc from 1 to 0.95\,\gevcc. As we decrease the invariant mass, the 
pole amplitude becomes more influenced by the rising $\pi\pi$ propagator until it starts to fall as we move
further away from the pole mass. Above 1.2\,\gevcc the production amplitude tends to follow the undulations
of the $\pi\pi$ and $K\bar{K}$ propagators, producing broad local peaks centred on 1.3 and 1.9\,\gevcc
as well as a more narrow one at 1.5\,\gevcc. The modulation of the third pole ($m_3 = 1.55817$\,\gevcc) 
amplitude is dominated by the $4\pi$ channel, although the other channels give significant 
contributions, varying from 33\% ($\eta\eta$) up to 66\% $(\pi\pi$). The $4\pi$ propagator has a very broad
maximum centred very close to the pole mass, and so we would expect the production pole peak to remain very close to
1.56\,\gevcc (with an asymmetric width that is slightly narrower on the low side). However, as we decrease the 
invariant mass, the rising contributions from the $\pi\pi$ and $K\bar{K}$ propagators effectively shift the 
production peak by 60\,\mevcc down to 1.5\,\gevcc. Below 1.2\,\gevcc, the modulations from the $\pi\pi$ and 
$K\bar{K}$ propagators become washed out since they are too far away from the pole position. Above 1.5\,\gevcc, 
the width of the production amplitude remains very wide owing to the dominant $4\pi$ propagator.
The fourth pole located at 1.21\,\gevcc is almost degenerate with the second pole (1.2036\,\gevcc) and so we 
would naively expect them to have essentially identical production shapes. However, the coupling coefficients 
are completely different, where now the $4\pi$ channel propagator dominates, with a factor of two or more 
reduction in the other contributions, leading to the production of the two broad peaks centred around 
1.4 and 1.7\,\gevcc. For invariant masses at 1\,\gevcc and below, the amplitude does indeed closely follow 
the shape of the second production pole since the $4\pi$ propagator becomes negligible and the $\pi\pi$ 
and $K\bar{K}$ contributions dominate. The fifth and last pole ($m_2 = 1.82206$\,\gevcc) has an amplitude
that is strongly influenced by the $4\pi$ channel, with much smaller contributions from the others.
The large, broad $4\pi$ propagator maximum near 1.6\,\gevcc shifts the production peak by around 70\,\mevcc down to 
1.75\,\gevcc, while the other smaller production peaks near 1.4 and 1.1\,\gevcc match those seen in the $4\pi$ shape.
The magenta lines in Fig.~\ref{fig:ProdPoleAmp} show the effect of multiplying the Adler zero suppression 
factor $f_{A0}(s)$ to the production pole amplitudes, where we can see that it only reduces the magnitudes
for invariant masses below 1\,\gevcc.

The mass distributions of the SVP production amplitudes ${\cal{A}}_{{\rm{SVP}},u=1,v}(s)$ defined in Eq.~(\ref{eq:prodSVP})
follows very closely the shape modulations of the propagator terms shown in Fig.~\ref{fig:PropAmp_0} 
(when $f_v^{\rm{prod}} \equiv 1$), since each SVP is simply equal to the propagator multiplied by the
common function $4/(s+3)$, obtained using $m_0^2 = 1.0$\,\gevgevcccc and 
$s_0^{\rm{prod}} = -3.0$\,\gevgevcccc, which enhances (suppresses) features at low (high) invariant 
mass.

\begin{figure}[!htb]
\centering
\includegraphics[width=0.49\textwidth]{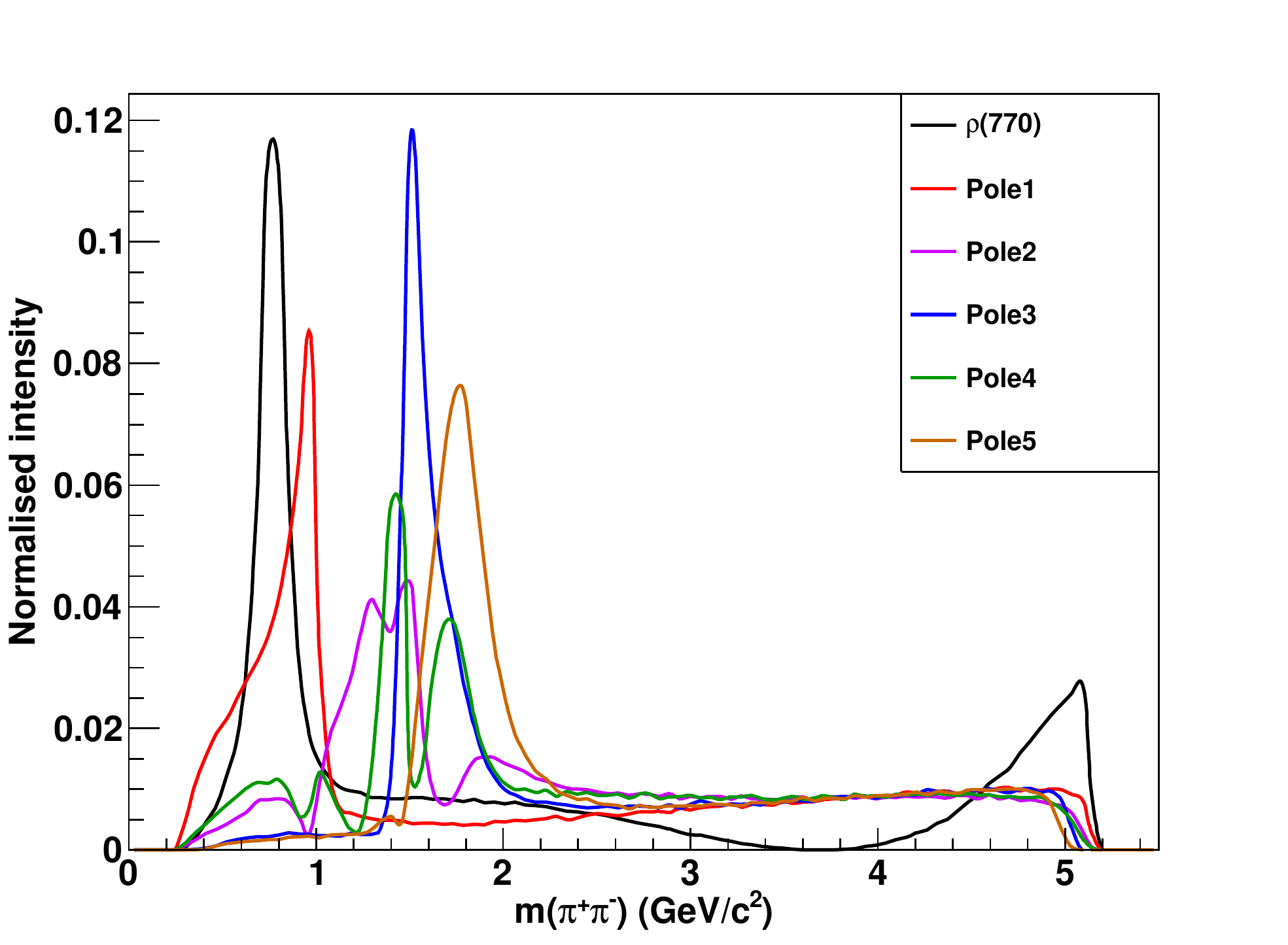}
\includegraphics[width=0.49\textwidth]{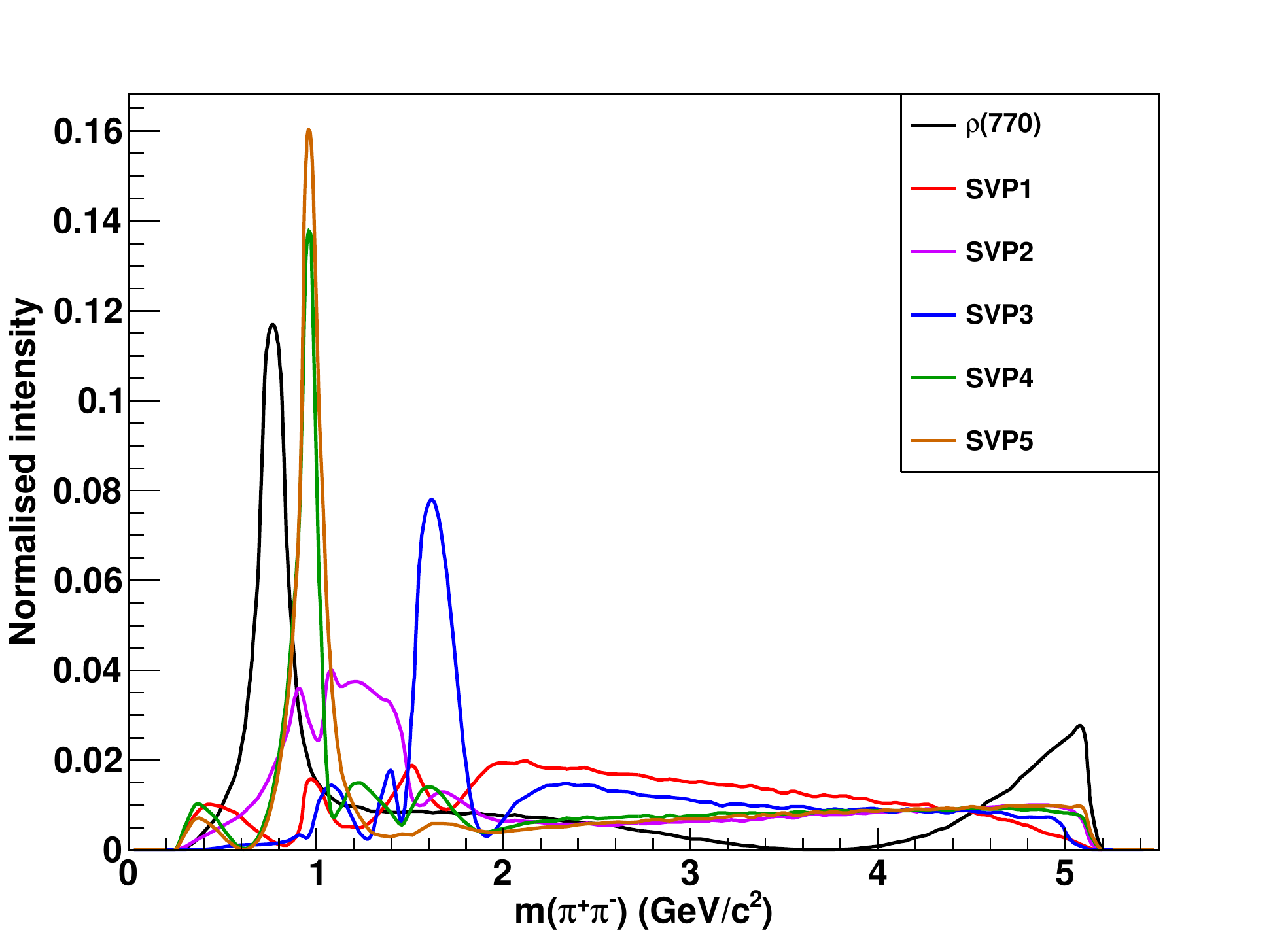}
\caption{\small Mass projections of the individual production pole (left) and SVP (right) amplitudes, with
the $\rho(770)$ resonance shown for comparison.}
\label{fig:B3piKMProj}
\end{figure}
The final set of pedagogical plots are given in Fig.~\ref{fig:B3piKMProj}, which 
show the normalised mass projections of the individual production pole and SVP amplitudes, as well
as the $\rho(770)$ resonance for comparison, using events generated uniformly across 
the Dalitz plot. In general, we see that the peaking structures observed in Fig.~\ref{fig:ProdPoleAmp} 
are replicated here, with a reduction in the intensity as the invariant mass
approaches the $\pi\pi$ threshold (fewer events are generated at the kinematic boundary), with an almost
flat intensity seen for invariant masses above 2\,\gevcc, which is the effective cut-off for
the $K$-matrix parameterisation owing to the fact that there are no bare poles defined in this region.
The first production pole generates the peak corresponding to the $f_0(980)$ along with a broad
shoulder on the low mass side, which can be referred to as the $f_0(500)$ or $\sigma$ resonance. The third pole
generates the $f_0(1500)$ peak, the fifth pole is the main contributor for the $f_0(1710)$, while a combination 
of the second and fourth poles (which have almost degenerate bare masses) produces peaks in the $f_0(1370)$ region.
All of these peaks are generated dynamically by the $K$-matrix amplitude.
The first two SVPs have rather oscillatory shapes in the region around 1 to 2\,\gevcc; when they are combined
we can approximately obtain a very broad bump between the $f_0(980)$ and $f_0(1370)$ peak locations. The third
SVP generates a large peak very near 1.6\,\gevcc, in between the $f_0(1500)$ and $f_0(1710)$ regions. Lastly,
the fourth and fifth SVPs are essentially degenerate, peaking at the same position of the $f_0(980)$
from the first production pole. This means that we can ignore these two contributions, or at least
remove the fifth SVP, since nothing is gained by their inclusion in the total amplitude description.


\setcounter{table}{0}
\renewcommand{\thetable}{B\arabic{table}}

\section{Formulae for available angular distributions}
\label{sec:angular-formulae}

The angular distributions and Blatt--Weisskopf form factors set out in Sec.~\ref{sec:angular} are the default settings in \laura.
However, other formalisms to describe the angular distributions are also implemented in the package and it is straightforward to switch between them.
This appendix details these alternative formalisms and illustrates the few additional lines of code required to use them.

The four spin-factor formalisms are defined in the enumeration
\texttt{LauAbsResonance::LauSpinType}, which can take the values
\texttt{Zemach\_P} (the default setting), \texttt{Zemach\_Pstar},
\texttt{Covariant}, and \texttt{Legendre}.
The simplest description of the spin factors is that of the \texttt{Legendre}
formalism, where the spin factors are simply the Legendre polynomials
(with some additional numerical constants in order to maintain consistency of
the phase conventions among the various formalisms)
\begin{eqnarray}
\label{eq:Legendre-TFactors}
L = 0 \ : \ T(\vec{p},\vec{q}) & = & \phantom{-}\,1\,,\\
L = 1 \ : \ T(\vec{p},\vec{q}) & = & -\,2\,\cos{\theta}\,,\\
L = 2 \ : \ T(\vec{p},\vec{q}) & = & \phantom{-}\,\frac{4}{3} \left[3\cos^2{\theta} - 1\right]\,,\\
L = 3 \ : \ T(\vec{p},\vec{q}) & = & -\,\frac{24}{15} \left[5\cos^3{\theta} - 3\cos{\theta}\right]\,,\\
L = 4 \ : \ T(\vec{p},\vec{q}) & = & \phantom{-}\,\frac{16}{35} \left[35\cos^4{\theta} - 30\cos^2{\theta} + 3\right]\,,\\
L = 5 \ : \ T(\vec{p},\vec{q}) & = & -\,\frac{32}{63} \left[63\cos^5{\theta} - 70\cos^3{\theta} + 15\cos{\theta}\right]\,.
\label{eq:Legendre-TFactors-end}
\end{eqnarray}
The spin factors for \texttt{Zemach\_P} are those given in Eqs.~(\ref{eq:ZTFactors})--(\ref{eq:ZTFactors-end}), which differ from the expressions of Eqs.~(\ref{eq:Legendre-TFactors})--(\ref{eq:Legendre-TFactors-end}) by factors of $(p\,q)^L$.
Similarly, those for \texttt{Zemach\_Pstar} are the same as those for
\texttt{Zemach\_P} but with the bachelor momentum evaluated in the rest frame
of the parent particle~($p^{\ast}$), rather than that of the resonance~($p$).
The angular distributions have been implemented in \laura\ up to $L=5$, which is two units larger than the maximum spin of any resonance observed to be produced in any Dalitz plot to date~\cite{Aaij:2014xza,Aaij:2014baa,Aaij:2015sqa}.

\begin{figure}[!htb]
\centering
\includegraphics[width=0.49\textwidth]{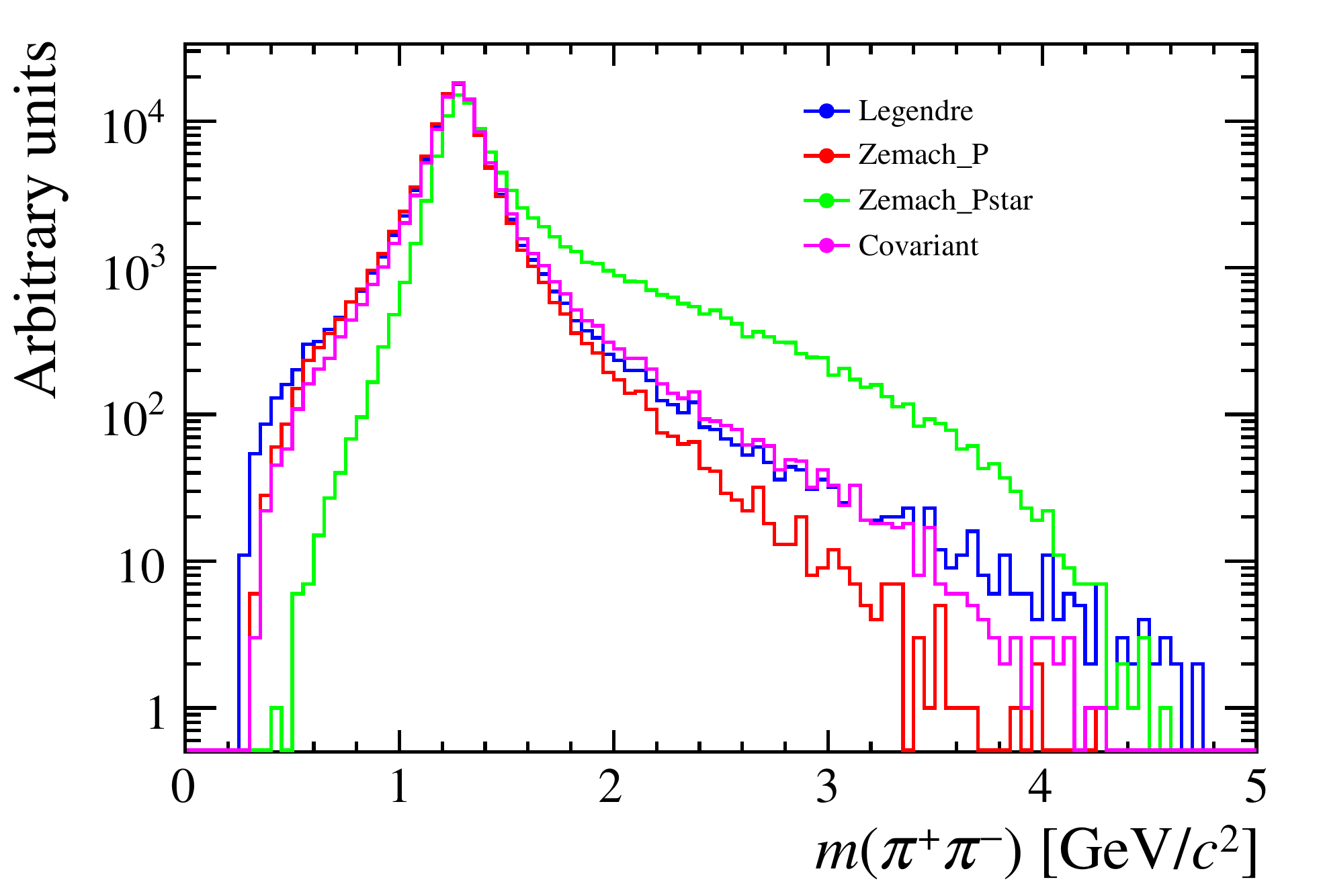}
\includegraphics[width=0.49\textwidth]{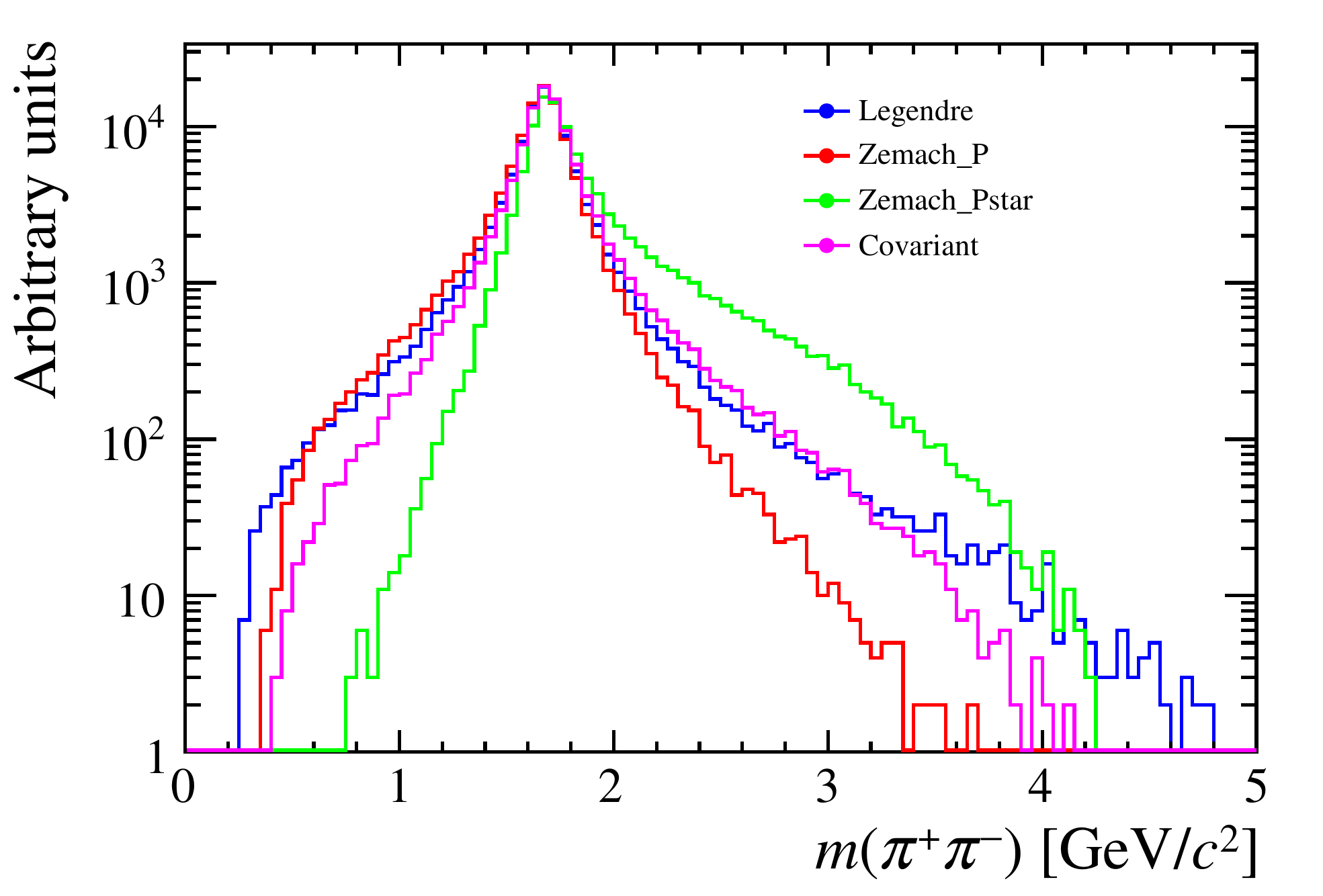}
\caption{
  Lineshapes for the (left) $f_2(1270)$ and (right) $\rho_3(1690)^0$
  resonances decaying to $\pip\pim$ (in the $\Bp\to\Kp\pip\pim$ Dalitz
  plot) with the (blue) \texttt{Legendre}, (red) \texttt{Zemach\_P},
  (green) \texttt{Zemach\_Pstar} and (magenta) \texttt{Covariant} spin
  formalisms.
  In all cases the relativistic Breit--Wigner description is used, with
  mass and width parameters as given in App.~\ref{sec:resNames} and the
  two Blatt--Weisskopf factors set to unity.
}
\label{fig:angular-formulae}
\end{figure}

The angular distributions discussed above are based on a non-relativistic assumption.  
For certain channels, this may not be sufficiently precise, and therefore the \texttt{Covariant} formalism is also made available.
This is given by
\begin{eqnarray}
\label{eq:Covariant-TFactors}
L = 0 \ : \ T(\vec{p},\vec{q}) & = & \phantom{-}\,1\,,\\
L = 1 \ : \ T(\vec{p},\vec{q}) & = & -\,2\,(p^{\ast}q)\sqrt{1+\frac{p^2}{m^2_P}}\,\cos{\theta}\,,\\
L = 2 \ : \ T(\vec{p},\vec{q}) & = & \phantom{-}\,\frac{4}{3}\,(p^{\ast}q)^2\left(\frac{3}{2}+\frac{p^2}{m^2_P}\right)\left[3\cos^2{\theta} - 1\right]\,,\\
L = 3 \ : \ T(\vec{p},\vec{q}) & = & -\,\frac{24}{15}\,(p^{\ast}q)^3\sqrt{1+\frac{p^2}{m^2_P}}\left(\frac{5}{2}+\frac{p^2}{m^2_P}\right)\left[5\cos^3{\theta} - 3\cos{\theta}\right]\,,\\
L = 4 \ : \ T(\vec{p},\vec{q}) & = & \phantom{-}\,\frac{16}{35}\,(p^{\ast}q)^4\left(\frac{8p^4}{m^4_P}+\frac{40p^2}{m^2_P}+35\right)\left[35\cos^4{\theta} - 30\cos^2{\theta} + 3\right]\,.
\label{eq:Covariant-TFactors-end}
\end{eqnarray}
The first three of these expressions are derived in
Ref.~\cite{Filippini:1995yc} and, based on that work, the last two were
derived in Ref.~\cite{Aaij:2015sqa}.
As can be seen from the expressions, the differences between formalisms are more significant for higher spin resonances, and particularly affect tails of the distributions.  
To give an idea of the effect, the lineshapes for the $f_2(1270)$ and $\rho_3(1690)^0$ resonances decaying to $\pip\pim$ are shown in Fig.~\ref{fig:angular-formulae}.

It is possible to switch between these different formalisms via a function of
the \texttt{LauResonanceMaker} factory object.
For example, to use the \texttt{Covariant} formalism one would do:
\begin{lstlisting}
LauResonanceMaker& resMaker = LauResonanceMaker::get();
resMaker.setSpinFormalism( LauAbsResonance::Covariant );
\end{lstlisting}
It is important to note that any such operation must be performed prior to
constructing any resonances, \ie before calling
\texttt{LauIsobarDynamics::addResonance} or
\texttt{LauIsobarDynamics::addIncoherentResonance} for the first time.

As the angular and Blatt--Weisskopf factors are strongly coupled, it is also
possible to straightforwardly modify the form of the Blatt--Weisskopf factors.
In particular, the momentum value used for the factor that is related to the
decay of the parent particle into the resonance and the bachelor can be
selected from the following options (defined in the
\texttt{LauBlattWeisskopfFactor::RestFrame} enumeration):
\begin{itemize}
\item
\texttt{LauBlattWeisskopfFactor::ResonanceFrame}, the momentum of the bachelor in the rest frame of the resonance, $p$ (the default setting),
\item
\texttt{LauBlattWeisskopfFactor::ParentFrame}, the momentum of the bachelor in the rest frame of the parent, $p^{\ast}$,
\item
\texttt{LauBlattWeisskopfFactor::Covariant}, the product of the momentum of the bachelor in the rest frame of the parent, $p^{\ast}$,
and a function of the ratio of the energy and mass of the resonance in the
rest frame of the parent, $\sqrt{1 + p^2/m^2_P}$.  More precisely, this
function is the expression in the middle term in
Eqs.~(\ref{eq:Covariant-TFactors}) to~(\ref{eq:Covariant-TFactors-end})
raised to the power of $1/L$.
\end{itemize}
This setting is changed as follows:
\begin{lstlisting}
LauResonanceMaker& resMaker = LauResonanceMaker::get();
resMaker.setBWBachelorRestFrame( LauBlattWeisskopfFactor::ParentFrame );
\end{lstlisting}
where in this example the momentum of the bachelor in the rest frame of the parent ($p^{\ast}$) is to be used.
Again, this operation must be performed before constructing any resonances.

In addition, it is possible to change the form of the Blatt--Weisskopf factors,
with the different types being defined by the \texttt{LauBlattWeisskopfFactor::BarrierType} enumeration.
The default setting, corresponding to Eqs.~(\ref{eq:BWFormFactors})--(\ref{eq:BWFormFactors-end}),
is given by \texttt{LauBlattWeisskopfFactor::BWPrimeBarrier} and is recommended when the angular terms contain momentum factors.
One possible alternative is to use the \texttt{LauAbsResonance::Legendre} angular terms and the \texttt{LauBlattWeisskopfFactor::BWBarrier} form for the Blatt--Weisskopf factors:
\begin{eqnarray}
\label{eq:BWFormFactors-NonPrimed}
L = 0 \ : \ X(z) & = & 1\,, \\
L = 1 \ : \ X(z) & = & \sqrt{\frac{2z^2}{1 + z^2}}\,, \\
L = 2 \ : \ X(z) & = & \sqrt{\frac{13z^4}{z^4 + 3z^2 + 9}}\,,\\
L = 3 \ : \ X(z) & = & \sqrt{\frac{277z^6}{z^6 + 6z^4 + 45z^2 + 225}}\,,\\
L = 4 \ : \ X(z) & = & \sqrt{\frac{12746z^8}{z^8 + 10z^6 + 135z^4 + 1575z^2 + 11025}}\,,\\
L = 5 \ : \ X(z) & = & \sqrt{\frac{998881z^{10}}{z^{10} + 15z^8 + 315z^6 + 6300z^4 + 99225z^2 + 893025}}\,.
\label{eq:BWFormFactors-NonPrimed-end}
\end{eqnarray}
An exponential form for these factors,
\texttt{LauBlattWeisskopfFactor::ExpBarrier},
which has been used in some analyses for virtual contributions has also been implemented,
\begin{equation}
X(z) = e^{-z^L}\,.
\end{equation}
To change the form of the barrier factors for all resonances, the following lines are required
\begin{lstlisting}
LauResonanceMaker& resMaker = LauResonanceMaker::get();
resMaker.setBWType( LauBlattWeisskopfFactor::BWBarrier );
\end{lstlisting}
where in this example the forms in Eqs.~(\ref{eq:BWFormFactors-NonPrimed})--(\ref{eq:BWFormFactors-NonPrimed-end})
are to be used.
Again, this operation should be performed before constructing any resonances.

\begin{figure}[!htb]
\centering
\includegraphics[width=0.49\textwidth]{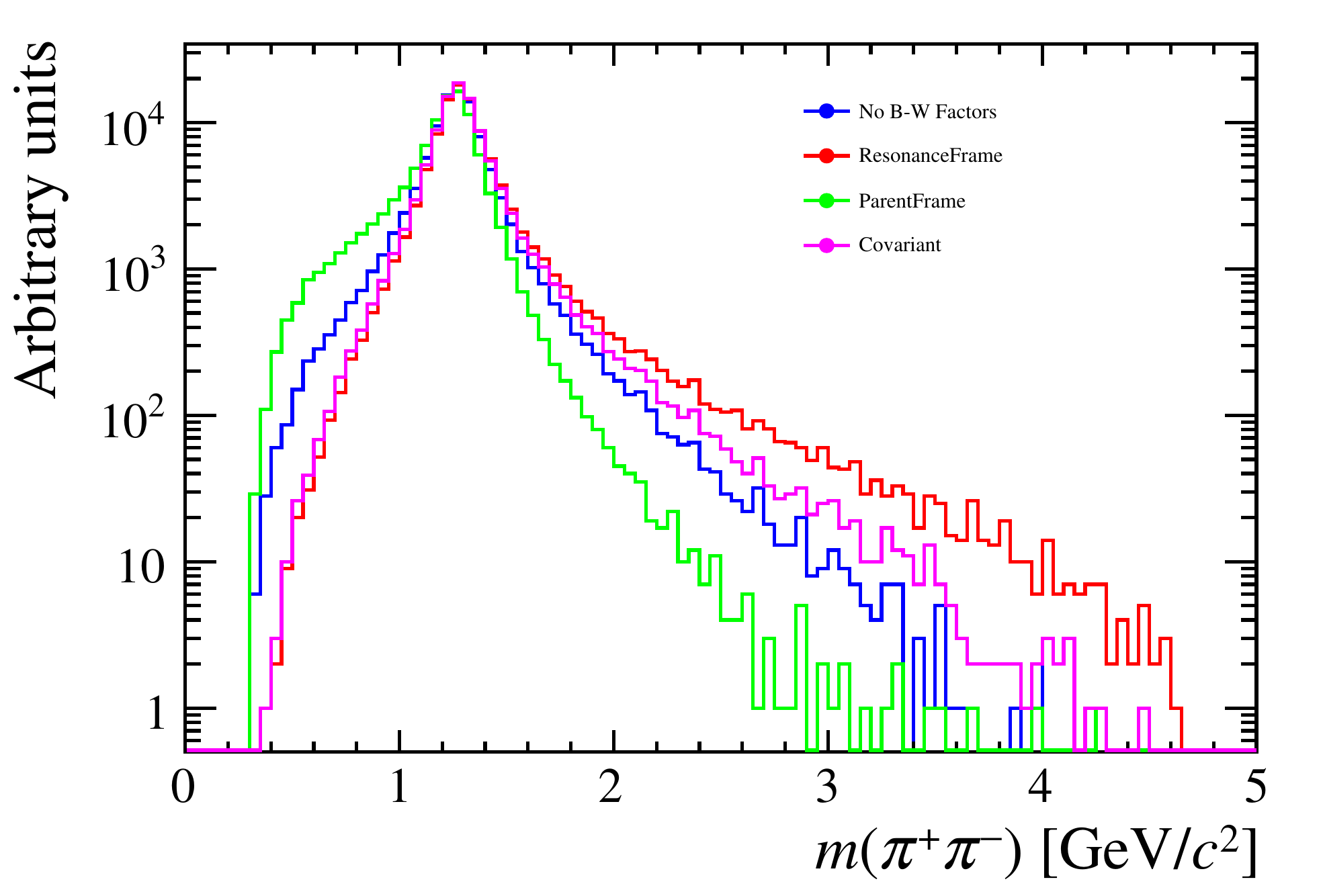}
\includegraphics[width=0.49\textwidth]{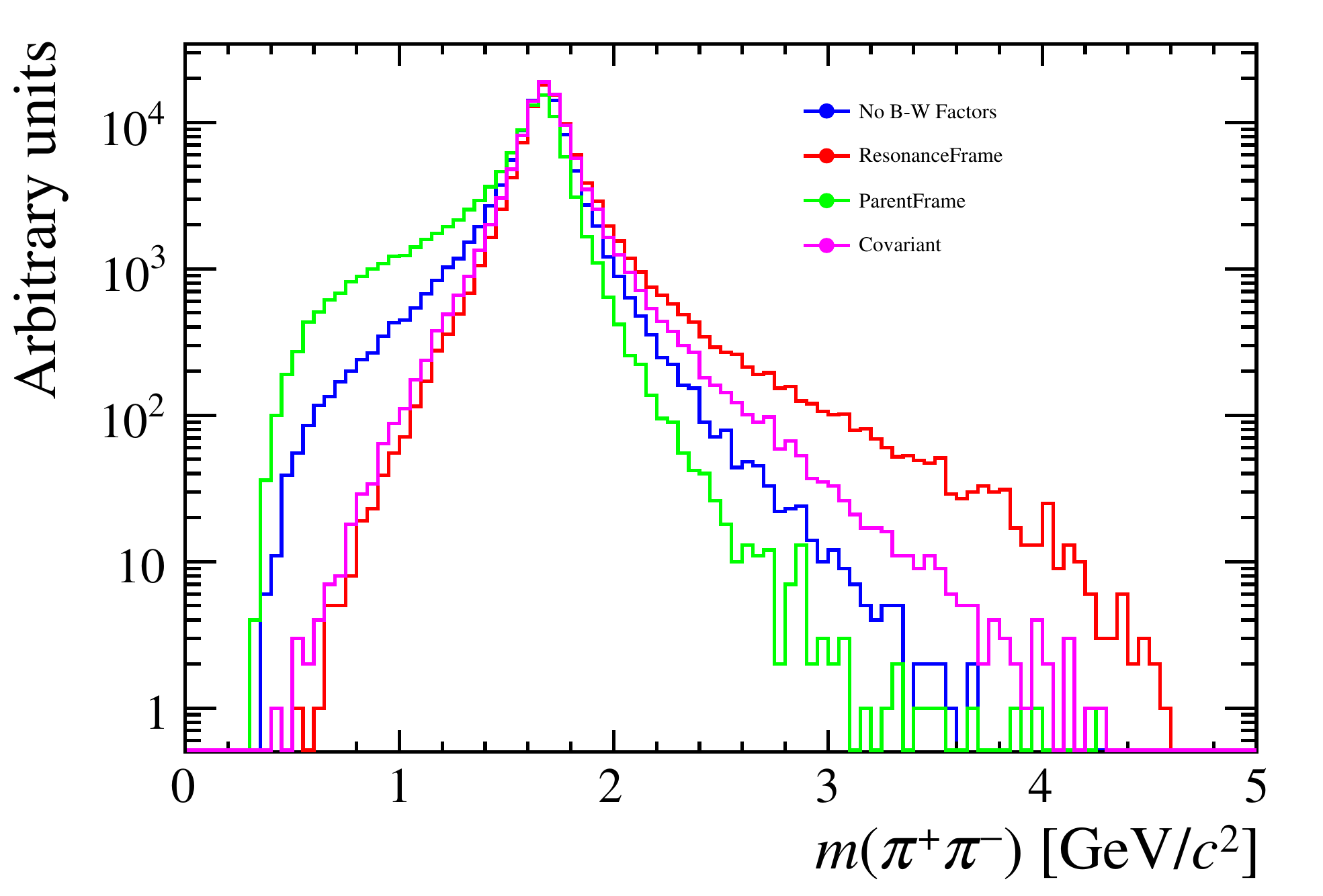}
\caption{
  Lineshapes for the (left) $f_2(1270)$ and (right) $\rho_3(1690)^0$
  resonances decaying to $\pip\pim$ (in the $\Bp\to\Kp\pip\pim$ Dalitz
  plot) with (blue) no Blatt--Weisskopf factors, and with the (red)
  \texttt{ResonanceFrame}, (green) \texttt{ParentFrame} and (magenta)
  \texttt{Covariant} settings for evaluating the momentum that enters the
  Blatt--Weisskopf factor associated with the decay of the parent particle.
  In all cases the relativistic Breit--Wigner description is used, with
  mass and width parameters as given in App.~\ref{sec:resNames} and the
  \texttt{Zemach\_P} formalism for the spin factors.
}
\label{fig:BWFF-formulae}
\end{figure}

As for the $T(\vec{p},\vec{q})$ terms, the differences between Blatt--Weisskopf form factor formalisms are more significant for higher spin resonances, and far from the peak of the resonance.  
An illustrative comparison of the shapes is given in Fig.~\ref{fig:BWFF-formulae}.

It is possible to make all of the changes discussed in this Appendix at the
level of individual resonances, using the functions
\texttt{LauAbsResonance::setSpinType} and
\texttt{LauAbsResonance::setBarrierRadii},
but this requires much care to be taken and is not generally recommended.


\setcounter{table}{0}
\renewcommand{\thetable}{C\arabic{table}}

\section{Standard resonances}
\label{sec:resNames}

This section provides the complete set of available resonances, indicating the
name, mass $m_0$, width $\Gamma_0$, spin, charge and Blatt--Weisskopf barrier radius $r^R_{\rm BW}$.
Table~\ref{tab:resNames1} contains information for light meson resonances, 
Table~\ref{tab:resNames2} for charm, charmonium, strange-charm, beauty and strange-beauty resonances,
Table~\ref{tab:resNames3} for $K^*$ resonances and
Table~\ref{tab:resNames4} for nonresonant terms.
Most data are taken from Ref.~\cite{PDG2016}.
The tables list the information contained in the information records for both neutral and positively-charged resonances.
Negatively-charged resonance records are implemented as charge-conjugates of
the positively charged ones; the plus sign in the name is replaced with a minus sign.

In case a user wishes to modify the values of the parameters from those given in the tables, the {\tt LauAbsResonance::changeResonance} function, which takes the mass, width and spin as arguments, can be used.
The Blatt--Weisskopf barrier radius can be changed with the {\tt LauAbsResonance::changeBWBarrierRadii} function, and other parameters specific to particular lineshapes can be changed with the {\tt LauAbsResonance::setResonanceParameter} function.
The same approach can be used to include a resonance that is not available in these tables, by using any of the existing states of appropriate charge and redefining its properties.

\begin{table}[!htb]
\caption{Standard light meson resonances defined in \laura.}
\label{tab:resNames1}
\centering
\begin{tabular}{lllccc}
\hline \\ [-2.5ex]
Name   & $m_0$ (\nbspgevcc) & $\Gamma_0$ (\nbspgevcc) & spin & charge & $r^R_{\rm BW}$ ($\nbspgev^{-1}$) \\ [0.1ex]
\hline
rho0(770)     & 0.77526  & 0.1478   & 1 & 0 & 5.3 \\
rho+(770)     & 0.77511  & 0.1491   & 1 & 1 & 5.3 \\
rho0(1450)    & 1.465    & 0.400    & 1 & 0 & 4.0 \\  
rho+(1450)    & 1.465    & 0.400    & 1 & 1 & 4.0 \\        
rho0(1700)    & 1.720    & 0.250    & 1 & 0 & 4.0 \\        
rho+(1700)    & 1.720    & 0.250    & 1 & 1 & 4.0 \\ 
rho0(1900)    & 1.909    & 0.130    & 1 & 0 & 4.0 \\        
rho+(1900)    & 1.909    & 0.130    & 1 & 1 & 4.0 \\ 
rho0\_3(1690) & 1.686    & 0.186    & 3 & 0 & 4.0 \\
rho+\_3(1690) & 1.686    & 0.186    & 3 & 1 & 4.0 \\
rho0\_3(1990) & 1.982    & 0.188    & 3 & 0 & 4.0 \\
rho+\_3(1990) & 1.982    & 0.188    & 3 & 1 & 4.0 \\
phi(1020)     & 1.019461 & 0.004266 & 1 & 0 & 4.0 \\      
phi(1680)     & 1.680    & 0.150    & 1 & 0 & 4.0 \\      
f\_0(980)     & 0.990    & 0.070    & 0 & 0 & --- \\      
f\_2(1270)    & 1.2751   & 0.1851   & 2 & 0 & 4.0 \\      
f\_0(1370)    & 1.370    & 0.350    & 0 & 0 & --- \\      
f'\_0(1300)   & 1.449    & 0.126    & 0 & 0 & --- \\      
f\_2(1430)    & 1.430    & 0.150    & 2 & 0 & 4.0 \\      
f\_0(1500)    & 1.505    & 0.109    & 0 & 0 & --- \\      
f'\_2(1525)   & 1.525    & 0.073    & 2 & 0 & 4.0 \\      
f\_2(1565)    & 1.562    & 0.134    & 2 & 0 & 4.0 \\      
f\_2(1640)    & 1.639    & 0.099    & 2 & 0 & 4.0 \\      
f\_0(1710)    & 1.722    & 0.135    & 0 & 0 & --- \\      
f\_2(1810)    & 1.816    & 0.197    & 2 & 0 & 4.0 \\      
f\_2(1910)    & 1.903    & 0.196    & 2 & 0 & 4.0 \\      
f\_2(1950)    & 1.944    & 0.472    & 2 & 0 & 4.0 \\      
f\_2(2010)    & 2.011    & 0.202    & 2 & 0 & 4.0 \\      
f\_0(2020)    & 1.992    & 0.442    & 0 & 0 & --- \\      
f\_4(2050)    & 2.018    & 0.237    & 4 & 0 & 4.0 \\      
f\_0(2100)    & 2.101    & 0.224    & 0 & 0 & --- \\      
omega(782)    & 0.78265  & 0.00849  & 1 & 0 & 4.0 \\      
a0\_0(980)    & 0.980    & 0.092    & 0 & 0 & --- \\      
a+\_0(980)    & 0.980    & 0.092    & 0 & 1 & --- \\      
a0\_0(1450)   & 1.474    & 0.265    & 0 & 0 & --- \\      
a+\_0(1450)   & 1.474    & 0.265    & 0 & 1 & --- \\      
a0\_2(1320)   & 1.3190   & 0.1050   & 2 & 0 & 4.0 \\      
a+\_2(1320)   & 1.3190   & 0.1050   & 2 & 1 & 4.0 \\      
sigma0        & 0.475    & 0.550    & 0 & 0 & --- \\        
sigma+        & 0.475    & 0.550    & 0 & 1 & --- \\      
\hline
\end{tabular}
\end{table}

\begin{table}[!htb]
\caption{Standard charm, charmonium, strange-charm, beauty and strange-beauty resonances defined in \laura.}
\label{tab:resNames2}
\centering
\begin{tabular}{lllccc}
\hline \\ [-2.5ex]
Name   & $m_0$ (\nbspgevcc) & $\Gamma_0$ (\nbspgevcc) & spin & charge & $r^R_{\rm BW}$ ($\nbspgev^{-1}$) \\ [0.1ex]
\hline
chi\_c0       & 3.41475 & 0.0105  & 0 & 0 & --- \\   
chi\_c1       & 3.51066 & 0.00084 & 1 & 0 & 4.0 \\   
chi\_c2       & 3.55620 & 0.00193 & 2 & 0 & 4.0 \\   
X(3872)       & 3.87169 & 0.0012  & 1 & 0 & 4.0 \\   
dabba0        & 2.098   & 0.520   & 0 & 0 & --- \\        
dabba+        & 2.098   & 0.520   & 0 & 1 & --- \\        
D*0           & 2.00696 & 0.0021  & 1 & 0 & 4.0 \\        
D*+           & 2.01026 & $83.4\times10^{-6}$ & 1 & 1 & 4.0 \\        
D*0\_0        & 2.318   & 0.267   & 0 & 0 & --- \\   
D*+\_0        & 2.403   & 0.283   & 0 & 1 & --- \\   
D*0\_2        & 2.4626  & 0.049   & 2 & 0 & 4.0 \\   
D*+\_2        & 2.4643  & 0.037   & 2 & 1 & 4.0 \\   
D0\_1(2420)   & 2.4214  & 0.0274  & 1 & 0 & 4.0 \\   
D+\_1(2420)   & 2.4232  & 0.025   & 1 & 1 & 4.0 \\   
D0(2600)      & 2.612   & 0.093   & 0 & 0 & --- \\   
D+(2600)      & 2.612   & 0.093   & 0 & 1 & --- \\   
D0(2760)      & 2.761   & 0.063   & 1 & 0 & 4.0 \\   
D+(2760)      & 2.761   & 0.063   & 1 & 1 & 4.0 \\   
D0(3000)      & 3.0     & 0.15    & 0 & 0 & --- \\   
D0(3400)      & 3.4     & 0.15    & 0 & 0 & --- \\   
Ds*+          & 2.1121  & 0.0019  & 1 & 1 & 4.0 \\   
Ds*+\_0(2317) & 2.3177  & 0.0038  & 0 & 1 & --- \\   
Ds*+\_2(2573) & 2.5719  & 0.017   & 2 & 1 & 4.0 \\   
Ds*+\_1(2700) & 2.709   & 0.117   & 1 & 1 & 4.0 \\   
Ds*+\_1(2860) & 2.862   & 0.180   & 1 & 1 & 4.0 \\   
Ds*+\_3(2860) & 2.862   & 0.058   & 3 & 1 & 4.0 \\   
B*0           & 5.3252  & 0.00    & 1 & 0 & 6.0 \\
B*+           & 5.3252  & 0.00    & 1 & 1 & 6.0 \\
Bs*0          & 5.4154  & 0.00    & 1 & 0 & 6.0 \\
\hline
\end{tabular}
\end{table}

\begin{table}[!htb]
\caption{Standard $K^*$ resonances defined in \laura.}
\label{tab:resNames3}
\centering
\begin{tabular}{lllccc}
\hline \\ [-2.5ex]
Name   & $m_0$ (\nbspgevcc) & $\Gamma_0$ (\nbspgevcc) & spin & charge & $r^R_{\rm BW}$ ($\nbspgev^{-1}$) \\ [0.1ex]
\hline
K*0(892)     & 0.89581 & 0.0474 & 1 & 0 & 3.0 \\ 
K*+(892)     & 0.89166 & 0.0508 & 1 & 1 & 3.0 \\
K*0(1410)    & 1.414   & 0.232  & 1 & 0 & 4.0 \\      
K*+(1410)    & 1.414   & 0.232  & 1 & 1 & 4.0 \\      
K*0\_0(1430) & 1.425   & 0.270  & 0 & 0 & --- \\      
K*+\_0(1430) & 1.425   & 0.270  & 0 & 1 & --- \\      
K*0\_2(1430) & 1.4324  & 0.109  & 2 & 0 & 4.0 \\      
K*+\_2(1430) & 1.4256  & 0.0985 & 2 & 1 & 4.0 \\      
K*0(1680)    & 1.717   & 0.322  & 1 & 0 & 4.0 \\      
K*+(1680)    & 1.717   & 0.322  & 1 & 1 & 4.0 \\      
K*0\_0(1950) & 1.945   & 0.201  & 0 & 0 & --- \\      
K*+\_0(1950) & 1.945   & 0.201  & 0 & 1 & --- \\      
kappa0       & 0.682   & 0.547  & 0 & 0 & --- \\        
kappa+       & 0.682   & 0.547  & 0 & 1 & --- \\        
\hline
\end{tabular}
\end{table}

\begin{table}[!hbt]
\caption{Standard nonresonant terms defined in \laura.}
\label{tab:resNames4}
\centering
\begin{tabular}{lllcc}
\hline \\ [-2.5ex]
Name   & $m_0$ (\nbspgevcc) & $\Gamma_0$ (\nbspgevcc) & spin & charge \\ [0.1ex]
\hline
NonReson        & 0.0 & 0.0 & 0 & 0 \\   
NRModel         & 0.0 & 0.0 & 0 & 0 \\   
BelleSymNR      & 0.0 & 0.0 & 0 & 0 \\   
BelleNR         & 0.0 & 0.0 & 0 & 0 \\   
BelleNR+        & 0.0 & 0.0 & 0 & 1 \\   
BelleNR\_Swave  & 0.0 & 0.0 & 0 & 0 \\   
BelleNR\_Swave+ & 0.0 & 0.0 & 0 & 1 \\ 
BelleNR\_Pwave  & 0.0 & 0.0 & 1 & 0 \\   
BelleNR\_Pwave+ & 0.0 & 0.0 & 1 & 1 \\   
BelleNR\_Dwave  & 0.0 & 0.0 & 2 & 0 \\   
BelleNR\_Dwave+ & 0.0 & 0.0 & 2 & 1 \\   
BelleNR\_Fwave  & 0.0 & 0.0 & 3 & 0 \\   
BelleNR\_Fwave+ & 0.0 & 0.0 & 3 & 1 \\   
NRTaylor        & 0.0 & 0.0 & 0 & 0 \\   
PolNR\_S0       & 0.0 & 0.0 & 0 & 0 \\   
PolNR\_S1       & 0.0 & 0.0 & 0 & 0 \\   
PolNR\_S2       & 0.0 & 0.0 & 0 & 0 \\   
PolNR\_P0       & 0.0 & 0.0 & 1 & 0 \\   
PolNR\_P1       & 0.0 & 0.0 & 1 & 0 \\   
PolNR\_P2       & 0.0 & 0.0 & 1 & 0 \\   
\hline
\end{tabular}
\end{table}

\clearpage

\setcounter{table}{0}
\renewcommand{\thetable}{D\arabic{table}}

\section{PDF classes}
\label{sec:pdfs}

This section details the formulae within classes that can
be used to parameterise additional PDFs ${\cal P}(x; p_1, p_2,...,p_n)$
for the likelihood function.
Here, $x$ denotes the dependent variable, while $p_1,p_2,...,p_n$ is 
the list of parameters in the form of a vector of \texttt{LauParameter} 
objects, each containing a descriptive name (which must contain the 
case-sensitive word shown in quotes), the value of the parameter and 
optionally its validity range, uncertainty and constantness.
All PDFs used in \laura\ are normalised to unity, although the normalisation
factors are omitted in many equations in this section for brevity.

\subsection{LauArgusPdf}
\label{sec:argus-pdf}

The ARGUS threshold function~\cite{argus} can be used to parameterise the shape
of combinatorial or partially-reconstructed backgrounds of the invariant
mass $m$ of parent candidates:
\begin{equation}
\label{eq:argus}
{\cal P}(x; m_0, \xi) = x \sqrt{1 - x^2} e^{-\xi (1 - x^2)} \theta(x) \,,
\end{equation}
where $x = m/m_0$, $m_0$ is the end-point of the curve (``m0''), $\xi$ is the shape
parameter (``xi''), while $\theta(x \leq m_0) = 1$ and $\theta(x > m_0) = 0$.

\subsection{LauBifurcatedGaussPdf}

This PDF is the bifurcation of two Gaussians having different
widths, $\sigma_L$ (``sigmaL'') or $\sigma_R$ (``sigmaR''), to the left or right of the
``mean'' $\mu$:
\begin{equation}
\label{eq:bifgauss}
{\cal P}(x; \mu, \sigma_L, \sigma_R) = 
\begin{cases}
e^{-(x-\mu)^2/(2\sigma_L^2)} & \text{for } x \leq \mu \\
e^{-(x-\mu)^2/(2\sigma_R^2)} & \text{for } x > \mu \,.
\end{cases}
\end{equation}

\subsection{LauChebychevPdf}

This class implements a sum of Chebyshev polynomials of the first kind $T_i$ 
(up to seventh order):
\begin{equation}
\label{eq:chebyshev}
{\cal P}(x; c_i) = 1 + \sum_{i=1}^{n \leq 7} c_i T_i(x) \,,
\end{equation}
where $c_i$ is the parameter coefficient (``c1'', ``c2'', etc.) and
\begin{equation}
\label{eq:chebyshev2}
T_n(y) = \frac{(y - \sqrt{y^2 - 1})^n + (y + \sqrt{y^2 -1})^n}{2} \,,
\hspace{0.5cm} y \equiv -1 + \frac{2(x - x_{\rm{min}})}{(x_{\rm{max}} - x_{\rm{min}})} \,.
\end{equation}

\subsection{LauCruijffPdf}

The Cruijff PDF is a bifurcated Gaussian, which has different widths
$\sigma_L$ (``sigmaL'') and $\sigma_R$ (``sigmaR'') to the left and right of the ``mean''
$\mu$, along with asymmetric tails $\alpha_L$ (``alphaL'') and $\alpha_R$ (``alphaR''):
\begin{equation}
\label{eq:cruijff}
{\cal P}(x; \mu, \sigma_L, \sigma_R, \alpha_L, \alpha_R) =
\begin{cases}
e^{-(x-\mu)^2/(2\sigma_L^2 + \alpha_L(x-\mu)^2)} & \text{for } x \leq \mu \\
e^{-(x-\mu)^2/(2\sigma_R^2 + \alpha_R(x-\mu)^2)} & \text{for } x > \mu \,.
\end{cases}
\end{equation}

\subsection{LauCrystalBallPdf}

The Crystal Ball function~\cite{Gaiser:1982yw} is a PDF that contains a Gaussian core with a 
continuous power-law tail on one side:
\begin{equation}
\label{eq:crystalball}
{\cal P}(x; \mu, \sigma, \alpha, n) =
\begin{cases}
e^{-\frac{1}{2}t^2} & \text{for } t \geq |\alpha| \\
\left(\frac{n}{|\alpha|}\right)^n e^{-\frac{1}{2}|\alpha|^2} \left(\frac{n}{|\alpha|} - |\alpha| - t \right)^{-n}
& \text{otherwise} \,,
\end{cases}
\end{equation}
where $\mu$ and $\sigma$ are the Gaussian ``mean'' and width (``sigma''), respectively, 
$\alpha$ (``alpha'') is the positive or negative 
distance from the mean in which the Gaussian and the tail 
parts match up, $n$ (``order'') is the power exponent for the tail,
while $t$ is equal to $(x-\mu)/\sigma$, which changes sign if $\alpha$ is negative.

\subsection{LauExponentialPdf}

This exponential function is simply given by
\begin{equation}
\label{eq:exp}
{\cal P}(x; \lambda) = e^{\lambda x} \,,
\end{equation}
where $\lambda$ is the ``slope'' parameter.

\subsection{LauGaussPdf}

The Gaussian PDF is defined by a ``mean'' $\mu$ and width $\sigma$ (``sigma''):
\begin{equation}
\label{eq:gauss}
{\cal P}(x; \mu, \sigma) = \frac{1}{\sigma \sqrt{2 \pi}} e^{-(x - \mu)^2/(2 \sigma^2)} \,.
\end{equation}

\subsection{LauLinearPdf}

This linear function only needs the gradient (``slope'') $\lambda$:
\begin{equation}
\label{eq:linear}
{\cal P}(x; \lambda) = \lambda x + c \,,
\end{equation}
where $c$ is the intercept and is evaluated using the range of the abscissa $x$.

\subsection{LauNovosibirskPdf}

The Novosibirsk PDF is a Gaussian with a logarithmic exponent~\cite{Ikeda2000401}:
\begin{equation}
\label{eq:novo}
{\cal P}(x; \mu, \sigma, \tau) = 
\text{exp}\left[-\frac{1}{2}
\left(\frac{\text{ln}^2[1 + \Lambda \tau (x - \mu)]}{\tau^2} + \tau^2 \right) \right]\,,
\hspace{0.25cm} \text{with } \Lambda \equiv \frac{\text{sinh }(\tau \sqrt{\text{ln }4})}{\sigma \tau \sqrt{\text{ln }4}} \,,
\end{equation}
where $\mu$ and $\sigma$ are the usual ``mean'' and width (``sigma'') values,
respectively, and $\tau$ is the ``tail'' parameter; as $\tau \rightarrow 0$, the PDF
converges to a normal Gaussian with width $\sigma$.

\subsection{LauParametricStepFuncPdf}

This parametric step function is a binned distribution whose parameters are
the contents of each bin (except one), essentially representing a histogram
with variable bin content.  The content of the remaining bin is determined
from that of the others and the requirement of normalisation.
The constructor requires two vectors of \texttt{LauParameter} objects; the
first stores the bin contents or weights (which are parameters that can be
fitted), while the second stores the lower edge abscissa limits of each bin
(in ascending order) as well as the upper edge limit of the last bin.
It can also be specified whether the normalisation bin is the first or
last; it is advisable to use the bin with the larger content.

\subsection{LauSigmoidPdf}

The sigmoid function follows an ``S'' shape and is defined as
\begin{equation}
\label{eq:sigmoid}
{\cal P}(x; a, b) = \frac{1}{1 + e^{b - ax}} \,,
\end{equation}
with parameters ``a'' and ``b'' defining the steepness of the slope
and the shift of the distribution, respectively. A negative value
of $a$ will flip the distribution around the ordinate axis.

\subsection{LauSumPdf}

This class implements the summation of two PDFs, ${\cal P}_1$ 
and ${\cal P}_2$, with a relative fraction (``frac'') $f$:
\begin{equation}
\label{eq:pdfsum}
{\cal P}(x) = f {\cal P}_1(x) + (1 - f){\cal P}_2(x) \,.
\end{equation}

\subsection{Dalitz-plot-dependent PDFs}
\label{sec:pdfs-DPdep}

What follows are descriptions of PDFs with parameters that can vary across the Dalitz plot via
power-law scaling with Laurent polynomials containing either positive or negative exponents.
Here, the parameterisation of a given PDF parameter $g$ is given by
\begin{equation}
\label{eq:dppdfpar}
g = \sum_{i=1}^{n}{c_i D^i} \hspace{1cm} \text{or }  \hspace{1cm} g = \sum_{i=1}^{n}{c_i D^{-i}} \,,
\end{equation}
where $D$ is the invariant mass-squared variable representing the DP position
and $c_i$ is the coefficient of expansion for the power term $i$. The variable $D$ can be either
$m^2_{13}$, $m^2_{23}$, or $m^2_{12}$, the minimum or maximum values of $m^2_{13}$ or $m^2_{23}$,
or the distance from the DP centre. The constructors of these PDFs require
vectors of the coefficients $c_i$ for each function parameter $g$, as well as 
the pointer to the \texttt{LauDaughters} object in order to find $D$ from the kinematics.
The \texttt{LauDPDepBifurGaussPdf, LauDPDepCruijffPdf} and \texttt{LauDPDepGaussPdf} classes 
represent bifurcated Gaussian, Cruijff and normal Gaussian PDFs, given in
Eqs.~(\ref{eq:bifgauss}),~(\ref{eq:cruijff}) and~(\ref{eq:gauss}) respectively, with parameters that can vary according to Eq.~(\ref{eq:dppdfpar}). 

The \texttt{LauDPDepMapPdf} class can be used to define
a PDF that requires different functions depending on the DP region.
It uses a \root\ histogram that divides the DP, or its projection onto one axis,
into ascending, numbered regions. The region number (starting from zero) for a given value 
of $D$ is then used to choose the corresponding PDF from the ordered list of functions 
provided as a vector in the constructor. 

The \texttt{LauDPDepSumPdf} class implements the sum of two PDFs,
defined by Eq.~(\ref{eq:pdfsum}), in which the fraction $f$ (``frac'') depends on the 
variable $D$, using either the contents from a two-dimensional histogram directly 
or a vector of coefficients $c_i$ for the positive-power Laurent polynomial 
shown in Eq.~(\ref{eq:dppdfpar}).


\clearpage
\addcontentsline{toc}{section}{References}
\setboolean{inbibliography}{true}
\bibliographystyle{LHCb}
\bibliography{references}

\end{document}